\documentclass[superscriptaddress,pre,onecolumn,preprintnumbers,amsmath,amssymb]{revtex4-1}
\usepackage{graphicx}
\usepackage{epstopdf, epsfig}
\usepackage{amsmath}
\usepackage{array}
\usepackage{multirow}
\usepackage{psfrag}
\usepackage[font=footnotesize,format=plain,labelsep=period,justification=raggedright]{caption}
\usepackage{subcaption}

\usepackage[utf8]{inputenc}
\usepackage{combelow}
\usepackage{amssymb}
\usepackage{braket}
\usepackage{bm}
\usepackage{color}

\def\feq{\ensuremath{f^{(\mathrm{eq})}}}

\def\hata{{\hat{a}}}
\def\hatb{{\hat{b}}}
\def\hatc{{\hat{c}}}
\def\hatd{{\hat{d}}}
\def\hvarphi{{\hat{\varphi}}}
\def\htheta{{\hat{\theta}}}

\begin{document}

\title{Multicomponent Flow on Curved Surfaces: A Vielbein Lattice Boltzmann Approach}
\author{Victor E. Ambru\cb{s}}
\affiliation{Department of Physics, West University of Timi\cb{s}oara, 300223 Timi\cb{s}oara, Romania}
\email{victor.ambrus@e-uvt.ro}
\author{Sergiu Busuioc}
\affiliation{Department of Physics, West University of Timi\cb{s}oara, 300223 Timi\cb{s}oara, Romania}
\author{Alexander J. Wagner}
\affiliation{Department of Physics, North Dakota State University, Fargo, North Dakota 58108, USA}
\author{Fabien Paillusson}
\affiliation{School of Mathematics and Physics, University of Lincoln, Lincoln LN6 7TS, UK}
\author{Halim Kusumaatmaja}
\email{halim.kusumaatmaja@durham.ac.uk}
\affiliation{Department of Physics, Durham University, Durham, DH1 3LE, UK}

\begin{abstract}
We develop and implement a novel lattice Boltzmann scheme to study multicomponent flows
on curved surfaces, coupling the continuity and Navier-Stokes equations with the Cahn-Hilliard
equation to track the evolution of the binary fluid interfaces. 
Standard lattice Boltzmann method relies on regular Cartesian grids, 
which makes it generally unsuitable to study flow problems on curved surfaces. 
To alleviate this limitation, 
we use a vielbein formalism to write down the Boltzmann equation on an arbitrary geometry, 
and solve the evolution of the fluid distribution functions using a finite difference method.
Focussing on the torus geometry as an example of a curved surface, we demonstrate
drift motions of fluid droplets and stripes embedded on the surface of a torus. Interestingly,
they migrate in opposite directions: fluid droplets to the outer side while fluid stripes 
to the inner side of the torus. 
For the latter we demonstrate that the global minimum configuration is unique 
for small stripe widths, but it becomes bistable for large stripe widths. Our simulations
are also in agreement with analytical predictions for the Laplace pressure of the fluid stripes,
and their damped oscillatory motion as they approach equilibrium configurations, 
capturing the corresponding decay timescale and oscillation frequency.
Finally, we simulate the coarsening dynamics of phase separating binary fluids in the
hydrodynamics and diffusive regimes for tori of various shapes, and compare the results
against those for a flat two-dimensional surface.
Our lattice Boltzmann scheme can be extended to other surfaces and 
coupled to other dynamical equations,
opening up a vast range of applications involving 
complex flows on curved geometries.
\end{abstract}

\maketitle

\section{Introduction}

Hydrodynamics on curved manifolds is relevant for a wide range of physical phenomena. Examples range from the motion of electrons in graphene at the micro-scale \citep{Giordanelli18}, through thin liquid films \citep{Schwartz95, Howell03}, confined active matter \citep{Keber14,Henkes18,Janssen17} and bio-membranes \citep{Henle10,Arroyo09} at the meso-scale, to relativistic flows in astrophysics \citep{Marti15} and at the cosmological scale \citep{Ellis12}. However, despite its importance, the study of flows on curved space has received much less attention when compared to corresponding investigations on two- and three-dimensional flat space. Suitable numerical approaches to study these problems are also still limited, especially when the flow phenomena of interest involve several fluid components.

Here our focus is on multicomponent flow on curved two-dimensional surfaces. An important motivation to study such problem arises from biological membranes and their synthetic counterparts. Experimentally it has been observed that self-assembled lipid and polymer membranes can adopt an astonishing range of shapes and morphologies \citep{Seddon95}, from single bilayers to stacks and convoluted periodic structures. Moreover, these membranes are usually comprised of several species, which can mix or demix depending on the thermodynamic conditions under which they are prepared \citep{Baumgart03,Bacia05,Roberts17}. The interplay between curvature and composition is a ubiquitous structural feature for bio-membranes, and they are key to biological functions and synthetic membrane-based applications \citep{McMahon05,Groves06,Pontani13}.

There is much interest to understand this interplay between membrane curvature and composition. However, to date continuum modelling of membranes with several lipid components have largely focussed on their equilibrium configurations \citep{Julicher96,Hu11,Paillusson17,Fonda2018}. Several dynamic studies of phase separation on curved surfaces have been carried out in the literature. However, apart from a few exceptions \citep{Nitschke12}, they usually involve diffusive dynamics and ignore the importance of hydrodynamics \citep{Marenduzzo13,Jeong15,Gera17}. The aim of this paper is to develop a flexible lattice Boltzmann framework to simulate multicomponent flow on arbitrary curved surfaces. For simplicity, here we will assume the two-dimensional flow is Newtonian. For lipid membranes, this assumption is supported by both Molecular Dynamics simulations and experimental observations \citep{Dimova06,Otter07,Cicuta07,Camley11}.

Our approach is based on the lattice Boltzmann method (LBM) \citep{KrugerBook,Succi01}, which has recently become increasingly popular to study multicomponent flow phenomena, with good agreement against experiments and other simulation methods, including for drop dynamics, liquid phase separation, microfluidics and porous media \citep{Liu16, Sadullah18, Varagnolo2013, Liu2015}. Within the lattice Boltzmann literature, there are several models for multicomponent flow, including the so-called free energy \citep{Swift96}, pseudo-potential \citep{Shan93} and color \citep{Gunstensen91} models. In this work, we have chosen to employ the free energy model, though our framework can be adapted to account for the pseudo-potential and color models. Our approach can also be extended to account for more fluid components \citep{Semprebon16,Wohrwag18,Abadi18,Liang16,RidlWagner2018}, as well as coupled to other dynamical equations, including those for liquid crystals \citep{Spencer06,Denniston01} and viscoelastic fluids \citep{Malaspinas10,Gupta15}. 

Standard lattice Boltzmann method is based on regular Cartesian grids, and this makes it unsuitable for use on curved surfaces. Thus, an important contribution of this work is the use of vielbein formalism \citep{busuioc19} to solve the hydrodynamics equations of motion on curved surfaces, which we combine with the free energy binary fluid model. To our knowledge, this is the first time this has been done. Previous lattice Boltzmann simulations on curved surfaces have been carried out for single-component flows \citep{Mendoza13,busuioc19,hejranfar17pre}. Since it is not always possible to have a regular grid on curved surfaces and for the discrete velocity sets to coincide with the lattice discretizations, here we solve the discrete Boltzmann equation using a finite difference approach, rather than a collision-propagation scheme. The latter is usually the case in standard lattice Boltzmann implementation.

The capabilities of our new method are demonstrated using several problems. Firstly, 
we study drift motion of fluid droplets and stripes when placed on the surface of a torus. 
This drift is due to non-uniform curvature, and as such, is not present on flat space, or for surfaces with uniform curvature (e.g. a sphere). For the stripes, analytical results are available for their equilibrium configuration, Laplace pressure, and relaxation dynamics \citep{Busuioc19bench}, thus providing an excellent platform to systematically examine the accuracy of our method. We demonstrate that these predictions are accurately captured in our simulations. Secondly, we simulate binary phase separation on the surface of a torus for equal and unequal compositions, both in diffusive and hydrodynamic regimes. We compare and contrast the results for tori of various shapes against those for flat two-dimensional surface \citep{Bray02,Kendon01,Wagner97}.

\section{Computational Model and Method}
\label{sec:gen}

In this section we develop a framework that allows simulations of multicomponent 
flow on arbitrary curved surfaces. Our vielbein lattice Boltzmann approach has
three key features. Firstly, similar to standard lattice Boltzmann method, we exploit the Boltzmann 
equation to solve the continuum equations of motion, and we use a discrete and finite set of
fluid distribution functions. Secondly, unlike standard lattice Boltzmann method, the discrete
velocity sets do not coincide with the neighbouring lattice points. Thus, rather than solving
the Boltzmann equation using a sequence of collision and propagation steps, we
take advantage of a finite difference method. Thirdly, to describe the curved surface,
we employ a vielbein field, which decouples the velocity space from the 
coordinate space \citep{cardall13,busuioc19}. This simplifies the formulation and computation
of the governing Boltzmann equation.

\subsection{Brief Introduction to Vielbein Fields}\label{sec:gen:vielb}

Let us begin by considering a two-dimensional curved surface embedded in three dimensions. 
Vector fields, such as the velocity field $\bm{u}(\bm{x})$, on the two-dimensional surface 
can be expressed in the curvilinear coordinate system using 
\begin{equation}
 \bm{u}(\bm{x}) = u^a(q^b) \bm{\partial_a },
\end{equation}
where $u^{a}(q^b)$ represent the components of the velocity field
on a manifold parametrised using the coordinates $q^b$
($1 \le a, b\le 2$ for two-dimensional manifolds).
Furthermore, the squared norm of the velocity field $\bm{u}$ can be computed as 
\begin{equation}
 \bm{u}^2 = g_{ab} u^{a} u^{b}.
\end{equation}
$g_{ab}$ is called the metric tensor. 
This description of vector fields in curvilinear 
coordinates can become inconvenient for practical computations.
This is because the elements of the metric tensor $g_{ab}$ 
may become singular at various points 
due to the choice of surface parametrisation. In such instances, the 
contravariant components $u^a$ of the velocity must diverge in order for 
squared norm $\bm{u}^2$ to remain finite.

The difficulty described above can be alleviated by introducing, as an 
interface between the coordinate space and the velocity space, 
the vielbein vector fields (frame) $\bm{e_{\hata} }= e_{\hata}^a \bm{\partial_a}$.
Dual to the vielbein vector fields are the vielbein one-forms (co-frame)
$\bm{\omega^\hata} = \omega^\hata_a \bm{dq^a}$. 
We reserve the hatted indices to denote the vielbein framework.
The vielbein frame and co-frame have to satisfy the following relations
\begin{equation}
\braket{\bm{\omega^\hata}, \bm{e_\hatb}} \equiv \omega^\hata_a e^a_\hatb = 
 \delta^\hata{}_{\hatb}, \qquad 
 \omega^\hata_a e_\hata^b = \delta^b{}_a, \qquad 
 g_{ab} e^a_\hata e^b_\hatb = \delta_{\hata\hatb}.
 \label{eq:vielb_contr}
\end{equation}

With the above vielbein frame and co-frame, the vector field $\bm{u}$ can be 
written as
\begin{equation}
\bm{u} = u^\hata \bm{e_{\hata}},
\end{equation}
where the vector field components are
\begin{equation}
 u^\hata = \omega^\hata_a u^a, \qquad 
 u^a = e^a_\hata u^\hata, 
\end{equation}
and the squared norm
\begin{equation}
 \bm{u}^2 = \delta_{\hata\hatb} u^\hata u^\hatb.
\end{equation}
In the vielbein framework, the information on the metric tensor
is effectively absorbed in the components of the vector field,
which makes the formulation and derivation of the lattice Boltzmann approach
significantly less cumbersome. 

In the lattice Boltzmann implementation used in this paper, we need to introduce two more geometrical objects.
First, the Cartan coefficients $c_{\hata\hatb}{}^\hatc$ are defined as
\begin{equation}
 c_{\hata\hatb}{}^\hatc = \braket{\bm{\omega^\hatc}, [\bm{e_\hata}, \bm{e_\hatb}]} = \omega^\hatc_a ([e_\hata, e_\hatb])^a,\label{eq:cartan}
\end{equation}
with the commutator $([\bm{e_\hata}, \bm{e_\hatb}])^a = e_\hata^b \partial_{b} e_\hatb^a - 
 e_\hatb^b \partial_b e_\hata^a$. 
Second, $\Gamma^\hata{}_{\hatb\hatc}$ and $\Gamma_{\hata\hatb\hatc}$
represent the connection coefficients, which are defined as
\begin{equation}
 \Gamma^\hatd{}_{\hatb\hatc} = \delta^{\hatd\hata} \Gamma_{\hata\hatb\hatc}, \qquad 
 \Gamma_{\hata\hatb\hatc} = \frac{1}{2} (c_{\hata\hatb\hatc} + c_{\hata\hatc\hatb} - 
 c_{\hatb\hatc\hata}).\label{eq:conn}
\end{equation}

In Appendix \ref{app:curved}, we detail the application of the vielbein formalism for a torus.
It is worth noting that our approach is general and
other curved geometries can be handled in a similar way. 

\subsection{Binary Fluid Model and Equations of Motion}

We consider a binary mixture of fluids $A$ and $B$, characterised by an order 
parameter $\phi$, such that
$\phi = 1$ corresponds to a bulk $A$ fluid and $\phi = -1$ to a bulk $B$ fluid.
A simple free energy model that allows the coexistence of these two bulk fluids
is given by the following Landau free energy  \citep{Briant04,KrugerBook}
\begin{equation}
\Psi =  \int_V \left[ \frac{A}{4} (1-\phi^2)^2 + \frac{\kappa}{2} (\nabla \phi)^2 \right] 
dV, \label{eq:Landau}
\end{equation}
where 
% the volume element in the case of the torus is 
% $dV = \sqrt{g} d\theta d\varphi$, where $\sqrt{g} = r(R + r \cos\theta)$,
% while 
$A$ and $\kappa$ are free parameters, which are related to the 
interface width $\xi_0$ and surface tension $\gamma$ through %\citep{Briant04,KrugerBook}
%in the case of a planar interface 
\begin{equation}
 \xi_0 = \sqrt{\frac{\kappa}{A}}, \qquad 
 \gamma = \sqrt{\frac{8 \kappa A}{9}}.
 \label{eq:xi_gamma}
\end{equation}
The chemical potential can be derived by taking the functional derivative of 
the free energy with respect to the order parameter, giving
\begin{equation}
\mu({\bm x}) = \frac{\delta \Psi}{\delta \phi({\bm x})} =  -A\phi(1- \phi^2) - \kappa \Delta \phi.
\label{eq:mu}
\end{equation}

The evolution of the order parameter $\phi$ is specified by the Cahn-Hilliard equation.
In covariant form it is given by
\begin{equation}
 \partial_t \phi + \nabla_\hata (u^\hata \phi) =  \nabla_\hata (M \nabla^\hata \mu),
 \label{eq:CH}
\end{equation}
where the hatted indices are taken with respect to the orthonormal vielbein basis. 
Equivalently, indices with respect to the coordinate basis can be used, e.g. 
$\nabla_{\hata} (u^\hata \phi) = \nabla_a (u^a \phi)$.
In the above, $M$ is the mobility parameter, $\mu$ is the chemical potential,
and the fluid velocity $\bm{u}$ is a solution of the continuity and Navier-Stokes equations
\begin{equation}
 \partial_t n + \nabla_\hata (u^\hata n) = 0, \qquad 
 n m \frac{Du^\hata}{Dt} = - \nabla_{\hatb} T^{\hata\hatb} + 
 n F^\hata,
 \label{eq:hydro}
\end{equation}
where $D/Dt = \partial_t + u^\hatb \nabla_\hatb$ is the material (convective) derivative, 
$m$ is the particle mass, $n$ is the number density and
$T^{\hata\hatb} = p_{\rm i} \delta^{\hata\hatb} + \sigma^{\hata\hatb}$ is the
ideal gas stress tensor. $T^{\hata\hatb}$ consists of the ideal gas pressure $p_{\rm i} = n k_B T$ and
the viscous stress for the Newtonian fluid $\sigma^{\hata\hatb} = 
-\eta(\nabla^\hata u^\hatb + \nabla^\hatb u^\hata - \delta^{\hata\hatb} 
\nabla_\hatc u^\hatc) - \eta_v \delta^{\hata\hatb} \nabla_\hatc u^\hatc$.
The latter is written in terms of the dynamic (shear) and volumetric (bulk) 
viscosities $\eta$ and $\eta_v$.
The thermodynamic force term $F^\hata$ 
takes the following form %\citep{Briant04,KrugerBook}
\begin{eqnarray}
 & n F^{\hata} = - \phi \nabla^\hata \mu = - \nabla^\hata p_{\rm binary} + 
 \kappa \phi \nabla^\hata \Delta \phi, \nonumber \\
 & p_{\rm binary} = A\left(-\frac{1}{2} \phi^2 + \frac{3}{4} \phi^4\right).
 \label{eq:pbin}
\end{eqnarray}
A summary on how the differential operators must be applied for the cases of the 
Cartesian and torus geometries is provided in the Supplementary Information \citep{suppl}.
%In the Supplementary Information \citep{suppl} we also demonstrate that we
%satisfactorily reproduce known results for two-dimensional flat simulations \citep{KrugerBook}. 
%In particular, we capture correct interfacial profiles and Laplace pressures for a droplet of 
%one fluid component surrounded by another component. 

\subsection{The Vielbein Lattice Boltzmann Approach}
\label{sec:gen:boltz}

In this paper, we employ the lattice Boltzmann approach to solve the
hydrodynamics equations [Eq.~\eqref{eq:hydro}], while the Cahn-Hilliard equation [Eq.~\eqref{eq:CH}]
is solved directly using a finite difference method. The details of the numerical
implementation are discussed in the Supplementary Information \citep{suppl}.
It is possible to solve the Cahn-Hilliard equation using a lattice Boltzmann
scheme, and on flat manifolds, it has been suggested that extension to more fluid components
is more straightforward in this approach \citep{li2007symmetric,RidlWagner2018}.
However, for our purpose here, it is more expensive and require us to use a
higher order quadrature.

We use a discretised form of the Boltzmann equation that reproduces the fluid 
equations of motion in the continuum limit. In covariant form, the Boltzmann equation
on an arbitrary geometry is given by \citep{busuioc19}:
\begin{equation}
\frac{\partial f}{\partial t} + 
 \frac{1}{\sqrt{g}} \frac{\partial}{\partial q^b} 
 \left(v^\hata e_\hata^ b f \sqrt{g}\right) + 
 \frac{\partial}{\partial v^\hata} \left[\left(\frac{F^\hata}{m} - 
 \Gamma^\hata{}_{\hatb\hatc} v^\hatb v^\hatc\right) f\right] 
 = J[f],
 \label{eq:boltz_cons}
\end{equation}
where $\sqrt{g}$ is the square root of the determinant of the metric tensor, 
and $J[f]$ is the collision operator. 

For the specific case of a torus, the Boltzmann equation reads %(cf. Appendix~\ref{app:curved})
\begin{multline}
\frac{\partial f}{\partial t} + \frac{v^\hvarphi}{R + r\cos\theta} \frac{\partial f}{\partial \varphi}  
+ \frac{v^\htheta}{r(1 + a\cos\theta)} \frac{\partial [f(1 + a\cos\theta)]}{\partial\theta} 
+ \frac{F^\hvarphi}{m} \frac{\partial f}{\partial v^\hvarphi}
+ \frac{F^\htheta}{m} \frac{\partial f}{\partial v^\htheta} \\
- \frac{\sin\theta}{R + r\cos\theta} 
 \left[v^\hvarphi \frac{\partial (f v^\hvarphi)}{\partial v^\htheta} - 
 v^\htheta \frac{\partial (fv^\hvarphi)}{\partial v^\hvarphi} \right]
 = -\frac{1}{\tau}[f - \feq].\label{eq:boltz_tor}
\end{multline}
The steps needed to derive Eq.~\eqref{eq:boltz_tor} from Eq.~\eqref{eq:boltz_cons}
are summarised in Appendix~\ref{app:curved}.
Here $r$ and $R$ represent the inner (small) and outer (large) radii, 
$a = r/R$ is the radii ratio, while
the angle $\theta$ goes round the inner circle and $\varphi$ covers the large
circle. The range for both $\theta$ and $\varphi$ is $[0, 2\pi)$ and the 
system is periodic with respect to both these angles. The last term on the 
left hand side of Eq. \eqref{eq:boltz_tor} corresponds to inertial and reaction forces
that arise when we have flow on curved surfaces, since fluid motion is constrained
on the surface. 

As commonly the case in the lattice Boltzmann literature, we employ 
the BGK approximation for the collision operator,
\begin{equation}
 J[f] = -\frac{1}{\tau} [f - f^{\rm eq}].
\end{equation}
The relaxation time $\tau$ is related to the fluid kinematic viscosity $\nu$,
dynamic viscosity $\eta$ and volumetric viscosity $\eta_v$
by \citep{KrugerBook,Dellar2001}
\begin{equation}
\nu = \frac{\eta}{n m} = \frac{\eta_v}{n m} = 
\frac{\tau k_B T}{m},
\label{eq:nu}
\end{equation}
such that $\sigma^{\hata\hatb} = -\eta(\nabla^{\hata} u^\hatb
+ \nabla^{\hatb} u^\hata)$.
% where $m$ is the particle mass and $n$ is the particle number density. 

\begin{figure}
\begin{center}
 \includegraphics[scale=0.85,angle=0]{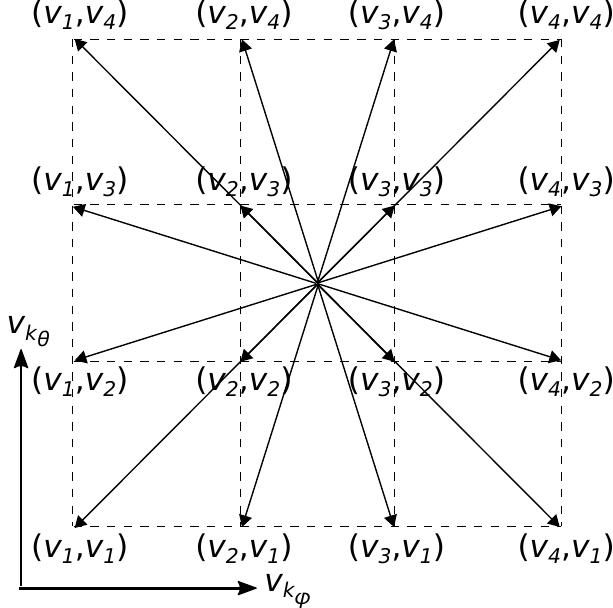} 
\caption{A fourth order quadrature ($Q = 4$) lattice Boltzmann model with $16$ velocities. 
The filled black circle in the centre of the figure corresponds to a lattice point in space. 
Here we have an off-lattice implementation, where 
the velocity directions do not coincide with neighbouring lattice points.
\label{fig:veldir}}
\end{center}
\end{figure}

Rather than considering fluid distribution functions $f(\bm v)$
with continuous velocity space $\bm{v} = (v^{\htheta}, v^{\hvarphi})$, we discretise the velocity
space using $\bm{v}_{\bm{k}} = (v_{k_\theta}, v_{k_\varphi})$. Due to the inertial and reaction force terms in Eq.~\eqref{eq:boltz_cons},
we need at least a fourth order quadrature ($Q = 4$) when non-Cartesian coordinates are employed.
We obtain inaccurate simulation results when third order quadrature (or lower) is used.
The 16 velocity directions, corresponding to $Q = 4$, are illustrated in Fig. \ref{fig:veldir} \citep{sofonea18pre}.
The possible values of $v_{k_\theta}$ and $v_{k_\varphi}$ ($1 \le k_\theta, k_\varphi \le 4$) are given as the roots of 
the fourth order Hermite polynomial, 
\begin{equation}
 \begin{pmatrix}
  v_1 \\ v_2 \\ v_3 \\ v_4 
 \end{pmatrix} = 
 \begin{pmatrix}
  -\sqrt{3 + \sqrt{6}} \\ 
  -\sqrt{3 - \sqrt{6}} \\ 
  \sqrt{3 - \sqrt{6}} \\ 
  \sqrt{3 + \sqrt{6}}
 \end{pmatrix}. \label{eq:vmatrix}
\end{equation}
We use Hermite polynomials orthogonal with respect to the weight function $e^{-v^2/2}/\sqrt{2\pi}$, 
which are described in detail, 
e.g. in the Appendix of Ref.~\citep{sofonea18pre}.
It is worth noting that, unlike standard lattice Boltzmann algorithm, in general the velocity directions do not coincide with the neighbouring lattice points.
For simplicity, we have set $k_BT = 1$ and $m = 1$, such that the reference scale for the velocity $v_{\rm ref} = (k_BT/m)^{1/2} = 1$,
which is also the sound speed in an isothermal fluid.

%The values provided in 
%Eq. \eqref{eq:vmatrix} also serve as quadrature points in the Gauss-Hermite 
%quadrature method for the evaluation of the integral
%\begin{equation}
 %\int_{-\infty}^\infty \frac{dv}{\sqrt{2\pi}} e^{-v^2/2} v^s \simeq
% \sum_{k = 1}^Q w_k v_k^s,
%\end{equation}
%where the equality is exact when $2Q > s$. 
The particle number density $n$ and velocity ${\bm u}$ can be computed as zeroth and first 
order moments of the distribution functions
\begin{equation}
n = \sum_{\bm k} f_{\bm k}, \qquad
n {\bm u} = \sum_{\bm k} f_{\bm k} {\bm v}_{\bm k}.
\end{equation}
With the discretisation of the velocity space, we also replace the Maxwell-Boltzmann equilibrium 
distribution with a set of distribution functions $f^{\rm eq}_{\bm{k}}$ corresponding to the discrete 
velocity vectors $\bm{v}_{\bm{k}}$.
Due to the use of the vielbein formalism, the expression for 
$f^{\rm eq}_{\bm{k}}$ coincides with the one employed on the flat Cartesian geometry
\citep{sofonea18pre} 
\begin{equation}
 f^{\rm eq}_{\bm{k}} = n w_{k_\theta} w_{k_\varphi} 
 \left\{1 + \bm{v}_{\bm{k}} \cdot \bm{u} + 
 \frac{1}{2} [(\bm{v}_{\bm{k}} \cdot \bm{u})^2 - \bm{u}^2] 
 + \frac{1}{6} \bm{v}_{\bm{k}} \cdot \bm{u} [
 (\bm{v}_{\bm{k}} \cdot \bm{u})^2 - 3 \bm{u}^2]\right\}.
 \label{eq:feq}
\end{equation}
The quadrature weights $w_k$ can be computed using the formula
\begin{equation}
 w_k = \frac{Q!}{H_{Q+1}^2(v_k)},
\end{equation}
where $Q$ is the order of the quadrature and $H_n(x)$ is the Hermite polynomial of order $n$.
For $Q = 4$, the weights have the following values \citep{sofonea18pre}
\begin{equation}
 w_1 = w_4 = \frac{3-\sqrt{6}}{12}, \qquad 
 w_2 = w_3 = \frac{3 + \sqrt{6}}{12}.
\end{equation}

To compute the force terms in the Boltzmann
equation, Eq. \eqref{eq:boltz_tor}, we consider a unidimensional expansion of the distribution 
with respect to the velocity space degrees of freedom~\citep{busuioc19,busuioc17arxiv}. 
In particular, the velocity derivatives appearing 
in Eq.~\eqref{eq:boltz_tor} can be computed as 
\begin{align}
& \left(\frac{\partial f}{\partial v^\htheta}\right)_{k_\theta k_\varphi} = 
 \sum_{k_{\theta}' = 1}^Q \mathcal{K}^H_{k_\theta,k_\theta'}
 f_{k_\theta', k_\varphi}, \quad
 \left(\frac{\partial f}{\partial v^\hvarphi}\right)_{k_\theta k_\varphi} = 
 \sum_{k_\varphi' = 1}^Q \mathcal{K}^H_{k_\varphi,k_\varphi'}
 f_{k_\theta, k_\varphi'}, \nonumber\\
&  \left(\frac{\partial (fv^\hvarphi)}{\partial v^\hvarphi}\right)_{k_\theta k_\varphi} = 
 \sum_{k_\varphi' = 1}^Q \widetilde{\mathcal{K}}^{H}_{k_\varphi,k_\varphi'}
 f_{k_\theta, k_\varphi'}, 
\end{align}
where the kernels $\mathcal{K}^H_{i,m}$ and
$\widetilde{\mathcal{K}}^H_{i,m}$ can be written in terms of 
Hermite polynomials \citep{busuioc19}
\begin{align}
 \mathcal{K}^{H}_{i,m} =& -w_i \sum_{\ell = 0}^{Q - 1} 
 \frac{1}{\ell!} H_{\ell+1}(v_i) H_{\ell}(v_{m}), \\
 \widetilde{\mathcal{K}}^{H}_{i,m} =& -w_i \sum_{\ell = 0}^{Q - 1} 
 \frac{1}{\ell!} H_{\ell+1}(v_i) [H_{\ell+1}(v_{m}) + 
 \ell H_{\ell-1}(v_{m})]. \nonumber
\end{align}
We list below the components of the above matrices for the case of $Q = 4$:
\begin{align}
 \mathcal{K}^H_{i,m} =& 
 \begin{pmatrix}
  \frac{1}{2} \sqrt{3 + \sqrt{6}} & 
  \frac{\sqrt{3 + \sqrt{3}}}{2(3 + \sqrt{6})} & 
  -\frac{\sqrt{3 - \sqrt{3}}}{2(3 + \sqrt{6})} &
  \frac{1}{2} \sqrt{1 - \sqrt{\frac{2}{3}}}\\
  -\sqrt{\frac{5 + 2\sqrt{6}}{2(3-\sqrt{3})}} & 
  \frac{1}{2}\sqrt{3-\sqrt{6}} & \frac{1}{2}\sqrt{1 + \sqrt{\frac{2}{3}}} & 
  -\frac{\sqrt{27+11\sqrt{6}}-\sqrt{3+\sqrt{6}}}{2\sqrt{6}}\\
  \frac{\sqrt{27+11\sqrt{6}}-\sqrt{3+\sqrt{6}}}{2\sqrt{6}} &
  -\frac{1}{2} \sqrt{1 + \sqrt{\frac{2}{3}}} & 
  -\frac{1}{2}\sqrt{3 - \sqrt{6}} &
  \frac{\sqrt{27+11\sqrt{6}} + \sqrt{3+\sqrt{6}}}{2\sqrt{6}} \\
  -\frac{\sqrt{3-\sqrt{6}}}{2\sqrt{3}} & 
  \frac{\sqrt{3-\sqrt{3}}}{2(3 + \sqrt{6})} & 
  -\frac{\sqrt{3+\sqrt{3}}}{2(3 + \sqrt{6})} &
  -\frac{1}{2} \sqrt{3 + \sqrt{6}}  
 \end{pmatrix}, \nonumber\\
 \widetilde{\mathcal{K}}^H_{i,m} =& 
 \begin{pmatrix}
  -\frac{3+\sqrt{6}}{2} & 
  \frac{2 - 5\sqrt{2} + \sqrt{6(9-4\sqrt{2})}}{4} &
  \frac{2 + 5\sqrt{2} - \sqrt{6(9+4\sqrt{2})}}{4} &
  \frac{1}{2} \\
  \frac{2 + 5\sqrt{2} + 4\sqrt{3} + \sqrt{6}}{4} &
  -\frac{3 - \sqrt{6}}{2} & 
  \frac{1}{2} &
  \frac{2 - 5\sqrt{2} - 4\sqrt{3} + \sqrt{6}}{4} \\
  \frac{2 - 5\sqrt{2} - 4\sqrt{3} + \sqrt{6}}{4} &
  \frac{1}{2} &
  -\frac{3 - \sqrt{6}}{2} & 
  \frac{2 + 5\sqrt{2} + 4\sqrt{3} + \sqrt{6}}{4} \\
  \frac{1}{2} &
  \frac{2 + 5\sqrt{2} - \sqrt{6(9+4\sqrt{2})}}{4} &
  \frac{2 - 5\sqrt{2} + \sqrt{6(9-4\sqrt{2})}}{4} &
  -\frac{3+\sqrt{6}}{2}
 \end{pmatrix}.
\end{align}

\section{Drift Dynamics of Fluid Stripes and Droplets}\label{sec:drift}

In this section we begin by studying the behaviour of fluid stripes on the torus geometry.
By minimising the interface length subject to area conservation, we find there is a second
order phase transition in the location of the equilibrium position as we vary the stripe area.
In particular we observe bistability when the stripe area exceeds a critical value. We validate the ability 
of our method to capture this effect in Subsec.~\ref{sec:drift:stripeeq}.
We then consider the Laplace pressure test in Subsec.~\ref{sec:drift:laplace}.
The Laplace pressure takes a different form on a curved torus geometry compared to 
that on a flat geometry \citep{Busuioc19bench}. Furthermore,
the approach to equilibrium configuration through a damped harmonic motion 
is investigated in Subsec.~\ref{sec:drift:damp}. We show that we recover 
the damping coefficient and the angular frequency as derived in 
\cite{Busuioc19bench}.
Finally, we contrast the drift dynamics of fluid stripes with droplets on the torus 
in section \ref{sec:drift:drop}. While the former drift to the inside of the torus, 
the latter move to the outside of the torus.

\subsection{Equilibrium positions of fluid stripes} \label{sec:drift:stripeeq}

The basic idea behind establishing the equilibrium position of fluid stripes 
is that the interface length must attain a minimum for a fixed stripe area. 
We consider a stripe of angular width $\Delta \theta$, centred on $\theta = \theta_c$,
such that its interfaces are located at
\begin{equation}
 \theta_- = \theta_c - \Delta \theta / 2, \qquad
 \theta_+ = \theta_c + \Delta \theta / 2. \label{eq:stripe_thetapm}
\end{equation}
As a convention, here the stripe is identified with the minority, rather than the majority,
fluid component. 
The area $\Delta A$ enclosed between the upper and lower interfaces 
can be obtained as follows
\begin{equation}
 \Delta A = 2\pi r R \int_{\theta_-}^{\theta_+} d\theta (1 + a \cos\theta) 
 = 2\pi r R [\Delta \theta + 2a \sin(\Delta \theta / 2) \cos \theta_c],
 \label{eq:stripe_area}
\end{equation}
where $a = r / R$. The preservation of the area allows the variation of the stripe width 
$\Delta \theta$ to be related to a variation of the stripe centre 
$\theta_c$. Setting $d \Delta A = 0$,
\begin{equation}
 d\frac{\Delta \theta}{2} = \frac{a \sin (\Delta \theta/2)}
 {1 + a \cos(\Delta\theta/2)} \cos\theta_c \sin \theta_c d\theta_c.
 \label{eq:stripe_dA0}
\end{equation}

The total interface length $\ell_{\rm total} = \ell_+ + \ell_-$ 
can be computed as
\begin{equation}
 \ell_{\rm total} = 4\pi R \left(1 + a \cos\theta_c \cos \frac{\Delta \theta}{2}\right).
 \label{eq:stripe_ltot}
\end{equation}
Imposing $d\ell_{\rm total} = 0$ yields an equation involving
the stripe width $\Delta \theta_{eq}$ and stripe centre 
$\theta_c^{eq}$ at equilibrium
\begin{equation}
 \left(a \cos\theta_c^{eq} + \cos\frac{\Delta \theta_{eq}}{2}\right) 
 \sin\theta_c^{eq}  = 0.
 \label{eq:minimaring}
\end{equation}
The above equation has different solutions depending on the stripe width. 
For narrow stripes, the equilibrium position is located
at $\theta_c^{eq} = \pi$. 
There is a critical point corresponding to stripe width
$\Delta \theta_{eq} = \Delta \theta_{\rm crit} = 2\arccos(a)$, or alternatively
stripe area
\begin{equation}
 \Delta A_{\rm crit} = 4\pi r R (\arccos a - a\sqrt{1 - a^2}).
 \label{eq:stripe_dA_crit}
\end{equation}
For stripes with areas larger than this critical value,
two equilibrium positions are possible, namely
\begin{equation}
 \theta^{eq}_{c} = \pi \pm \arccos\left[\frac{1}{a} 
 \cos\frac{\Delta \theta_{eq}}{2}\right].
 \label{eq:stripe_thceq}
\end{equation}

%\begin{figure}
%\begin{center}
%\begin{tabular}{cc}
% \includegraphics[width=0.4\columnwidth]{data/stripe/stripe-F-eps-converted-to.pdf} & 
% \includegraphics[width=0.4\columnwidth]{data/stripe/stripe-thc-eps-converted-to.pdf} \\
% \includegraphics[width=0.4\columnwidth]{data/stripe/stripe01-t0.png} &
% \includegraphics[width=0.4\columnwidth]{data/stripe/stripe01-t725.png} \vspace{1 mm}\\
% (c) $t = 0$ & (d) $t = 725$ \vspace{2 mm}\\
% \includegraphics[width=0.4\columnwidth]{data/stripe/stripe01-t1195.png} &
% \includegraphics[width=0.4\columnwidth]{data/stripe/stripe01-t2070.png} \vspace{1 mm}\\
% (e) $t = 1195$ & (f) $t = 2070$
%\end{tabular}
%\caption{
%Time evolution (a) of the total free energy $\Psi$ and (b) of 
%$1 - \theta_c / \pi$, indicating the centre position of the ring, 
%for rings initialised according to 
%Eq.~\eqref{eq:stripe_tanh}  with $(\theta_c/\pi, \Delta \theta/2\pi) \in 
%\{(0.1, 0.102), (0.5,0.140), (0.9,0.216)\}$.
%(c--f) Snapshots of the evolution of the ring initialised with 
%$\theta_c = \pi / 10$ corresponding to 
%the initial state (c) and to the first (d), second (e)
%and fourth (f) maxima of $1 - \theta_c / \pi$, occurring at $t = 0$, $725$, $1195$ and $2070$, 
%respectively.
%\label{fig:stripe}}
%\end{center}
%\end{figure}

We now reproduce the above phenomenon using our lattice Boltzmann approach.
Unless stated otherwise, in section \ref{sec:drift}, we use a torus with 
$r = 0.8$ and $R = 2$ ($a = r / R = 0.4$). We set the parameters in our free energy model,
Eq.\eqref{eq:Landau}, to $\kappa = 5 \times 10^{-4}$ and $A = 0.5$, and set the 
kinematic viscosity $\nu = 2.5 \times 10^{-3}$ and mobility parameter in the Cahn-Hilliard
equation $M = 2.5 \times 10^{-3}$. Due to its homogeneity with respect to 
$\varphi$, the system is essentially one dimensional, such that 
a single node is used on the $\varphi$ direction (i.e., $N_\varphi = 1$).
The discretisation along the $\theta$ direction is performed using 
$N_\theta = 320$ nodes. Throughout this paper we ensure that our
discretization is such that the spacing is always smaller than the interface width $\xi_0$,
as given in Eq.\eqref{eq:xi_gamma}.
The time step is set to $\delta t = 5\times 10^{-4}$.
%In the first set of simulations, we consider three stripes of 
%subcritical area $\Delta A < \Delta A_{\rm crit}$, 
%corresponding to width $r \Delta \theta = 16 \xi_0$ 
%for a stripe centred on $\theta_c = 0$:
%\begin{equation}
% \Delta A = 4\pi r R \left[\frac{8 \xi_0}{r} +  
% a \sin\left(\frac{8 \xi_0}{r}\right)\right] \simeq 8.859.
%\end{equation}

\begin{figure}
\begin{center}
\begin{tabular}{cc}
 \includegraphics[width=0.45\columnwidth]{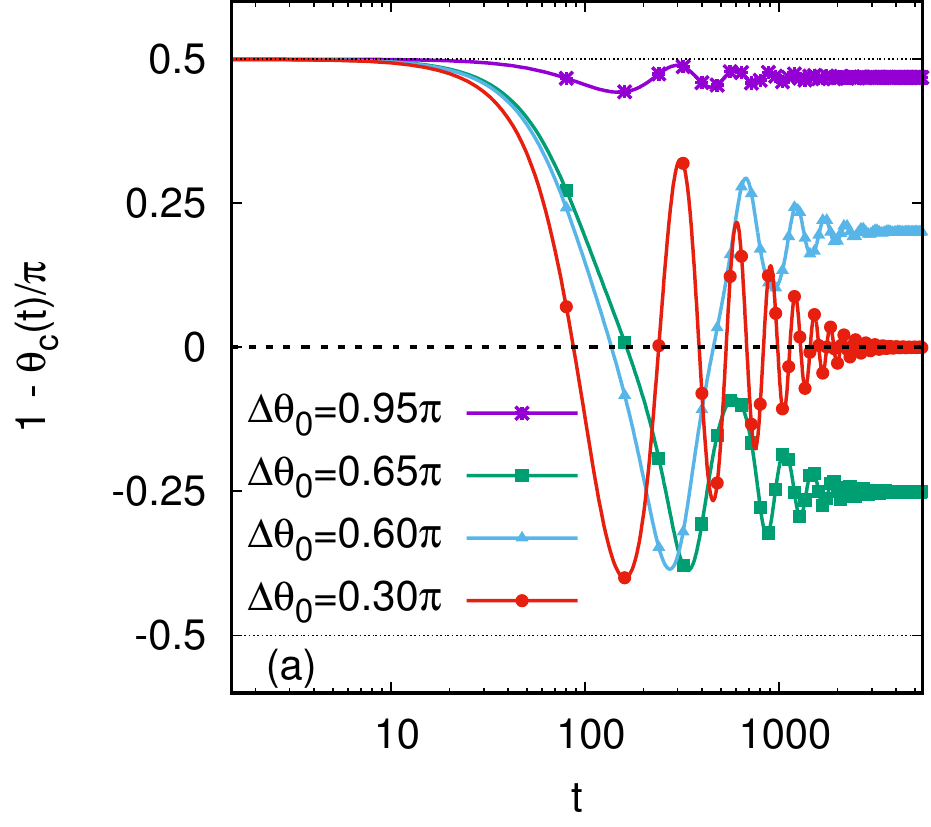} &
 \includegraphics[width=0.45\columnwidth]{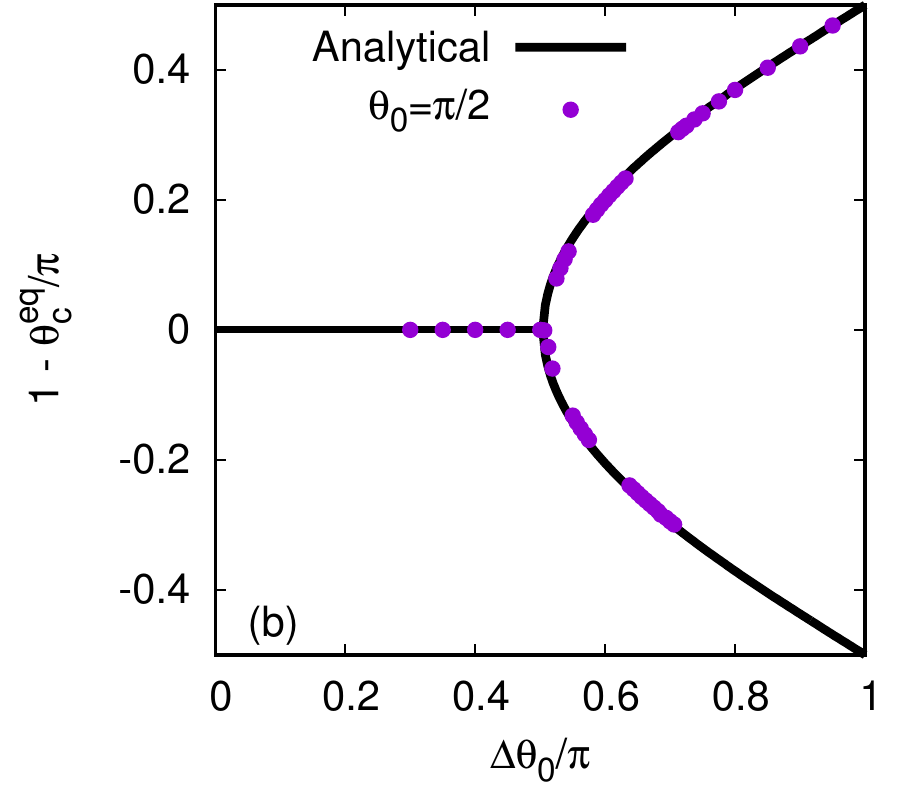}
  \vspace{2pt}
\end{tabular}
\begin{tabular}{ccc}
\hspace{5pt} \includegraphics[width=0.3\columnwidth]{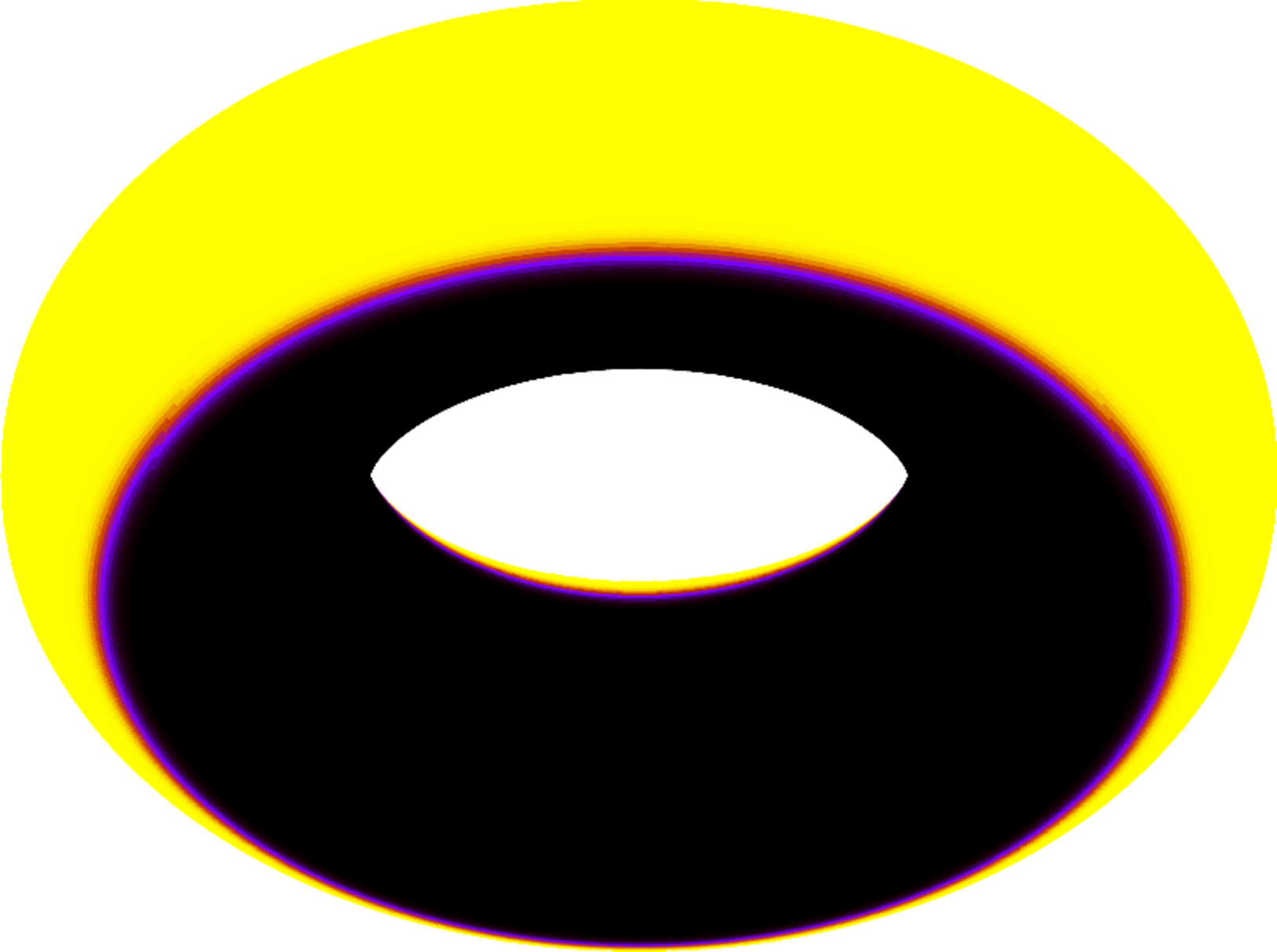}&
\hspace{5pt} \includegraphics[width=0.3\columnwidth]{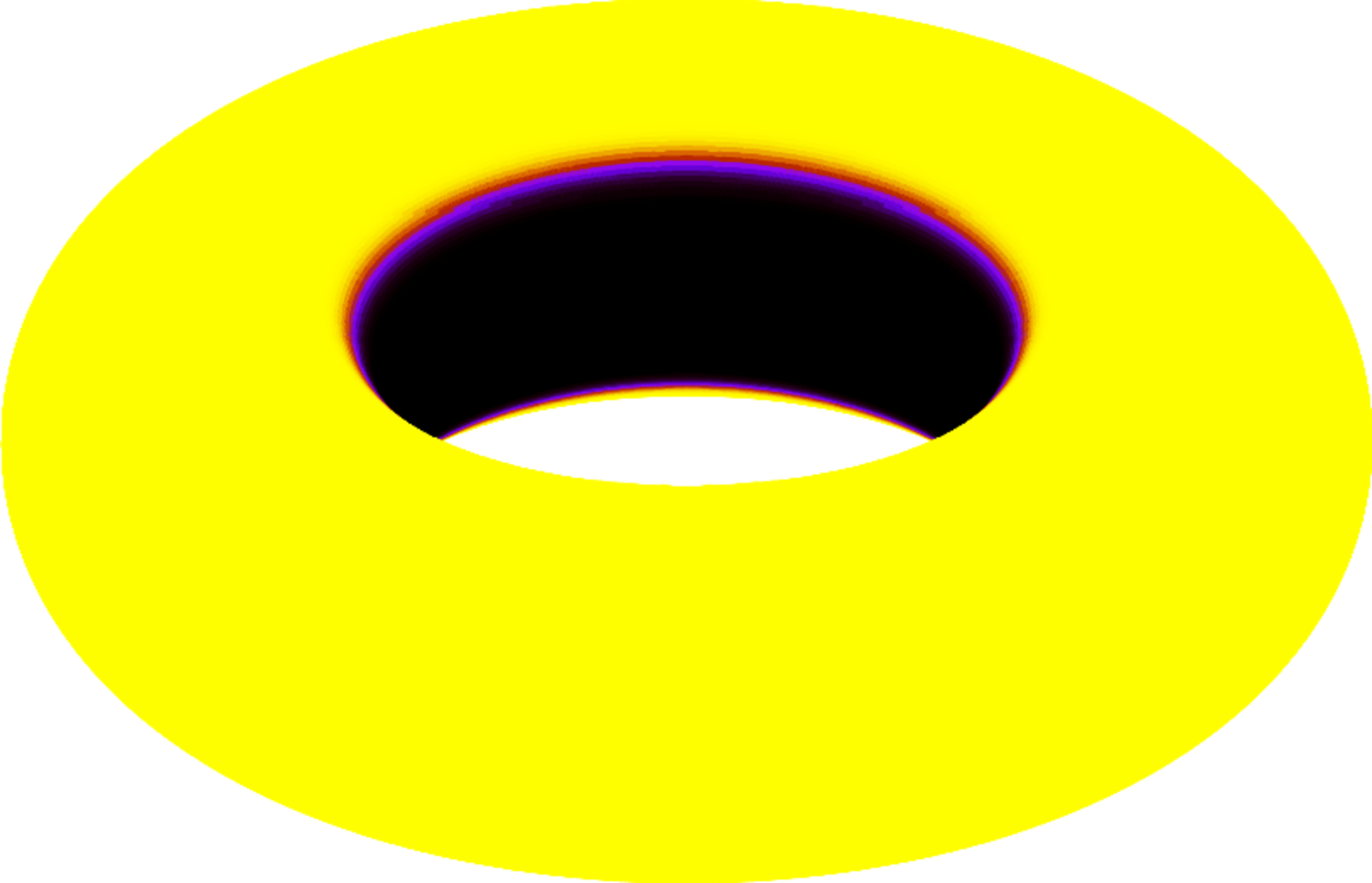}&
\hspace{5pt}  \includegraphics[width=0.3\columnwidth]{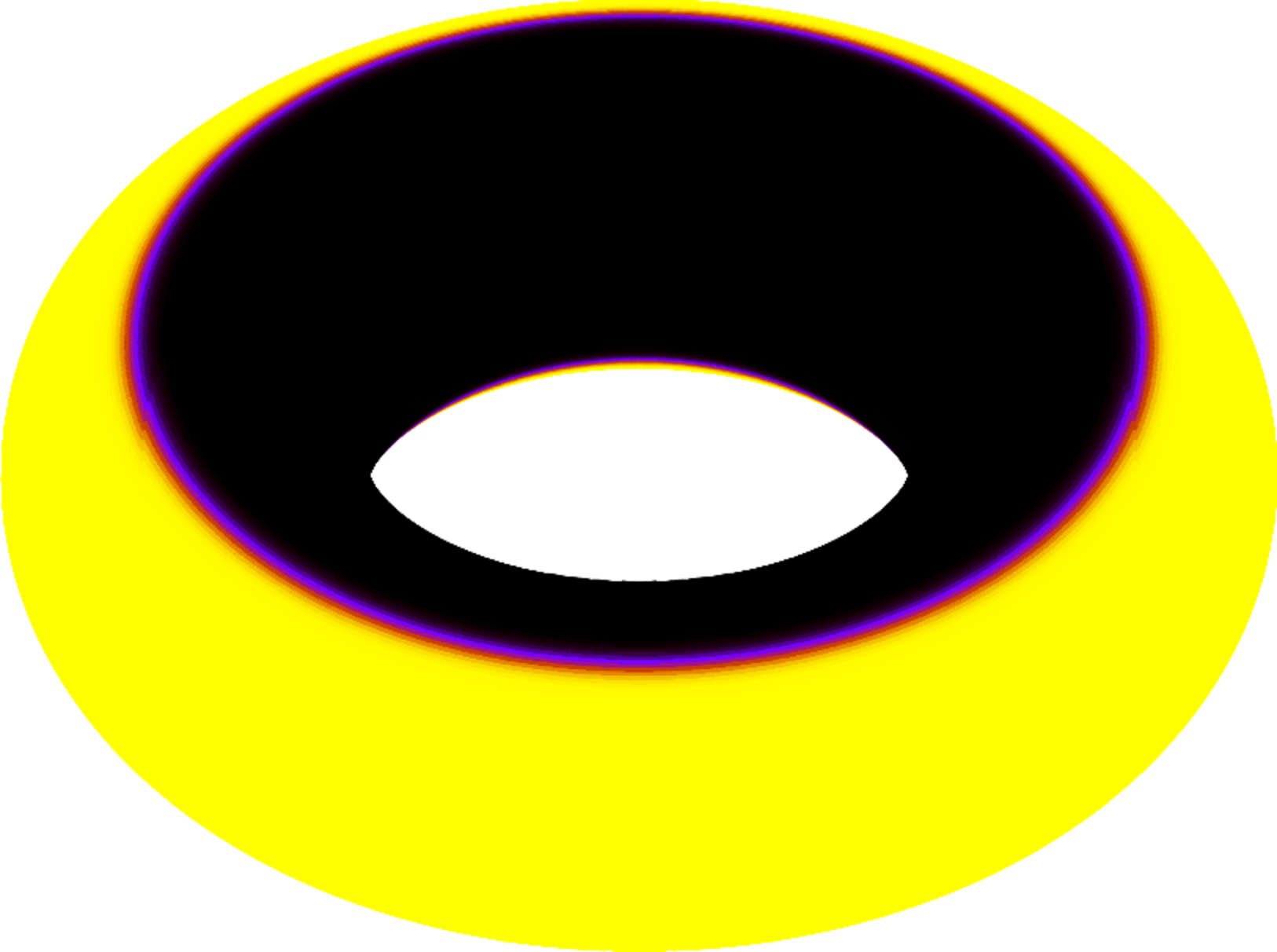}\vspace{2pt}\\
\hspace{15pt}(c)&\hspace{15pt}(d)&\hspace{15pt}(e)\vspace{2pt}\\
\hspace{-10pt} \includegraphics[width=0.31\columnwidth]{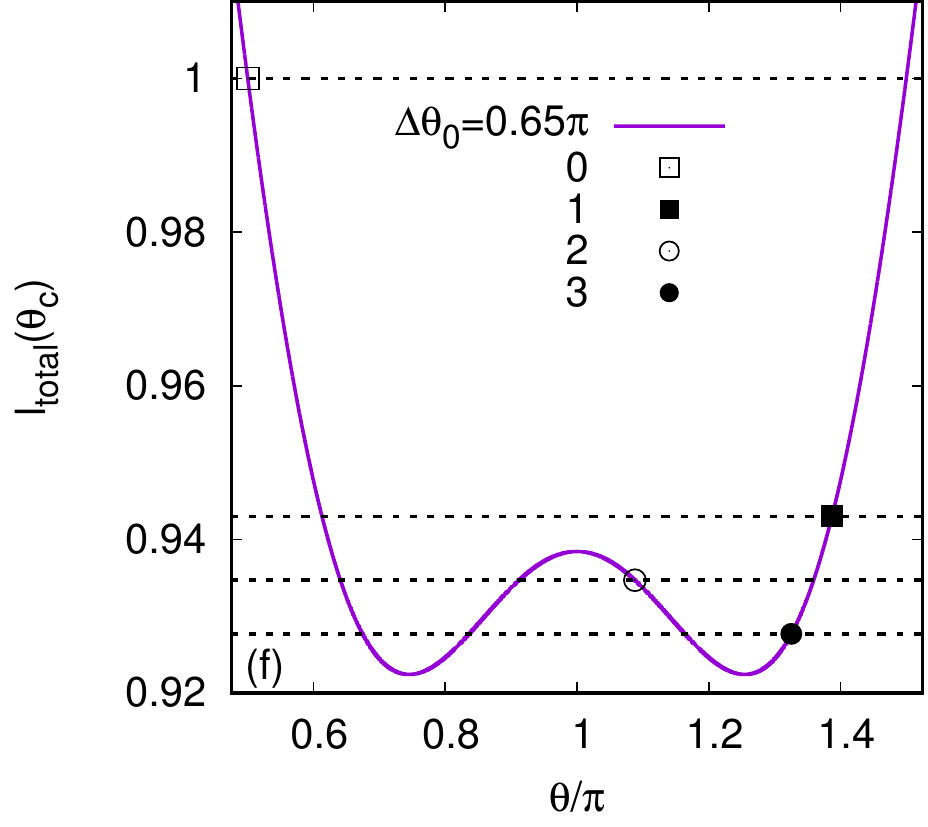} & 
\includegraphics[width=0.31\columnwidth]{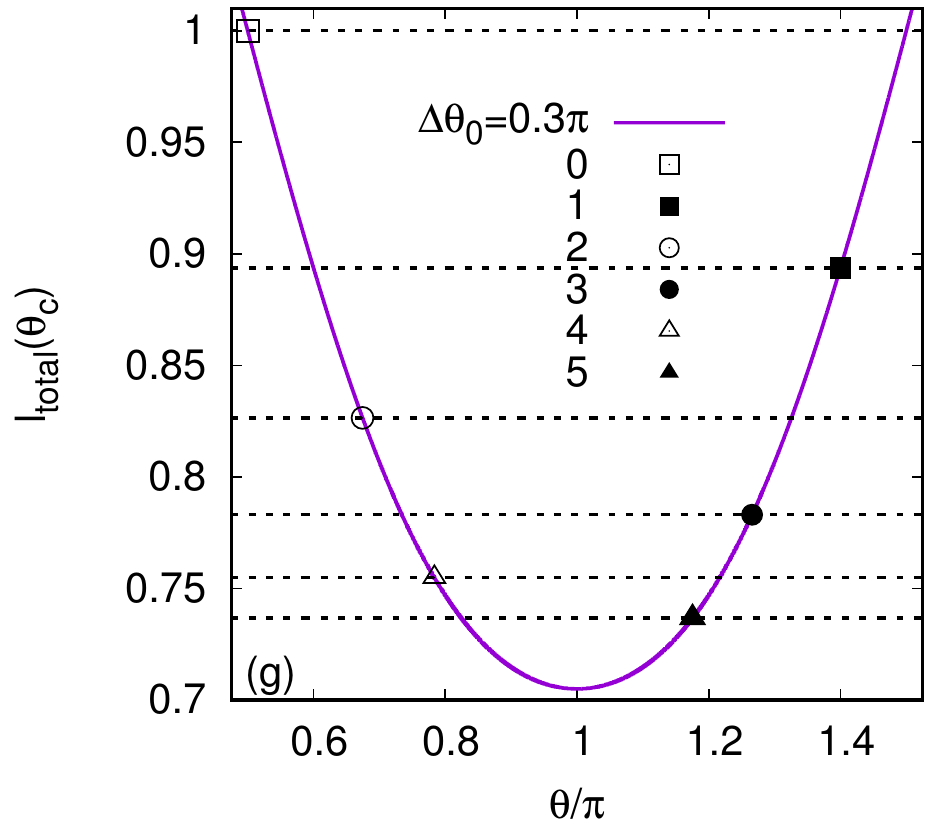} & 
\includegraphics[width=0.31\columnwidth]{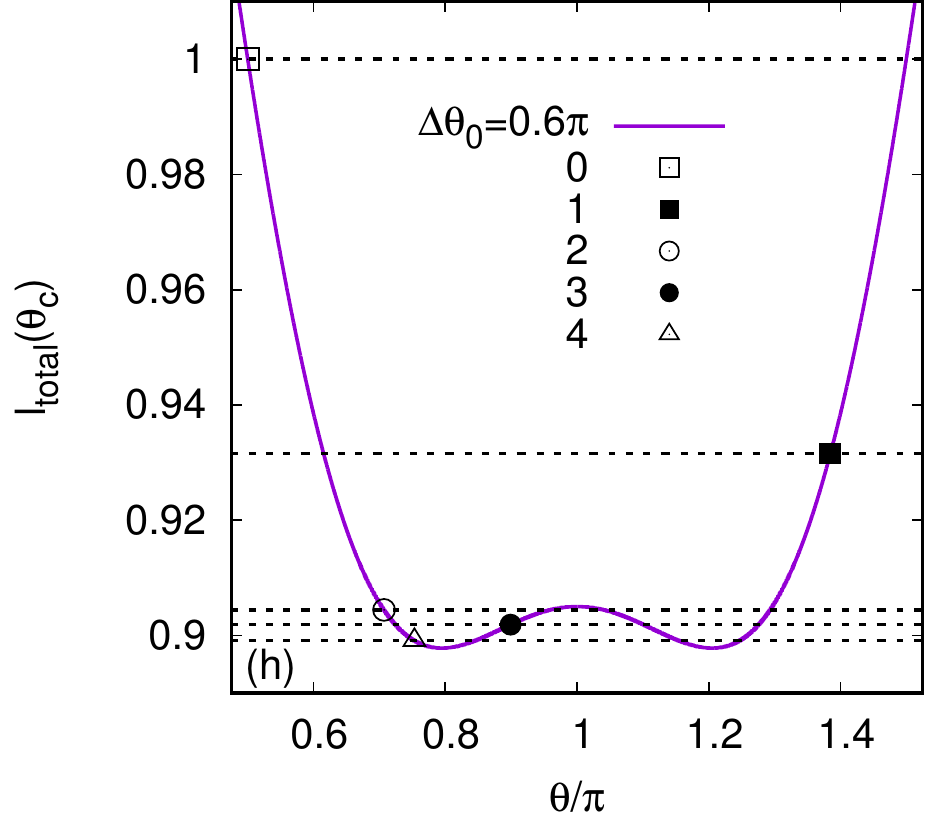} 
\end{tabular}
\end{center}
\caption{(a) Equilibrium position $\theta_c^{eq}$ for stripes 
initialised at $\theta_0 = \pi / 2$ on the torus with 
$r = 0.8$ and $R = 2$ ($a = 0.4$), as a function of the 
initial angular width $\Delta \theta_0$ in comparison 
with the analytical prediction.
(b) Diagram indicating the location of the equilibrium position $\theta_c^{eq}$
as a function of the stripe width $\Delta \theta_0$ and the radii ratio $a = r/R$,
for stripes initialised at $\theta_0 = \pi / 2$.
(c-e) Examples of stripes equilibrated at 
(c) $\theta_{c}^{eq} > \pi$ ($\Delta \theta_0 =0.65\pi$),
(d) $\theta_{c}^{eq} = \pi$ ($\Delta \theta_0 =0.3\pi$), and 
(e) $\theta_{c}^{eq} < \pi$ ($\Delta \theta_0 =0.6\pi$).
(f-h) Interface length $\ell_{\rm total}$ as a function of the 
stripe centre position (solid line) for the stripe parameters considered 
in (c-e). The symbols highlight the interface lengths at maximum 
oscillation amplitude at initialisation ($0$) and after each 
half period ($1$, $2$, etc).
\label{fig:tor_stripe_fork}}
\end{figure}

We initialise the fluid stripes using a hyperbolic tangent profile
\begin{equation}
 \phi_{\rm stripe}(\theta, t) = \phi_0 + 
 \tanh \left[\frac{r}{\xi_0 \sqrt{2}}\left(
 |\widetilde{\theta - \theta_c}| - \frac{\Delta \theta}{2}\right)\right], 
 \label{eq:stripe_tanh}
\end{equation}
where $\phi_0$ is an offset due to the Laplace pressure (see next subsection)
\begin{equation}
 \phi_0 = \frac{\xi_0}{3R \sqrt{2}} 
 \frac{\cos\theta_c \sin(\Delta \theta / 2)}
 {1 + a \cos\theta_c \cos (\Delta \theta/2)}. \label{eq:stripe_phi0}
\end{equation}
%We consider stripes initially centred at 
%$\theta_c = \pi /10$, $5\pi/10$ and $9\pi/10$. 
%Since we want them to have the same area, their initial widths 
%$\Delta \theta_0$ at the different locations are computed using 
%Eq.~\eqref{eq:stripe_area}, leading to 
%$\Delta \theta_0(\pi / 10) \simeq 0.204\pi$, 
%$\Delta \theta_0(\pi / 2) \simeq 0.281\pi$, 
%$\Delta \theta_0(9\pi / 10) \simeq 0.433\pi$.
%The evolution of the free energy $\Psi$ for these three stripes, 
%computed by numerically evaluating the integral in Eq.~\eqref{eq:Landau}, 
%is shown in Fig.~\ref{fig:stripe}(a) as a function of time.
%For all cases considered here, the dynamics of the system are 
%akin to a damped harmonic oscillator. The free energy oscillates 
%with a diminishing amplitude and the minimum of the free energy 
%corresponds to the stable equilibrium at $\theta_c = \pi$, as 
%predicted by Eq.~\eqref{eq:minimaring}.
%Generally the initial configuration provides a sufficient free energy 
%that the ring configuration overshoots the minimum free energy 
%configuration, as demonstrated in Fig.~\ref{fig:stripe}(b), where we 
%plot the centre of the stripe as a function of time. 
%As expected, the amplitude and lifetime of the oscillation are larger 
%the further the stripe is initialised from the equilibrium position. 
%In Fig.~\ref{fig:stripe}(c-f), we show the typical fluid morphologies as
%the stripe undergoes a damped oscillatory motion. 
We consider stripes having the same initial position centred 
at $\theta_0 = \pi / 2$, but initialised with different initial widths $\Delta \theta_0$. 
The area of these stripes is given by
\begin{equation}
 \Delta A = 2\pi r R \Delta \theta_0.
\end{equation}
%For $\Delta A < \Delta A_{\rm crit}$, the stripes migrate 
%towards $\theta_c = \pi$. When $\Delta A > \Delta A_{\rm crit}$, 
%there are two equilibrium positions that can be reached by the stripes.
The equilibrium positions $\theta_c^{eq}$ 
for four different stripes are shown in Fig.~\ref{fig:tor_stripe_fork}(a).
The first case corresponds to a very large stripe 
($\Delta \theta_0 = 0.95\pi$, $\Delta A \simeq 1.88 \Delta A_{\rm crit}$),
for which the possible equilibria $\theta^{eq}_{c}$ are close to
$\pi/2$ and $3\pi /2$. Due to the initial condition, the stripe is 
attracted by the equilibrium point on the upper side of 
the torus, where it will eventually stabilise. 
As the stripe size decreases, 
its kinetic energy as it slides towards the equilibrium point 
will be sufficiently large for it to go over the ``barrier''
at $\theta_c = \pi$ to the lower side of the torus. Because of energy 
loss due to viscous dissipation, its kinetic energy may be insufficient 
to overcome this barrier again, so the stripe remains trapped 
on the lower side. This is the case for the second stripe 
having $\Delta \theta_0 = 0.65\pi$ ($\Delta A \simeq 1.29 \Delta A_{\rm crit}$).
Further decreasing the stripe size 
causes the peak at $\theta_c = \pi$ 
to also decrease, allowing the stripe to overcome it a second time 
as it migrates back towards the upper side.
The third stripe, initialised with 
$\Delta \theta_0 = 0.6 \pi$ ($\Delta A \simeq 1.19 \Delta A_{\rm crit}$),
stabilises on the upper side of the torus.
Finally, the fourth stripe is initialised with 
$\Delta \theta_0 = 0.3 \pi$, such that its area $\Delta A \simeq 0.59 \Delta A_{\rm crit}$
is below the critical value. Thus, the fourth stripe will perform oscillations 
around the equilibrium at $\theta_c = \pi$, where it will eventually stabilise.

Judging by the number of times that the stripe centre $\theta_c$ crosses the barrier at
$\theta_c = \pi$, two types of stripes having $\Delta A > \Delta A_{\rm crit}$
can be distinguished: (i) the 
ones that cross the $\theta_c = \pi$ line an even number of times
stabilise on the upper half of the torus, while (ii) the ones that cross it an odd 
number of times stabilise on the lower half of the torus. 
This is presented in Fig. \ref{fig:tor_stripe_fork}(b), where
the equilibrium position $\theta_c^{eq}$ for stripes initialised at 
$\theta_0 = \pi / 2$ is represented as a function of $\Delta \theta_0$ 
in comparison with the analytical predictions in Eq.~\eqref{eq:stripe_thceq}. 

Panels (c-e) in Fig.~\ref{fig:tor_stripe_fork} illustrate the three scenarios where
the stripes are equilibrated at $\theta_{c}^{eq} > \pi$, $\theta_{c}^{eq} = \pi$, and 
$\theta_{c}^{eq} < \pi$ respectively. 
The total interface lengths $\ell_{\rm total}$ 
($\sim \Psi$) for the stripes shown in (c-e) are represented 
in panels (f-h) of Fig.~\ref{fig:tor_stripe_fork}. 
The interface lengths at the equilibrium positions corresponding 
to the initial state as well as to the turning points corresponding to 
half-periods are also shown using symbols, numbered sequentially 
in the legend ($0$ corresponds to the initial state). It can be seen that 
$\ell_{\rm total}$ measured at these turning points decreases monotonically.
When $\ell_{\rm total}$ decreases below its value at $\theta_c = \pi$,
the stripe centre can no longer cross the $\theta_c = \pi$ line and 
becomes trapped in one of the minima.

\begin{figure}
\begin{center}
\begin{tabular}{c}
 \includegraphics[width=0.5\columnwidth]{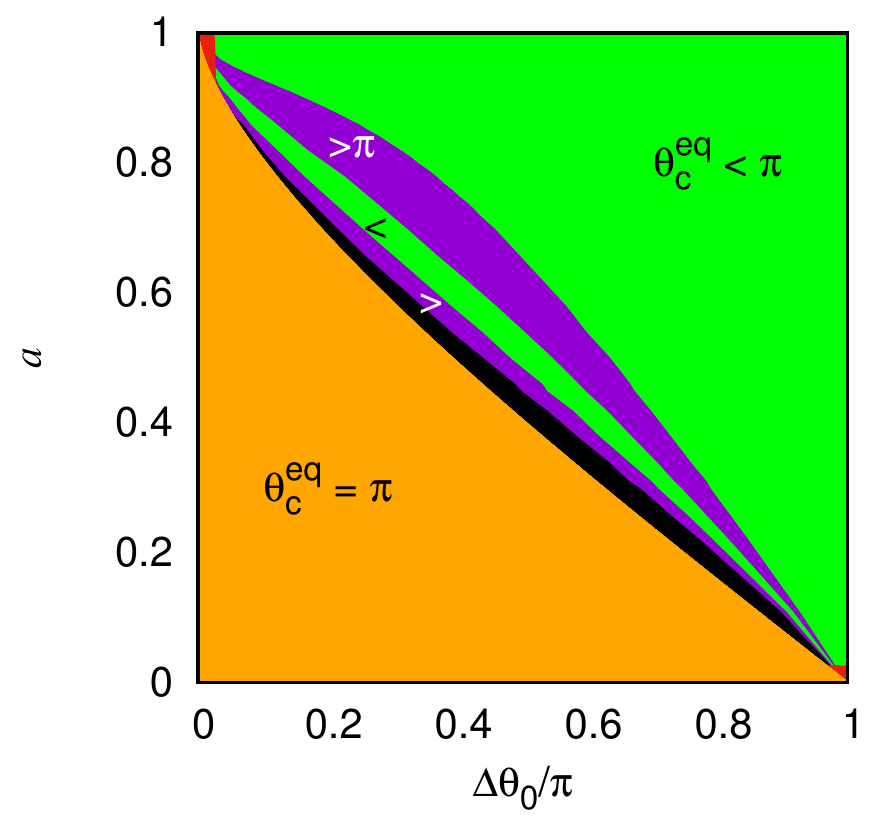}
\end{tabular}
\end{center}
\caption{Diagram indicating the location of the equilibrium position $\theta_c^{eq}$
as a function of the stripe width $\Delta \theta_0$ and the radii ratio $a = r/R$,
for stripes initialised at $\theta_0 = \pi / 2$.
\label{fig:tor_stripe_map}}
\end{figure}

Fig.~\ref{fig:tor_stripe_map} further summarises the location of the equilibrium 
stripe position as a function of the stripe width $\Delta \theta_0$ and 
the radii ratio $a = r / R$. Our simulations are performed by keeping $R = 2$ constant, 
such that the various values of $a$ are obtained by changing $r$.
As before, the stripe is initialised at $\theta_0 = \pi / 2$.
Moving from the top right corner of the diagram towards the bottom left corner, 
the subsequent regions distinguish between whether the stripes stabilise on 
the top half ($< \pi$) or on the bottom half ($> \pi$), depending on the number of times 
that $\theta_c$ crosses $\pi$. In the bottom left corner, the stripes 
stabilise at $\theta_c^{eq} = \pi$. The black region between the purple band and 
the lower left region corresponds to stripes that cross $\pi$ more 
than $3$ times but stabilise away from $\pi$ ($\theta_c^{eq} \neq \pi$).
Due to the diffuse nature of the interface, the stripes evaporate when 
$r \Delta \theta \lesssim 5 \xi_0$ 
($\xi_0 = \sqrt{\kappa / A} \simeq 0.031$).
These regions correspond to the top left and bottom right 
corners of the diagram and are shown in red.

\subsection{Laplace pressure test}
\label{sec:drift:laplace}

Since the stripe interfaces have a non-vanishing curvature, it can 
be expected that there will be a pressure difference across this interface.
This pressure difference is often termed the Laplace pressure.
This pressure difference was recently derived analytically on a torus and the result 
is \citep{Busuioc19bench}
\begin{equation}
 \Delta p = -\frac{\gamma}{R} \frac{\cos\theta_c \sin(\Delta \theta / 2)}
 {1 + a \cos\theta_c \cos(\Delta \theta / 2)}.\label{eq:stripe_laplace_gen}
\end{equation}
This expression can be simplified for the two types of minima highlighted in the
previous subsection
\begin{equation}
 \Delta p = 
 \begin{cases}
  {\displaystyle \frac{\gamma}{R} \frac{\sin(\Delta \theta_{eq} / 2)}
  {1 - a \cos(\Delta \theta_{eq} / 2)}}, & \Delta A < \Delta A_{\rm crit},\\
  {\displaystyle \frac{\gamma}{r} \cot \frac{\Delta \theta_{eq}}{2}}, 
  & \Delta A > \Delta A_{\rm crit},
 \end{cases}\label{eq:stripe_laplace_eq}
\end{equation}
We remind the readers that, on the first branch, $\theta_c^{eq} = \pi$.
On the second branch, the equilibrium position is determined via 
$a \cos \theta_c^{eq} + \cos(\Delta \theta_{eq} / 2) = 0$.

\begin{figure}
 \begin{center}
  \includegraphics[width=0.45\columnwidth]{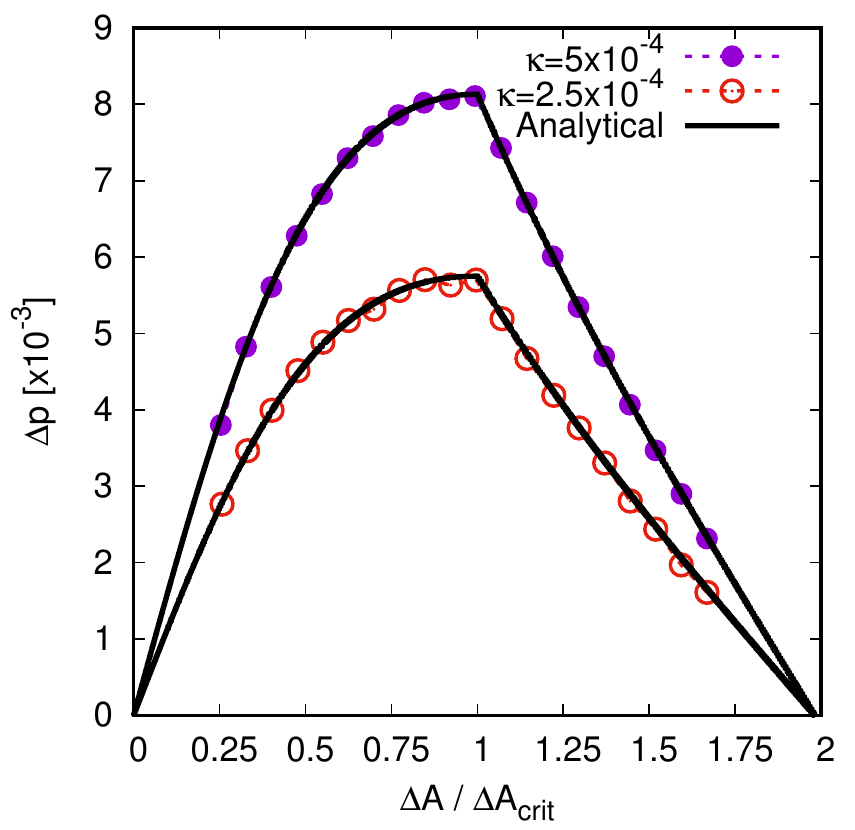}
 \end{center}
\caption{Comparison of the Laplace pressure obtained numerically (dashed lines and circles)
against the the analytic formula, Eq.~\eqref{eq:stripe_laplace_eq}, for $\kappa = 5\times 10^{-4}$
and $2.5 \times 10^{-4}$. The analytic prediction is almost everywhere 
overlapped with the numerical results.
\label{fig:stripe_laplace}}
\end{figure}

In order to validate our numerical scheme against the Laplace pressure test on 
the torus, we perform numerical simulations for two values of $\kappa$ in our free energy model,
 $\kappa = 2.5\times 10^{-4}$ and $5 \times 10^{-4}$. These effectively change
the surface tension and interface width in our simulations, see Eq.~\eqref{eq:xi_gamma}. 
All of the other simulation parameters are kept the same as in the previous subsection:
$R = 2$, $r = 0.8$, $A = 0.5$, $\nu = 2.5 \times 10^{-3}$ and $M = 2.5 \times 10^{-3}$.
We consider stripes of various areas $\Delta A$ in Fig.~\ref{fig:stripe_laplace}. 
%For each value of $\Delta A$, the equilibrium position $\theta_c^{eq}$ 
%is computed and the stripe is initialised at $\theta_c = \theta_c^{eq} - \delta \theta$, 
%where $\delta \theta = 0.05 \pi$. The stripes are allowed to accommodate from 
%an initial hyperbolic tangent profile of the form given in Eq.~\eqref{eq:stripe_tanh},
%where $\phi_0$ is given the value in Eq.~\eqref{eq:stripe_phi0} corresponding to the 
%initial stripe centre and width $\Delta \theta$. The initial stripe width 
%is obtained by numerically solving Eq.~\eqref{eq:stripe_area} for fixed 
%$\Delta A$ and $\theta_c$. In order to reach the stationary state, we performed 
%$5 \times 10^6$ iterations at $\delta t = 0.002$ (up to $t = 10^4$). 
After the stationary state is reached, we measure the total pressure 
$p = p_{\rm i} + p_{\rm binary} = n k_B T + A(-\frac{1}{2} \phi^2 + 
\frac{3}{4} \phi^4)$ in the interior and exterior of the stripe, and 
compute the difference $\Delta p$ between these two values. 
The simulation results are shown using dashed lines and symbols in Fig.~\ref{fig:stripe_laplace}.
We observe an excellent agreement with the analytic results, Eq.~\eqref{eq:stripe_laplace_eq}, 
which are shown using the solid lines.

\subsection{Approach to equilibrium}
\label{sec:drift:damp}

For stripes close to their equilibrium position, the 
time evolution of the departure $\delta\theta = \theta_c - \theta_c^{eq}$ 
can be described as a damped harmonic oscillation:
\begin{equation}
 \delta \theta \simeq \delta \theta_0 \cos(\omega_0 t + \varsigma) e^{-\alpha t},
 \label{eq:stripe_hydro_sol}
\end{equation}
where the damping coefficient $\alpha = \alpha_\nu + \alpha_\mu$ receives 
contributions from the viscous damping due to the fluid~\citep{Busuioc19bench}
\begin{equation}
 \alpha_\nu = \frac{\nu}{R^2 - r^2},\label{eq:stripe_alphanu}
\end{equation}
as well as from the diffusion due to the mobility of the order parameter, $\alpha_\mu$ \citep{Busuioc19bench}.
%It was shown in Ref.~\citep{Busuioc19bench} that 
%$\alpha_\mu \simeq \gamma M / r^3$. 
In the applications considered in 
this paper, $\alpha_\mu \ll \alpha_\nu$, such that 
we will only consider the approximation $\alpha \simeq \alpha_\nu$.
For the case of subcritical stripes ($\Delta A < \Delta A_{\rm crit}$), 
which equilibrate at $\theta_c^{eq}= \pi$, the oscillation frequency 
is \citep{Busuioc19bench}
\begin{equation}
 \omega_0^2 = \frac{\gamma \sqrt{1 - a^2}}{\pi r^2 R n m} 
 \frac{\cos(\Delta \theta_{eq} / 2) - a}{[1 - a \cos(\Delta \theta_{eq} / 2)]^3}.
 \label{eq:stripe_hydro_omega0_pi}
\end{equation}
For the supercritical stripes, ($\Delta A > \Delta A_{\rm crit}$),
when the equilibrium position is at $a \cos\theta_c^{eq} + \cos(\Delta \theta_{eq}/2) = 0$,
$\omega_0^2$ is given by
\begin{equation}
 \omega_0^2 = \frac{2\gamma}{\pi r R^2 n m (1 - a^2)^{3/2}} 
 \left[\frac{\sin \theta_c^{eq}}{\sin(\Delta \theta_{eq} / 2)}\right]^2.
 \label{eq:stripe_hydro_omega0_npi}
\end{equation}

\begin{figure}
 \begin{center}
 \begin{tabular}{cc}
  \includegraphics[width=0.48\columnwidth]{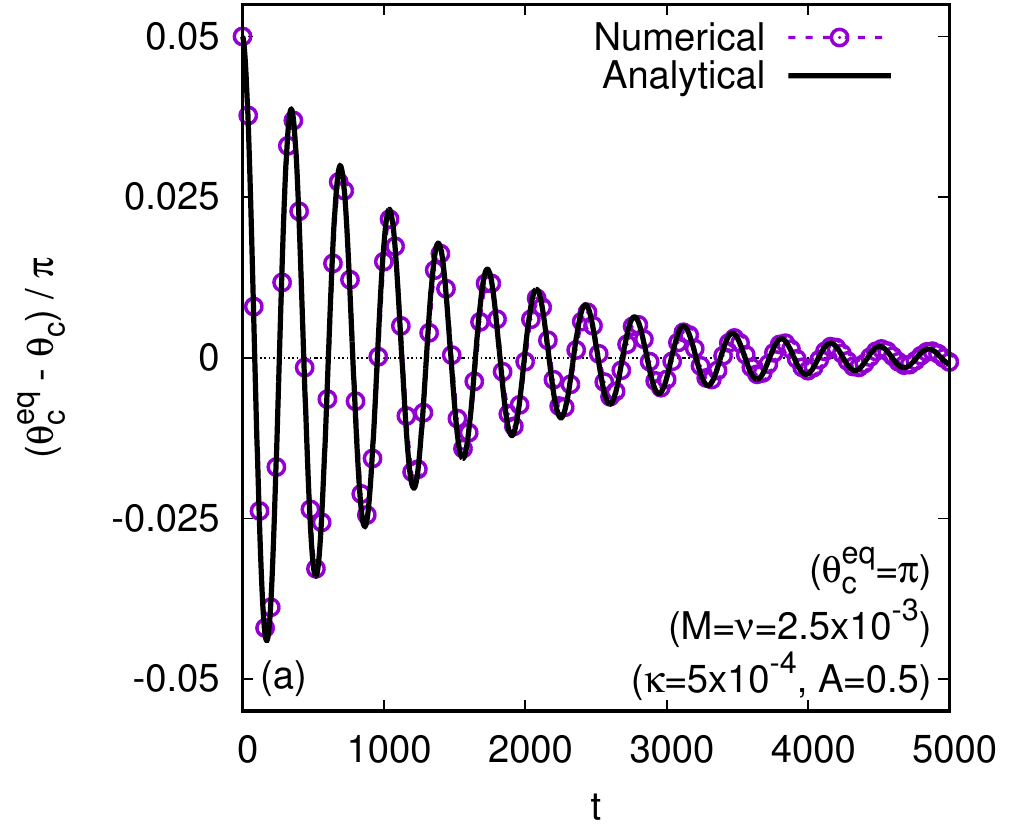} &
  \includegraphics[width=0.48\columnwidth]{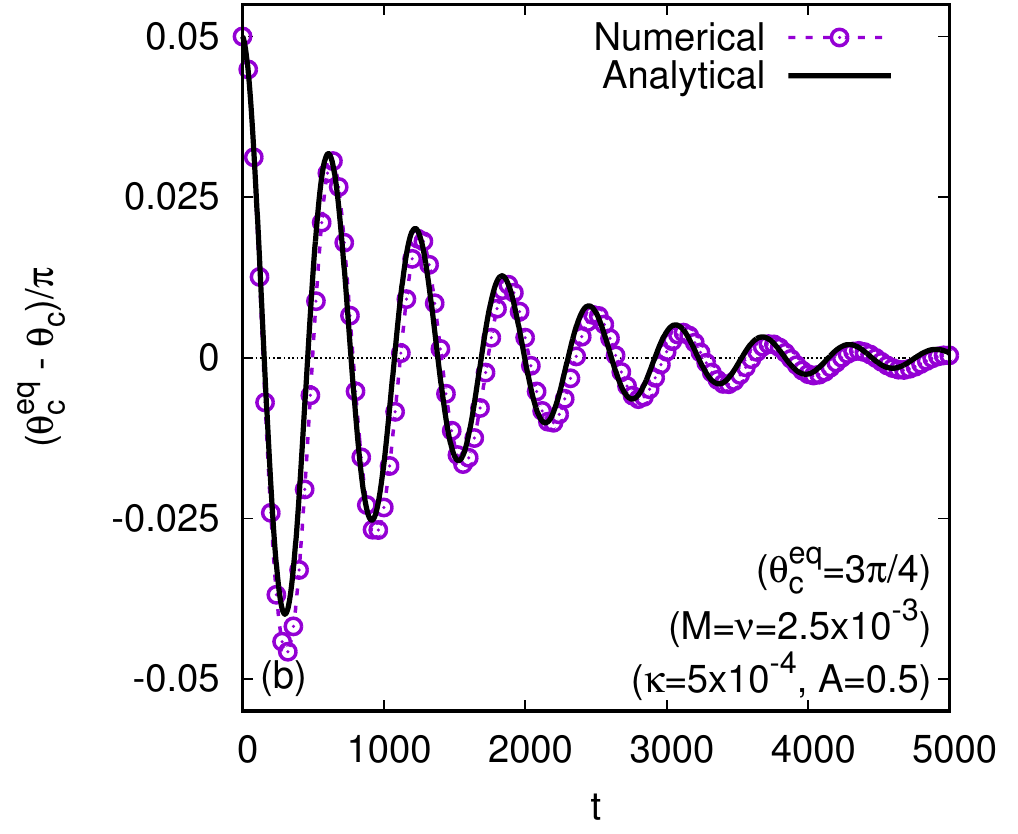}
 \end{tabular}
 \end{center}
\caption{Time evolution of the stripe center $\theta_c$ for stripes
initialised at (a) $\theta_0 = 0.95\pi$ with $\Delta \theta_0 = 0.280\pi$
(equilibrating at $\theta_c^{eq} = \pi$);
and (b) $\theta_0 = 0.7 \pi$ with $\Delta \theta_0 = 0.796\pi$ 
(equilibrating at $\theta_c^{eq} = 3\pi/4$).
The numerical results are shown using dotted lines and symbols, 
while the analytic solutions are shown 
using solid lines.
\label{fig:stripe_hydro_dth}}
\end{figure}

We will now demonstrate that our lattice Boltzmann implementation captures the dynamical 
approach to equilibrium as described by the analytical results.
First, we consider a torus with $r = 0.8$ and $R = 2$ ($a = 0.4$), and set
$\kappa = 5 \times 10^{-4}$, $A = 0.5$ and $\tau = M = 2.5 \times 10^{-3}$.
The number of nodes is $N_\theta = 320$, and the order parameter $\phi$ 
is initialised according to Eq.~\eqref{eq:stripe_tanh}, where the stripe centre is 
located at an angular distance 
$\delta \theta_0 = \theta_c - \theta_c^{eq} = -\pi / 20$ away 
from the expected equilibrium position. 
Fig.~\ref{fig:stripe_hydro_dth} shows a comparison between the 
numerical and analytical results for the time evolution of 
$(\theta_c^{eq} - \theta_c) / \pi$ for the cases 
(a) $\theta_c^{eq} = \pi$ with initial stripe width $\Delta \theta_0 = 0.28 \pi$,
and (b) $\theta_c^{eq} = 3\pi/4$ with $\Delta \theta_0 = 0.786 \pi$.
For the analytical solution, the angular velocity $\omega_0$ 
is computed using Eqs.~\eqref{eq:stripe_hydro_omega0_pi}
and \eqref{eq:stripe_hydro_omega0_npi} for cases (a) and (b)
respectively, and the damping factor $\alpha \simeq \alpha_\nu$ 
is computed using Eq.~\eqref{eq:stripe_alphanu}. We have also
set the offset to $\varsigma = 0$. It can be seen that the analytic expression provides 
an excellent match to the simulation results for the 
stripe that goes to $\theta_c^{eq} = \pi$. For the stripe  
equilibrating to $3\pi/4$, we observe a small discrepancy, 
especially  during the first oscillation period. 
However, the overall agreement is still very good.

\begin{figure}
\begin{center}
\begin{tabular}{cc}
 \includegraphics[width=0.48\columnwidth]{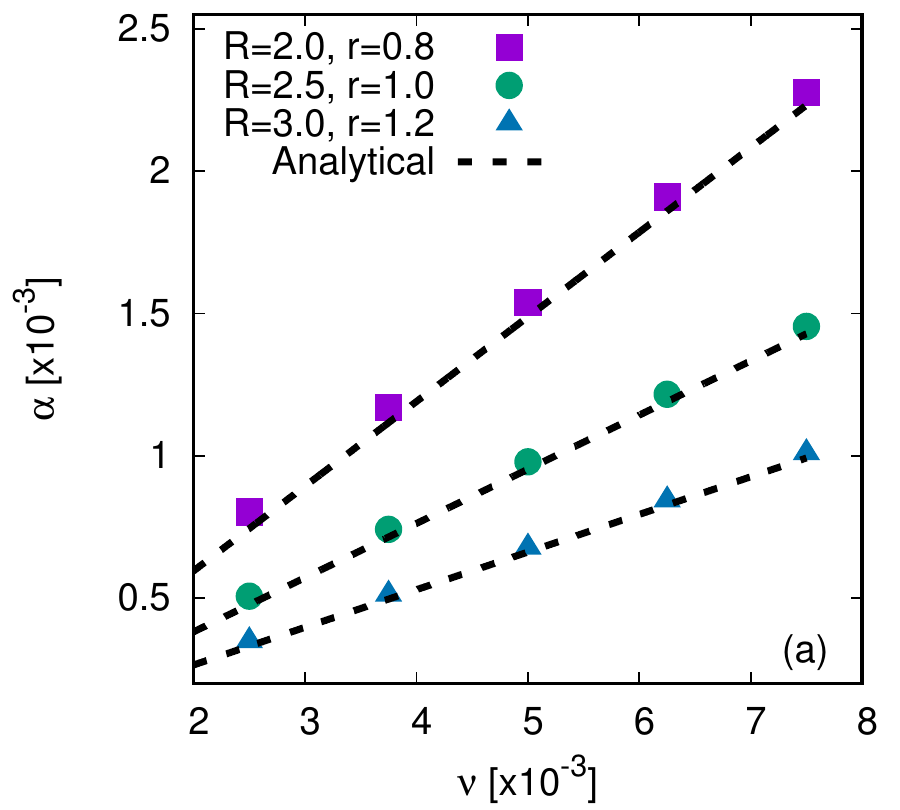} &
 \includegraphics[width=0.48\columnwidth]{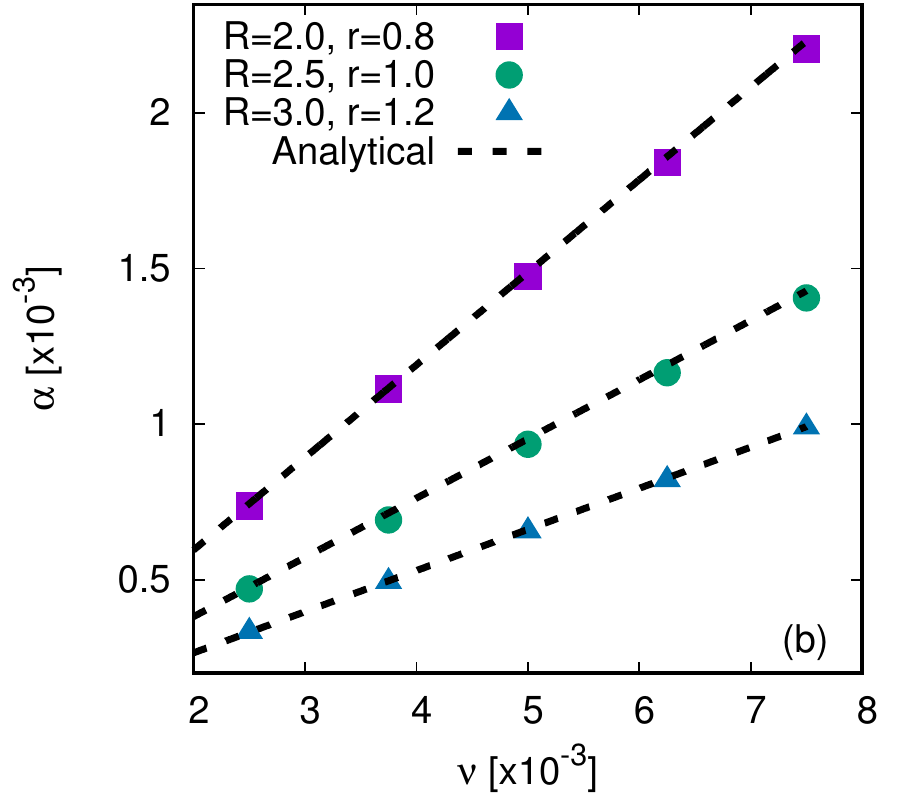} 
\end{tabular}
\end{center}
\caption{The damping coefficient $\alpha$
obtained by fitting Eq.~\eqref{eq:stripe_hydro_sol} to the simulation 
results (points), for stripes initialised at (a) $\theta_0=0.95\pi$ 
with $\theta_c^{eq}=\pi$; and (b) $\theta_0 = 0.7\pi$ with 
$\theta_c^{eq}=0.75\pi$.
The dashed lines represent the viscous damping coefficient 
$\alpha_\nu$, given in Eq.~\eqref{eq:stripe_alphanu}.\label{fig:stripe_hydro_alpha}
}
\end{figure}

\begin{figure}
\begin{center}
\begin{tabular}{cc}
 \includegraphics[width=0.48\columnwidth]{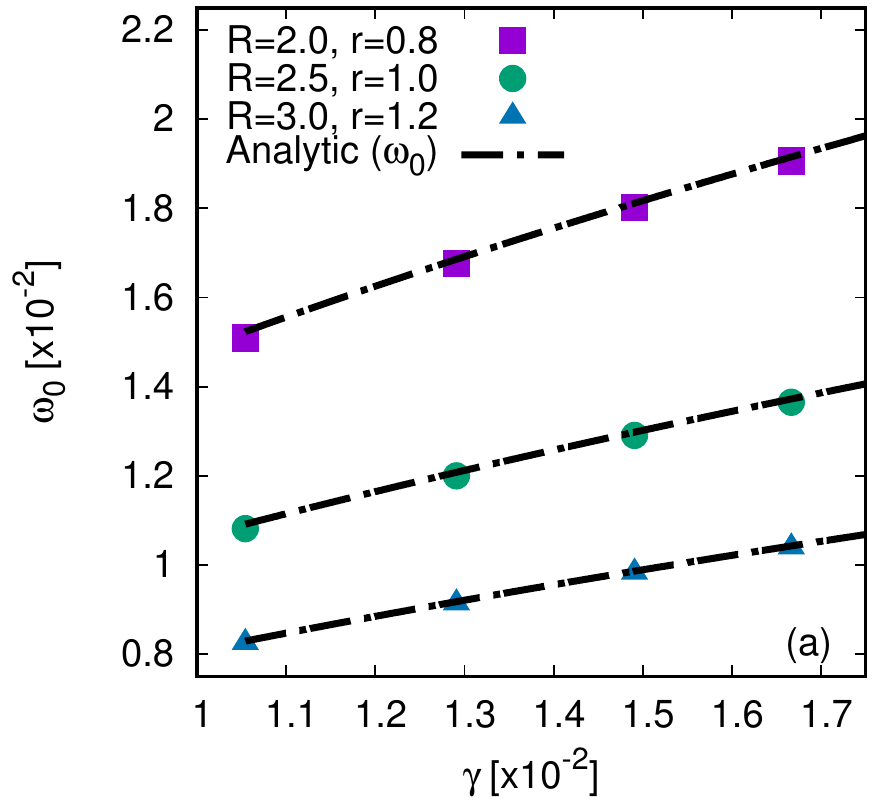} &
 \includegraphics[width=0.48\columnwidth]{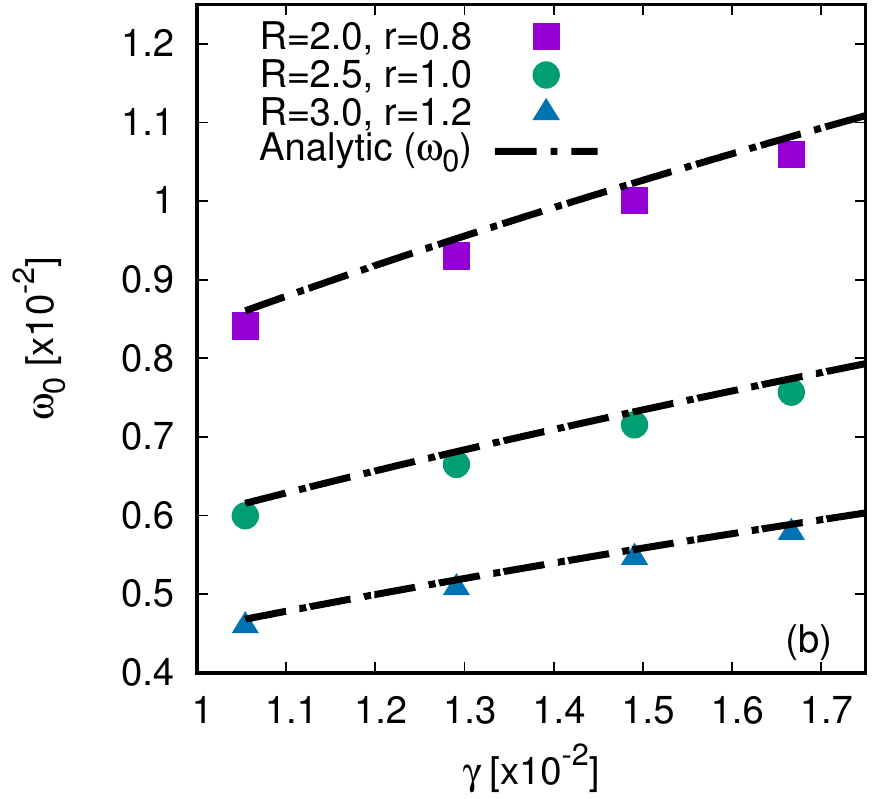}
\end{tabular}
\end{center}
\caption{The angular frequency $\omega_0$, obtained by fitting Eq.~\eqref{eq:stripe_hydro_sol} to the simulation 
results (points).
The black dash-dotted curves correspond to the analytic expressions, as given by 
Eq.~\eqref{eq:stripe_hydro_omega0_pi} for panel (a), when $\theta_c^{eq} = \pi$; and 
Eq.~\eqref{eq:stripe_hydro_omega0_npi} for panel (b), when $\theta_c^{eq} = 3\pi/4$.
\label{fig:stripe_hydro_w}}
\end{figure}

Next we consider three tori having radii ratio $a = r/R = 0.4$, with 
$r = 0.8$, $1$ and $1.2$, and perform two sets of simulations. 
In the first set of simulations, the 
initial configuration corresponds to a stripe centred on 
$\theta_0 = 0.95 \pi$, with initial width $\Delta \theta_0 = 0.28\pi$.
These stripes relax towards $\theta_c^{eq} = \pi$.
In the second set of simulations, the stripes are initially centred at  $\theta_0 = 0.7\pi$,
and they equilibrate at $\theta_c^{eq} = 3\pi/4$, 
with initial width $\Delta \theta_0 = 0.786\pi$.
The simulations are performed using 
$N_\theta = 320$, $400$ and $480$ nodes for $r = 0.8$, $1$ and $1.2$, 
respectively. The best-fit values of $\alpha$ and $\omega_0$ for the three torus geometries 
are shown in Fig.~\ref{fig:stripe_hydro_alpha} and 
Fig.~\ref{fig:stripe_hydro_w} respectively as function of the kinematic viscosity $\nu$ 
(varying between $2.5 \times 10^{-3}$ and $7.5 \times 10^{-3}$) at 
$\kappa = 5 \times 10^{-4}$;
and the surface tension parameter $\kappa$ (varying between $2.5 \times 10^{-4}$ and $6.25 \times 10^{-4}$) at $\nu = 2.5 \times 10^{-3}$.  
For each simulation, Eq.~\eqref{eq:stripe_hydro_sol} is fitted to the numerical data 
for the time evolution of the stripe centre as  it relaxes towards equilibrium,
using $\alpha$ and $\omega$ as free parameters, while 
$\varsigma = 0$. For simplicity, we have used $M = \nu$ and $A = 0.5$ in Fig.~\ref{fig:stripe_hydro_alpha} and 
Fig.~\ref{fig:stripe_hydro_w}.
Panels (a) in Fig.~\ref{fig:stripe_hydro_alpha} and 
Fig.~\ref{fig:stripe_hydro_w} correspond to stripes 
equilibrating at $\theta_c^{eq} = \pi$,
while panels (b)  in Fig.~\ref{fig:stripe_hydro_alpha} and 
Fig.~\ref{fig:stripe_hydro_w} are for $\theta_c^{eq} = 3\pi / 4$. 
It can be seen that the analytic expressions are in good agreement 
with the numerical data in all instances simulated.

\begin{figure}
\begin{center}
\begin{tabular}{c}
 \includegraphics[width=0.48\columnwidth]{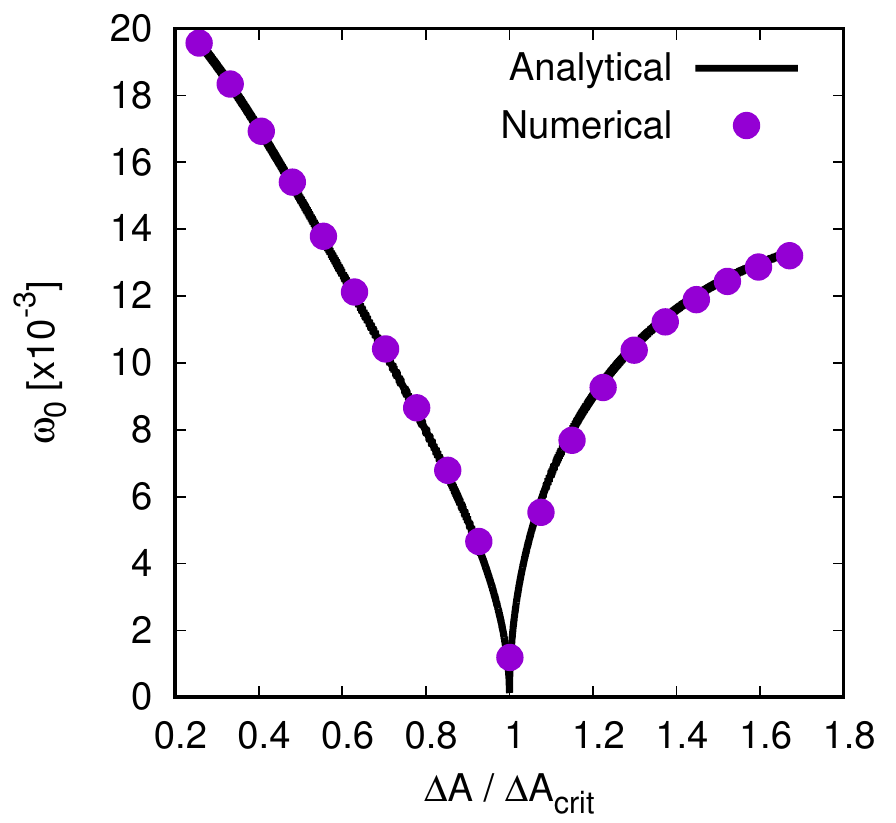}
\end{tabular}
\end{center}
\caption{
Comparison between the values of $\omega_0$ obtained by fitting 
Eq.~\eqref{eq:stripe_hydro_sol} to the numerical results, shown 
with points, and 
the analytic expressions, Eq.~\eqref{eq:stripe_hydro_omega0_pi} 
for $\Delta A< \Delta A_{\rm crit}$ and Eq.~\eqref{eq:stripe_hydro_omega0_npi} for $\Delta A > \Delta A_{\rm crit}$, shown with solid black lines.
\label{fig:stripe_hydro_w_all}}
\end{figure}

Finally we investigate the applicability of
Eqs.~\eqref{eq:stripe_hydro_omega0_pi} and \eqref{eq:stripe_hydro_omega0_npi} 
with respect to various values of the stripe area, $\Delta A$. The simulations 
are now performed on the torus with $r =0.8$ and $R = 2$, using 
$\kappa = 5 \times 10^{-4}$, $A = 0.5$, $\tau = M = 2.5 \times 10^{-3}$.
%$N_\theta = 320$ nodes and $\delta t = 2 \times 10^{-3}$.
Fig.~\ref{fig:stripe_hydro_w_all} shows the values of $\omega_0$
obtained by fitting Eq.~\eqref{eq:stripe_hydro_sol} to the numerical 
data (points) and the analytic expressions (solid lines). 
As before, for the fitting, we set $\varsigma = 0$, and use $\alpha$ and $\omega_0$ as free 
parameters. An excellent agreement can be seen, even for 
the nearly critical stripe, for which $\omega_0$ is greatly decreased.

\subsection{Droplets on Tori} \label{sec:drift:drop}

\begin{figure}
\begin{center}
\begin{tabular}{c}
\hspace{-20pt} \includegraphics[width=0.5\columnwidth]{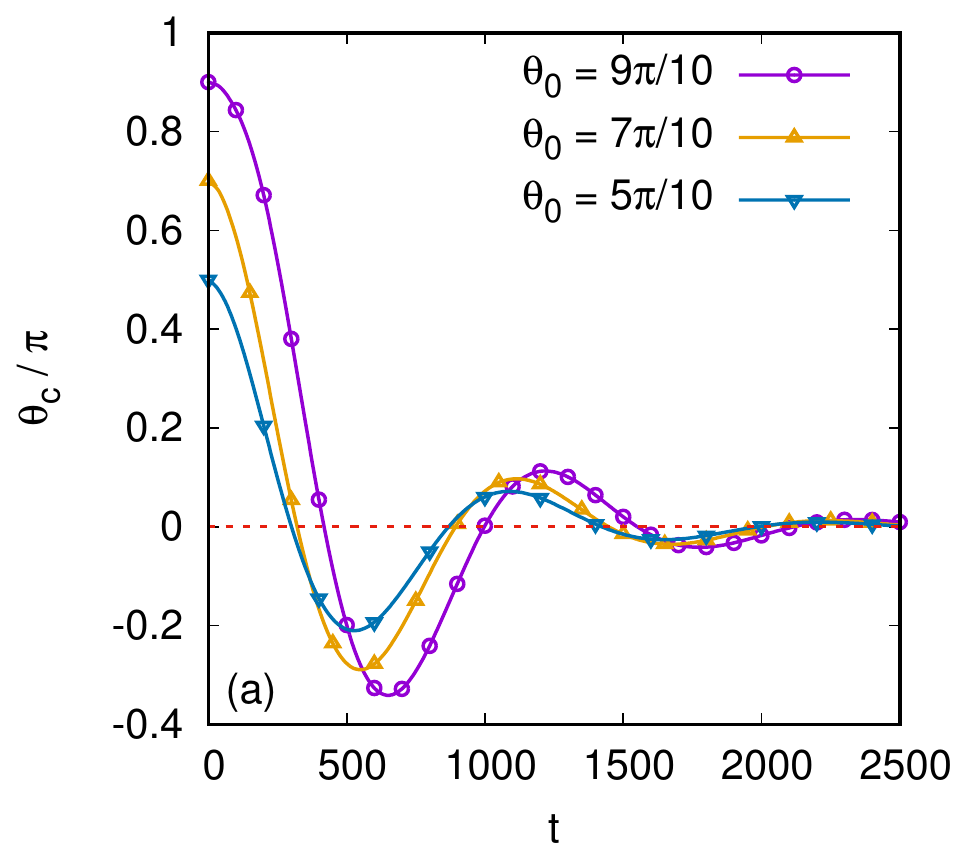} \\
 \vspace{2 mm}
\begin{tabular}{ccc}
 \includegraphics[width=0.31\columnwidth]{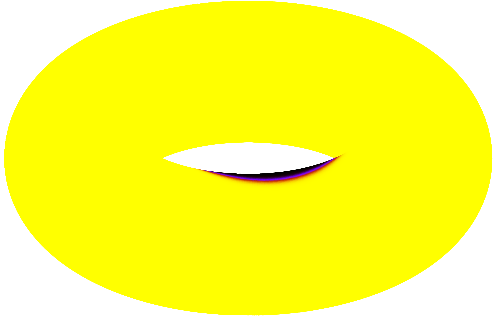}\vspace{2 mm}&
 \includegraphics[width=0.31\columnwidth]{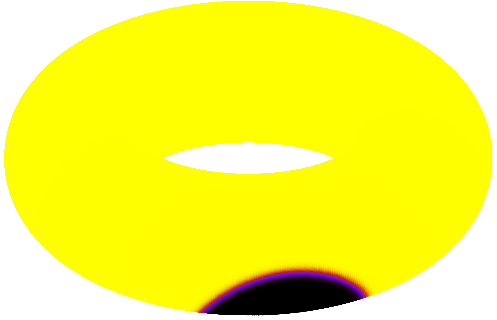} \vspace{2 mm}&
 \includegraphics[width=0.31\columnwidth]{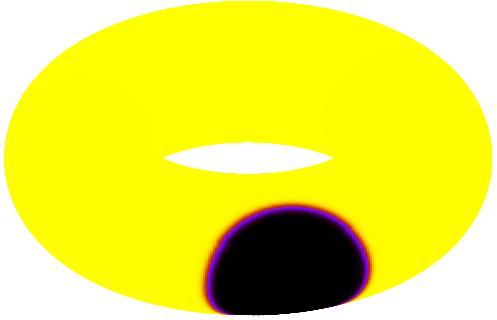} \vspace{2 mm} \\
 (b) $t = 0$ & (c) $t = 650$ & (d) $t = 1775$
\end{tabular}
\end{tabular}
\caption{
(a) Time evolution of the position of the center $\theta_c / \pi$ 
for drops initialised according to Eq.~\eqref{eq:drop_init} with 
$(\theta_0, R_0) \in \{(5\pi/10,0.938),
(7\pi/10,0.924), (9\pi/10,0.910)\}$.
(b--d) Snapshots of the evolution of the drop corresponding to 
$\theta_0= 9 \pi / 10$ for $t = 0$, $650$ and $1775$.
\label{fig:drops}}
\end{center}
\end{figure}

We will now show that, when placed on a torus, a fluid droplet will also exhibit
a drift motion. However, in contrast to stripes, the drops will move towards the outer 
rather than the inner side of the torus. To study this phenomenon quantitatively, 
we initialise drops on a torus using the following equation
\begin{equation}
 \phi_{\rm drop}(\theta_0, R_0; \theta, \varphi) = 
 \tanh \frac{r - R_0}{\xi_0 \sqrt{2}},
 \label{eq:drop_init}
\end{equation}
where $r = \sqrt{(x-x_c) + (y-y_c) + (z-z_c)}$ is the Euclidean distance  between 
the point with coordinates $(x,y,z)$ and the centre of the drop $(x_c, y_c, z_c)$,
corresponding to $(\theta, \varphi)$ and $(\theta_0, 0)$ in polar coordinates respectively.
The relation between the Cartesian and polar coordinates are given in Eq.~\eqref{eq:torus}
of Appendix \ref{app:curved}.
The parameter $\theta_0$ represents the center of the drop, while 
$R_0$ is a measure of its radius. $\xi_0$ is the interface width derived for the 
Cartesian case. In principle the interfacial profile will be different on a torus,
but currently we are not aware of a closed analytical formula. 
We also do not 
introduce in Eq.~\eqref{eq:drop_init} the offset $\phi_0$ responsible for the 
Laplace pressure difference, since the analysis of this quantity is less 
straightforward than for the azimuthally-symmetric stripe domains discussed 
in the previous subsections.

In order for the drops to have approximately the same areas, for a given value of $\theta_0$,
$R_0$ is obtained as a solution of 
\begin{equation}
 \int_0^{2\pi} d\varphi \int_0^{2\pi} d\theta (R + r\cos\theta)
 [\phi_{\rm drop}(0, 30\xi_0; \theta, \varphi) - 
 \phi_{\rm drop}(\theta_0, R_0; \theta, \varphi)] = 0,
\end{equation}
where the first term in the parenthesis corresponds to the configuration when the droplet
is centred on the outer equator and has $R_0 = 30\xi_0$. The drift phenomenon we report 
here is robust with respect to the drop size, but we choose a relatively large drop size because
small drops are known to evaporate in diffuse interface models. 
The simulation parameters are the same as in Sec.~\ref{sec:drift:damp}, 
namely $r =0.8$, $R = 2$, $\kappa = 5 \times 10^{-4}$, $A = 0.5$
and $\tau = M = 2.5 \times 10^{-3}$.

As shown in Fig.~\ref{fig:drops}(a), similar to the stripe configuration in the
previous sub-section, we observe a damped oscillatory motion. 
Here the three drops are initialised at different positions on the torus.
Moreover, as is commonly the 
case for an underdamped harmonic motion, the drops initially overshoot
the stable equilibrium position, but they eventually relax to the minimum energy configuration.
For the drops we find that all drops eventually drift to $\theta = 0$ (the outer side of the torus). Typical drop configurations
during the oscillatory motion are shown in Fig.~\ref{fig:drops}(b-d). Compared to the oscillatory 
dynamics for the stripe configurations, we also observe that the oscillation dies out quicker for the
drops. 
% This is due to a combination of larger viscous dissipation and smaller inertia for the drop motion, 
% compared to the fluid stripes in section \ref{sec:drift:damp}.

\section{Phase Separation}\label{sec:res:growth}

In this section we investigate binary phase separation on the torus 
and compare the results against those on flat surfaces. We consider 
hydrodynamics and diffusive regimes for both even (section 
\ref{sec:res:growth:even}) and uneven (section 
\ref{sec:res:growth:uneven}) mixtures.

The fluid order parameter at lattice point $(s,q)$ is initialised as 
\begin{equation}
 \phi_{s,q} = \overline{\phi} + (\delta \phi)_{s,q},\label{eq:phirand}
\end{equation}
where $\overline{\phi}$ is a constant and $(\delta \phi)_{s,q}$ is 
randomly distributed between $(-0.1, 0.1)$.
We characterise the coarsening dynamics using the instantaneous
domain length scale $L_d(t)$, computed using the following function:
\begin{equation}
 L_d(t) = \frac{A_{\rm total}}{L_I(t)},\label{eq:cart_scaling_L0}
\end{equation}
where $A_{\rm total}$ is the total area of the simulation domain.
The total interface length at time $t$, $L_I(t)$, is computed by 
visiting each cell $(s, q)$ exactly once, starting from the bottom left 
corner, where $s = q = 1$, and progressing towards the top right corner,
where $s = N_1$ and $q = N_2$. $N_1 = N_x$ and $N_2 = N_y$ 
for the Cartesian domains and $N_1 = N_\varphi$ and $N_2 = N_\theta$ 
for the torus domains. 
For each cell where $\phi_{s,q} \times \phi_{s+1,q} < 0$, 
the length of the vertical interface between the $(s,q)$ 
and $(s + 1, q)$ is added to $L_I$. 
In the case of the Cartesian geometry, this length is $\delta y$, while for 
the torus, the length is given by $r \delta \theta$.
Similarly, if $\phi_{s,q} \times \phi_{s,q+1} < 0$, the length of the horizontal 
interface ($\delta x$ for the Cartesian case and 
$(R + r \cos\theta_{q + 1/2}) \delta \varphi$ for the torus case,
where $\theta_{q+1/2} = \theta_q + \delta \theta / 2$ is the 
coordinate of the cell interface) is added to $L_I$. The periodic 
boundary conditions allow the cells with $(N_1 + 1, q)$ and 
$(s, N_1 + 1)$ to be identified with the cells $(1, q)$ and $(s, 1)$,
respectively.

Unless specified otherwise, we use the following 
parameters in this phase separation section: $M = \tau = 2.5 \times 10^{-3}$, 
$\delta t = 5 \times 10^{-4}$, $A = 0.5$ and $\kappa = 5 \times 10^{-4}$. 
In the initial state, the distributions for the 
LB solver are initialised using Eq.~\eqref{eq:feq} with 
a constant density $n_0 = 20$ and vanishing velocity.

\subsection{Even mixtures}\label{sec:res:growth:even}

\subsubsection{Cartesian Geometry}

\begin{figure}
\begin{center}
\begin{tabular}{cc}
 \includegraphics[width=0.4\columnwidth]{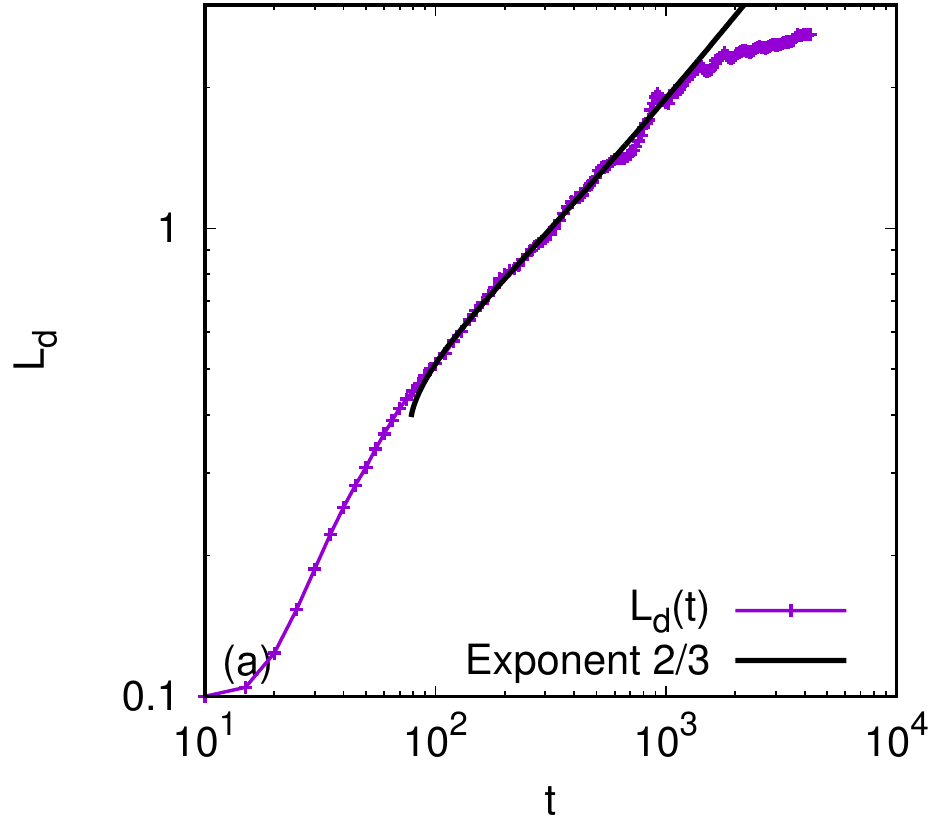} &
 \includegraphics[width=0.4\columnwidth]{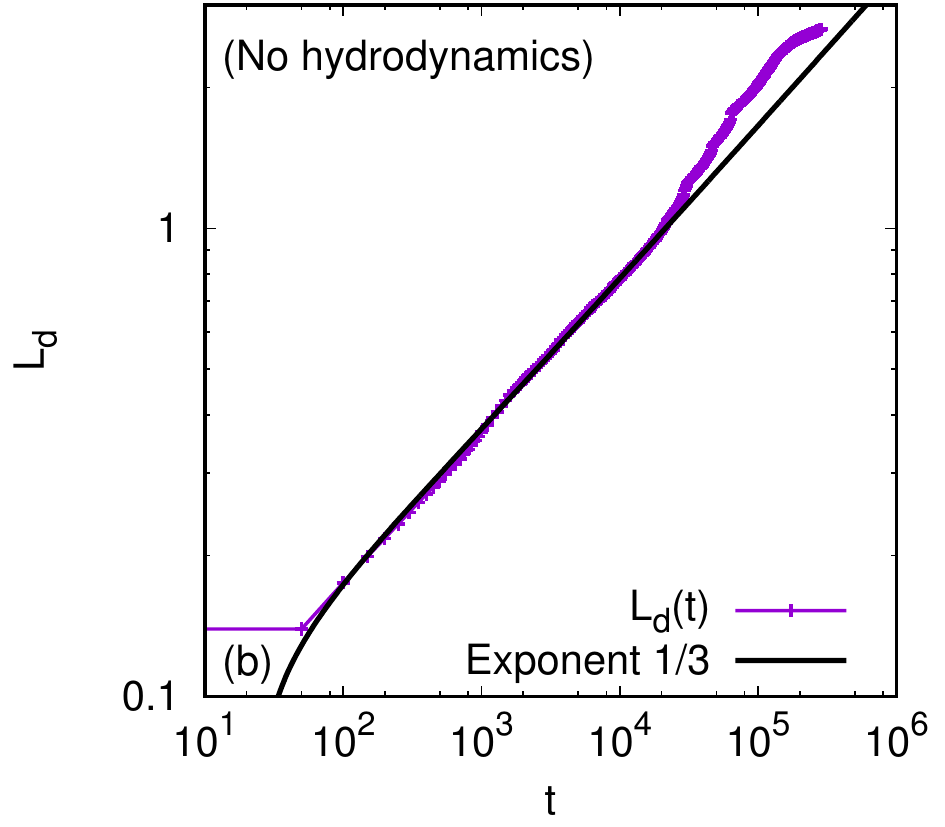} \\
\end{tabular}
\begin{tabular}{cccc}
 \includegraphics[width=0.225\columnwidth]{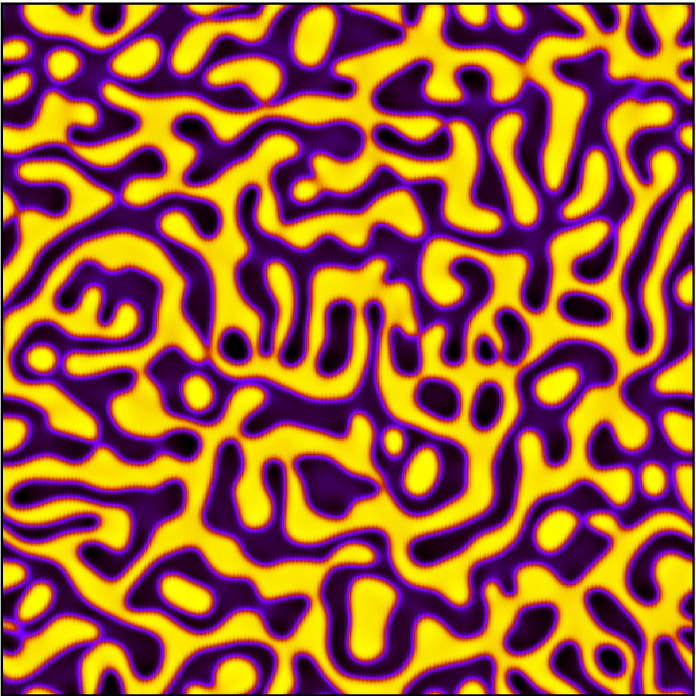} &
 \includegraphics[width=0.225\columnwidth]{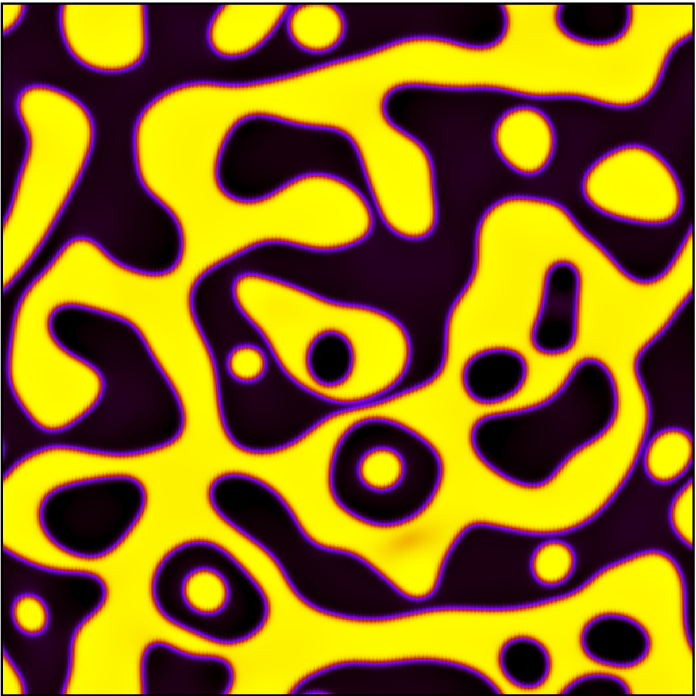}  & 
 \includegraphics[width=0.225\columnwidth]{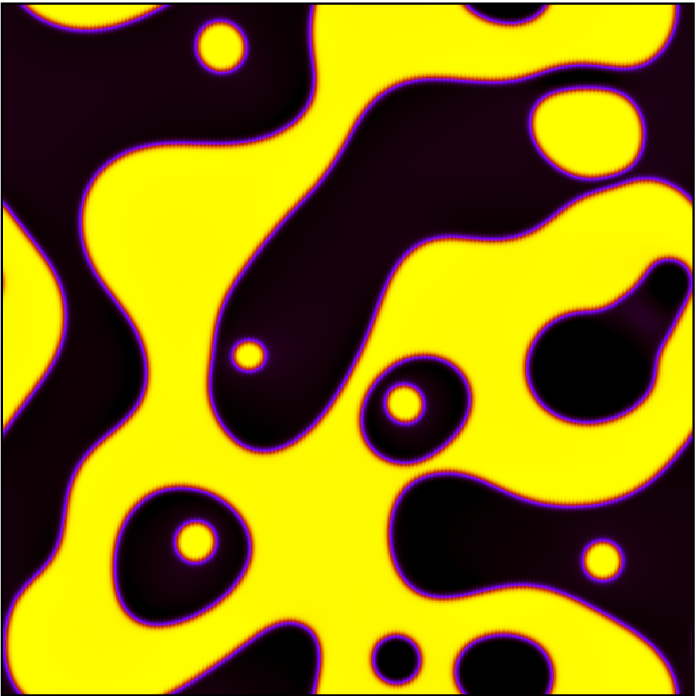}  & 
 \includegraphics[width=0.225\columnwidth]{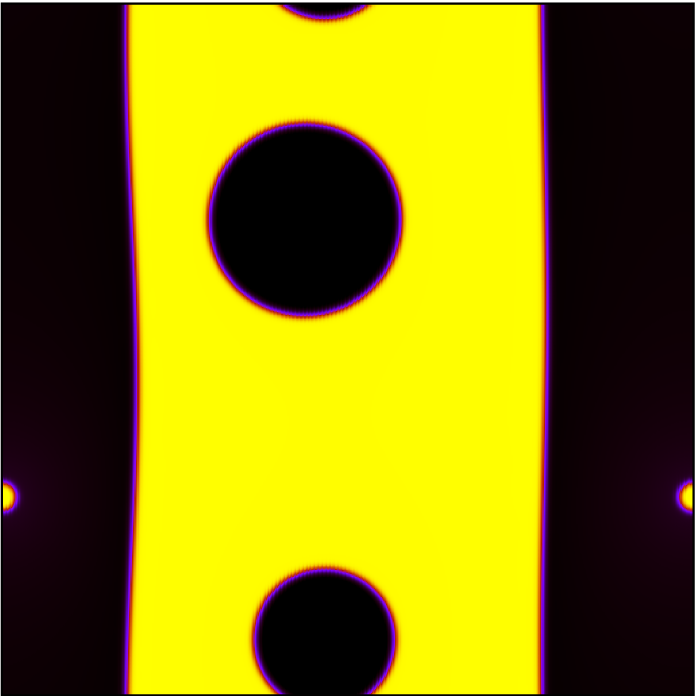}  \vspace{1 mm}\\
 (c) $t = 40$ & (d) $t = 100$  & (e) $t = 250$ & (f) $t = 3000$ \vspace{1 mm}\\
\end{tabular}
\begin{tabular}{cccc}
 \includegraphics[width=0.225\columnwidth]{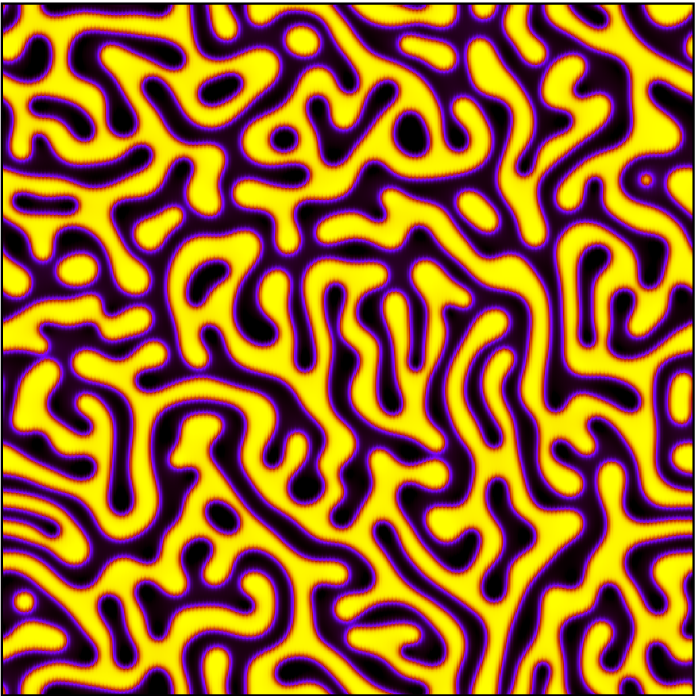} &
 \includegraphics[width=0.225\columnwidth]{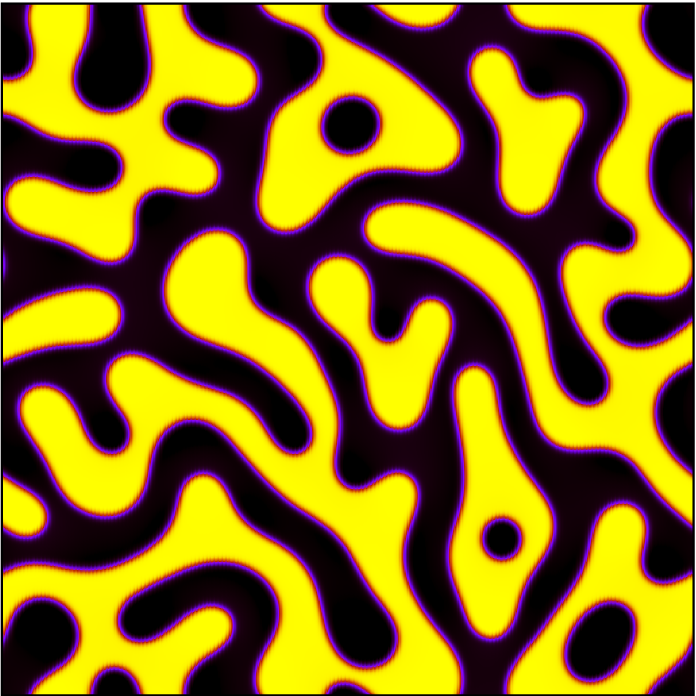}  & 
 \includegraphics[width=0.225\columnwidth]{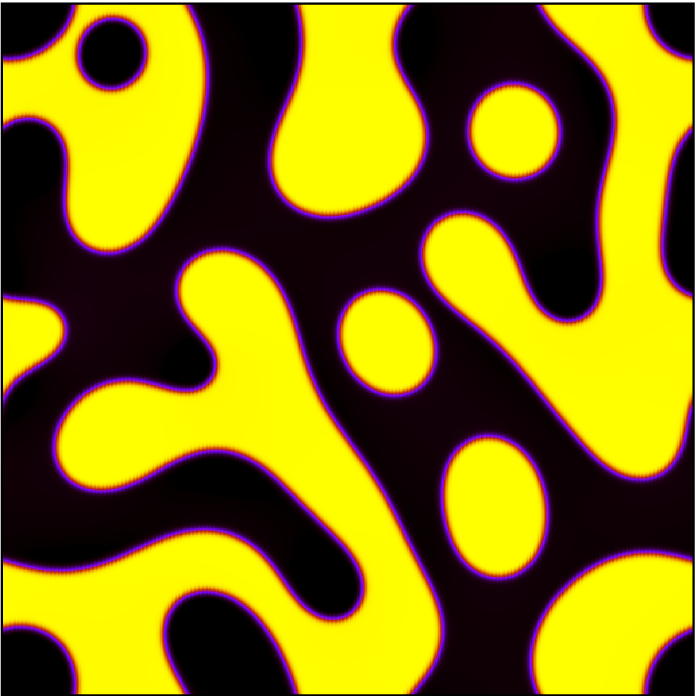} &
 \includegraphics[width=0.225\columnwidth]{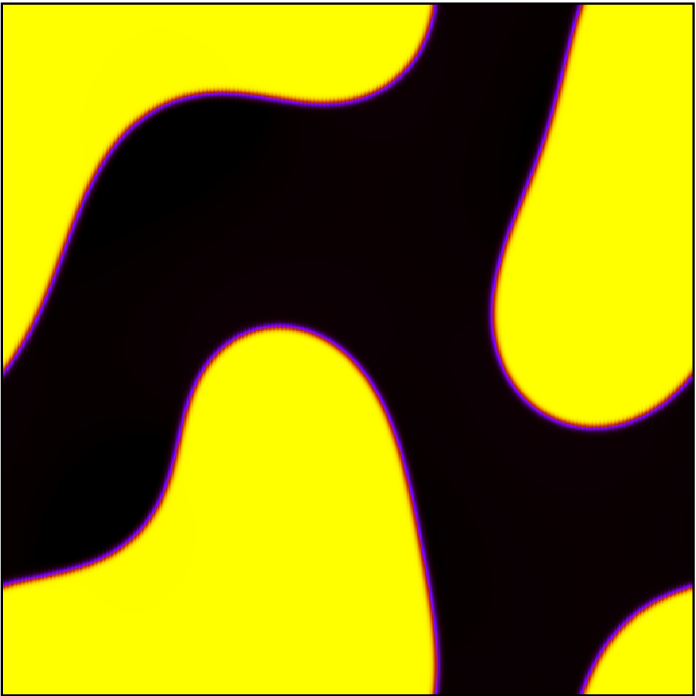} \vspace{1 mm}\\
 (g) $t = 300$ & (h) $t = 2700$  & (i) $t = 15000$  & (j) $t = 168000$
\end{tabular}
\caption{
Growth of the fluid domain size $L_d(t)$ for an even mixture in two dimensions in (a) the inertial-hydrodynamics
and (b) the diffusive regimes. For the diffusive regime, we remove the convective term in the Cahn-Hilliard
equation. 
(c-f) Snapshots of the typical fluid configurations at $t = 40$, $100$, $250$ and $3000$ 
corresponding to the case indicated in panel (a).
(g-j) Snapshots of the fluid configurations corresponding to the case indicated in panel (b),
at times $t = 300$, $2700$, $15000$, and $168000$. These are selected such that $L_d(t)$ 
matches the values corresponding to panels (c-f).
\label{fig:cart_scal}
}
\end{center}
\end{figure}

We begin by considering the coarsening dynamics of a phase separating binary fluid with 
even mixtures on a flat two-dimensional surface. We use a simulation domain of 
$N_x \times N_y = 512 \times 512$ with a grid spacing of $\delta x = \delta y = 0.02$. 
The linear size of the simulation domain is $L = 512 \times 0.02 = 10.24$ and its 
total area is $A_{\rm total} = L^2$. 

As shown in Fig. ~\ref{fig:cart_scal}(a), we observe that the fluid domain grows with an 
exponent of $2/3$. This exponent is often associated with the so-called inertial-hydrodynamics
regime for binary fluid phase separation in three dimensions \citep{Bray02,Kendon01}.
However, in two dimensions, it has been argued that self-similar growth in the 
inertial-hydrodynamics regime may be absent \citep{Wagner97}. The apparent exponent of $2/3$
is really due to a mixture of viscous exponent of $1$ for the growth of the connected domains and 
an exponent of $1/3$ for the diffusive dissolution of circular droplets.

Classical morphologies typical of a spinodal decomposition phenomenon are shown 
in Fig.~\ref{fig:cart_scal}(c-f). The deviation from this apparent scaling law is observed 
at early times when domains of fluid components A and B are formed from the initial perturbation,
and at late times when the domains become comparable in size to the simulation box. 
For the latter, there are very few domains left (see e.g. Fig.~\ref{fig:cart_scal}(f)) and coarsening 
slows down because of the lack of coalescence events between the fluid domains. 
%Similar finite-size effects have been observed for systems in planar geometries.

To access the diffusive regime, in this work we remove the advection term in 
the Cahn-Hilliard equation and decouple it from the Navier-Stokes equation.
In this case, coarsening can only occur via diffusive dynamics, and indeed we do observe 
a growth exponent of $1/3$, as shown in Fig.~\ref{fig:cart_scal}(b), as expected for diffusive 
dynamics \citep{Bray02,Kendon01}. Representative configurations from the coarsening
evolution are shown in  Fig.~\ref{fig:cart_scal}(g-j). These snapshots look somewhat similar to 
those shown in Fig.~\ref{fig:cart_scal}(c-f) for the apparent $2/3$ scaling regime. The key difference
between the morphologies is that more small droplets are accumulated during coarsening when
hydrodynamics is on. It is also worth noting that the coarsening dynamics are much slower
in the diffusive regime. At late times we see a deviation from the diffusive scaling exponent, where 
$L_d(t)$ appears to grow faster than $1/3$ exponent. In this limit, as illustrated in
 Fig.~\ref{fig:cart_scal}(j), the increase in $L_d(t)$ is primarily driven by finite size effects.

\subsubsection{Torus Geometry}

We now consider the coarsening dynamics of a phase separating binary fluid on the
surface of a torus. Initially we simulate a torus domain with $R = 2.5$ and $r = 1$
($a = r / R = 0.4$). These parameters are chosen such that the total area, 
$A_{\rm total} = 4\pi^2 r R$, is close to the one employed in the Cartesian case.
%having a total surface area of $4\pi^2 r R \simeq 98.7$. 
The $\varphi$ direction is discretised using $N_\varphi = 800$ nodes, 
%such that the effective spacing satisfies $2 \pi (R-r)/N_\varphi (=0.0117) <  \delta s < 2 \pi (R + r) / N_\varphi 
%(= 0.0275)$. 
while the $\theta$ direction is discretised using $N_\theta =400$ nodes.
%with an effective spacing $\delta s = 2\pi r / N_\theta \simeq 0.0157$. 
The fluid order parameter at lattice point $(s,q)$ is initialised 
according to Eq.~\eqref{eq:phirand} with $\overline{\phi} = 0$.
%The simulation parameters are  $M = \tau = 2.5 \times 10^{-3}$, $\delta t = 5 \times 10^{-4}$, $n = 20$, $A = 0.5$ and $\kappa = 5 \times 10^{-4}$.

\begin{figure}
\begin{center}
\begin{tabular}{cc}
 \includegraphics[width=0.4\columnwidth]{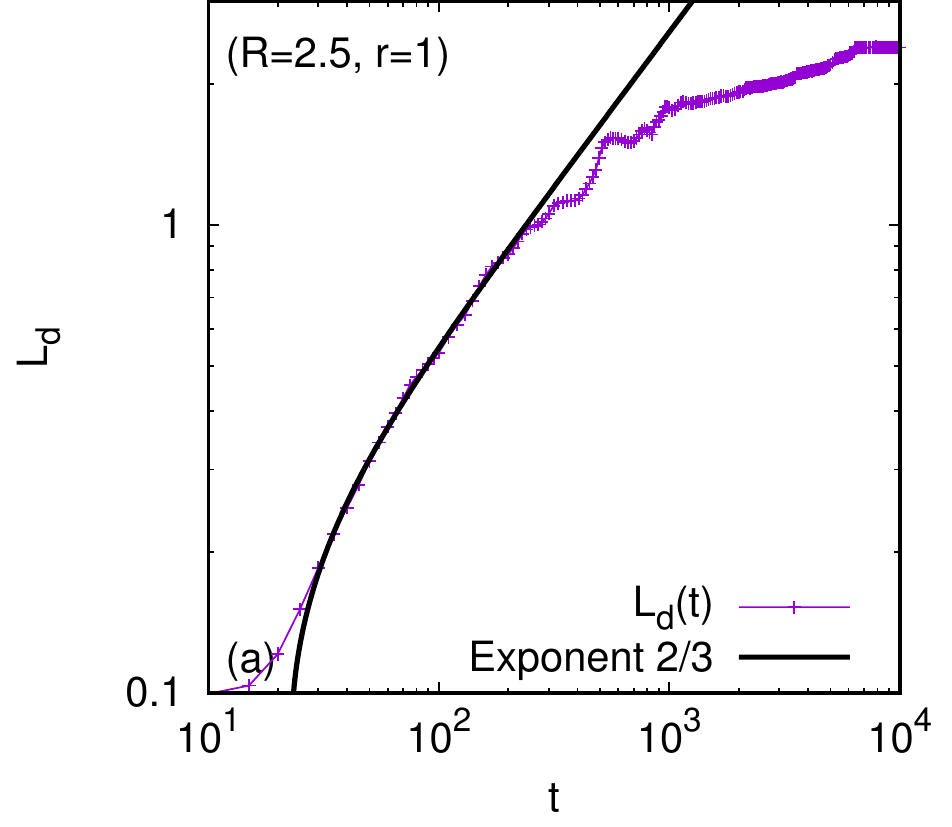} &
 \includegraphics[width=0.4\columnwidth]{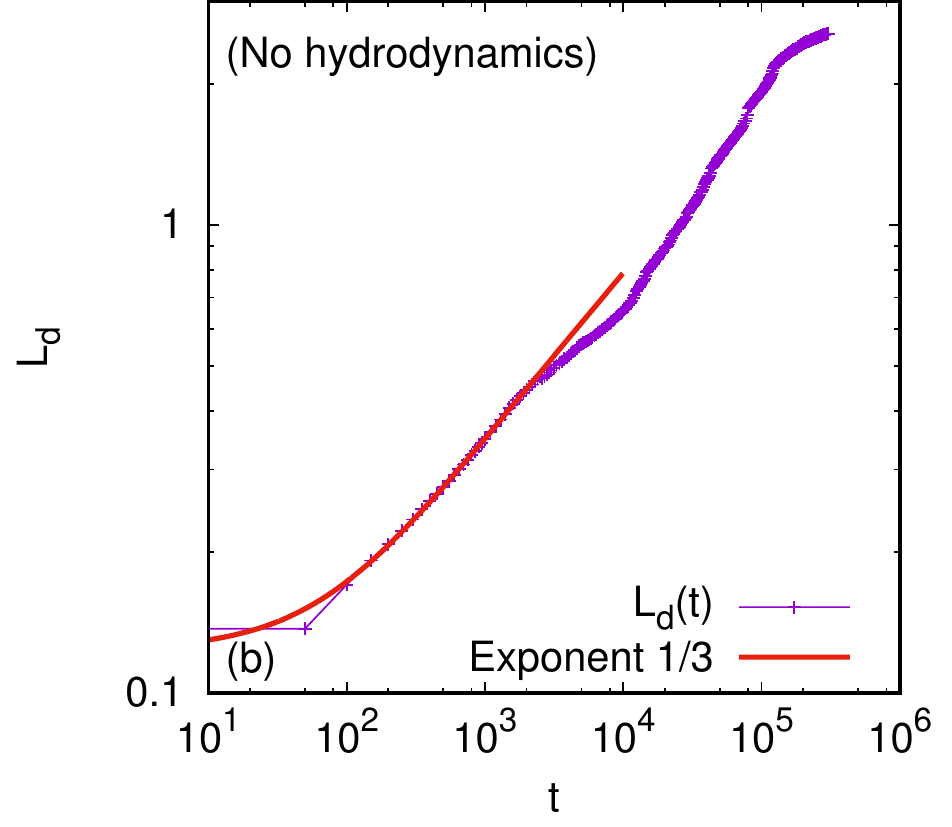} \\
\end{tabular}
\begin{tabular}{ccc}
 \includegraphics[width=.3\columnwidth]{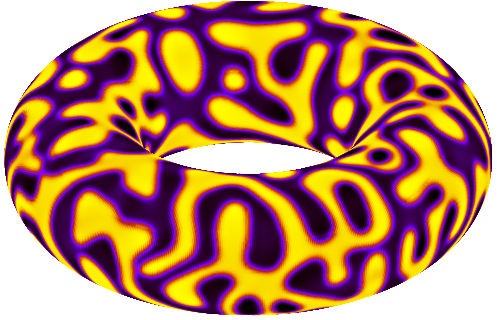} &
 \includegraphics[width=.3\columnwidth]{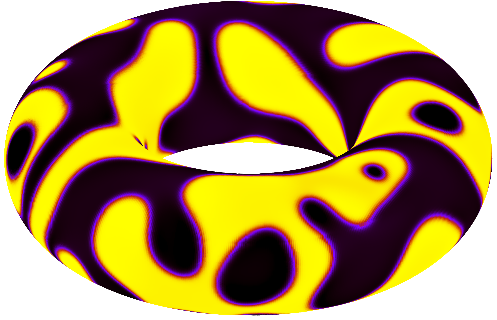} &
 \includegraphics[width=.3\columnwidth]{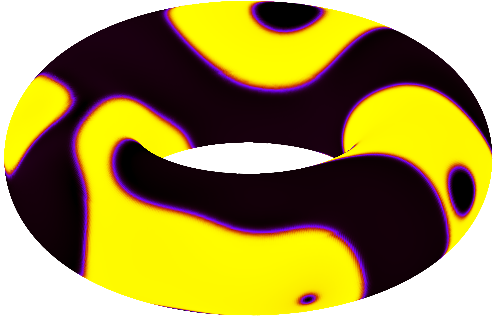} \vspace{1 mm}\\
 (c) $t = 40$ & (d) $t = 100$ & (e) $t = 250$ \vspace{1 mm}\\
 \includegraphics[width=.3\columnwidth]{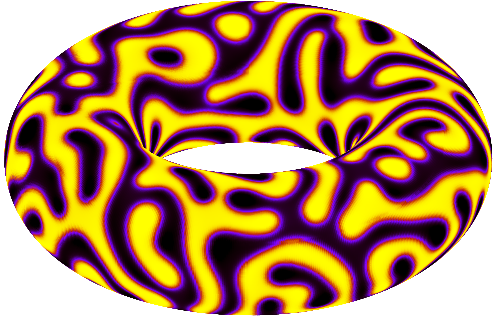} &
 \includegraphics[width=.3\columnwidth]{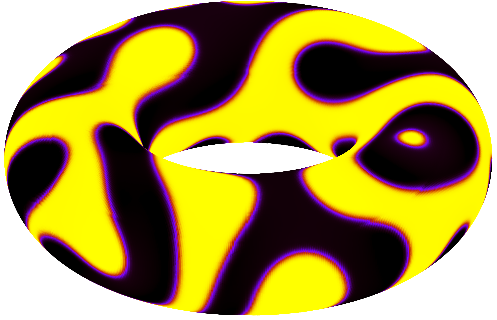} &
 \includegraphics[width=.3\columnwidth]{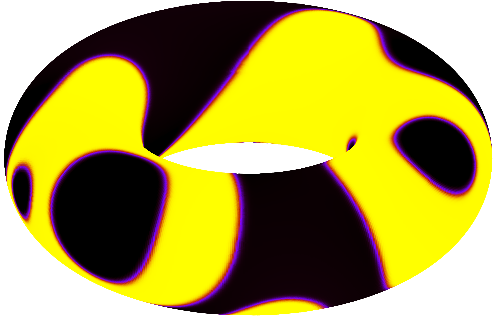} \vspace{1 mm}\\
 (f) $t = 350$ & (g) $t = 4250$ & (h) $t = 25500$ 
\end{tabular}

\caption{
Growth of the fluid domain size $L_d(t)$ for an even mixture on the surface of a torus 
with $R = 2.5$ and $r = 1$ in (a) the inertial-hydrodynamics
and (b) the diffusive regimes. For the diffusive regime, the convective term in the Cahn-Hilliard
equation is removed. 
(c-e) Snapshots of the typical fluid configurations at $t = 40$, $100$ and $250$
corresponding to the case indicated in panel (a).
(f-h) Snapshots of the typical fluid configurations at $t = 350$, $4250$ and $25500$
corresponding to the case indicated in panel (b). The times are chosen such that 
$L_d$ matches the ones corresponding to the panels (c-e).
\label{fig:tscal-ev}
}
\end{center}
\end{figure}

Our simulation results are shown 
in Figs.~\ref{fig:tscal-ev}(a) and \ref{fig:tscal-ev}(b) 
respectively for cases with and without coupling to
hydrodynamics. Qualitatively we find a similar behaviour to the results obtained
in the Cartesian case, Fig.~\ref{fig:cart_scal}. In panel (a), it can be seen that $L_d(t)$ grows with 
an apparent exponent of  $2/3$ when hydrodynamics is on.
Turning off the hydrodynamics, the $1/3$ diffusive exponent emerges, as demonstrated in panel (b). 
The coarsening dynamics is also much faster with hydrodynamics
on the torus. Snapshots of the order parameter configuration
at various times for the case of the even mixture with and without hydrodynamics are shown 
in panels (c-e) and (f-h) respectively. 

Quantitatively, we observe that finite size effects occur earlier (smaller $L_d$) for the torus considered in Fig.~\ref{fig:tscal-ev}
compared to the Cartesian case. This is expected since the effective length scale in the poloidal direction,
$2\pi r$, is smaller than the width of the simulation box in the Cartesian case, even though the total surface areas
are comparable. Indeed, we can observe that 
the departure from the $2/3$ (panel a) and $1/3$ (panel b) exponents
occur when the fluid domains start to wrap around the circle in the poloidal direction. 

\begin{figure}
\begin{center}
\begin{tabular}{cc}
 \includegraphics[width=0.4\columnwidth]{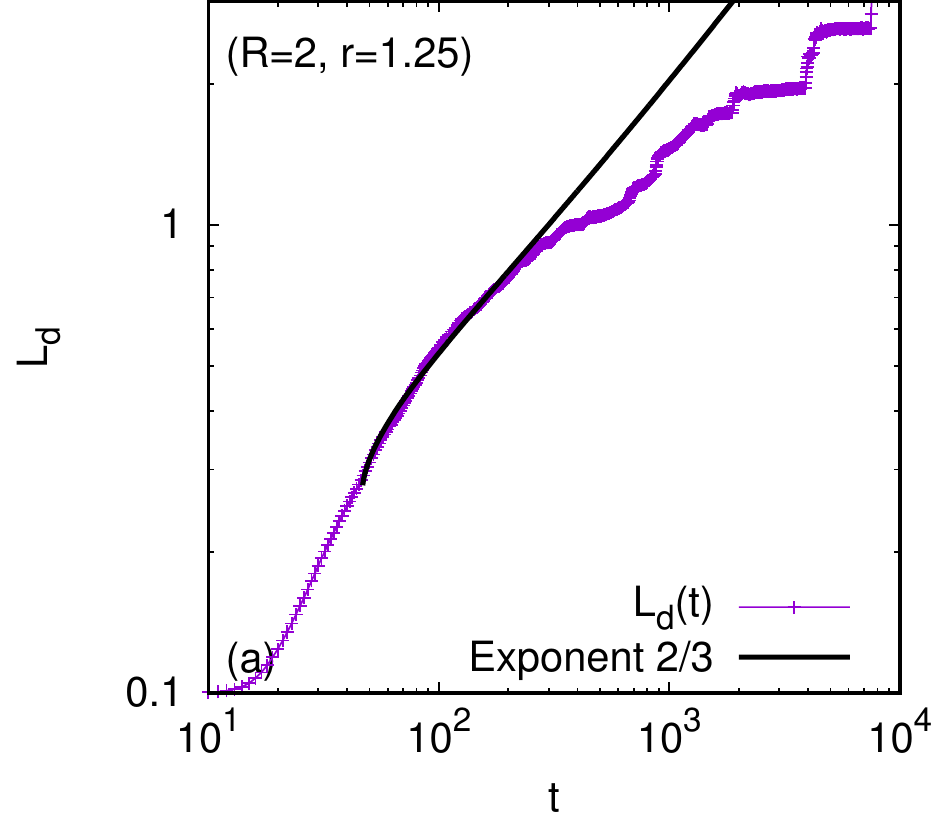} &
 \includegraphics[width=0.4\columnwidth]{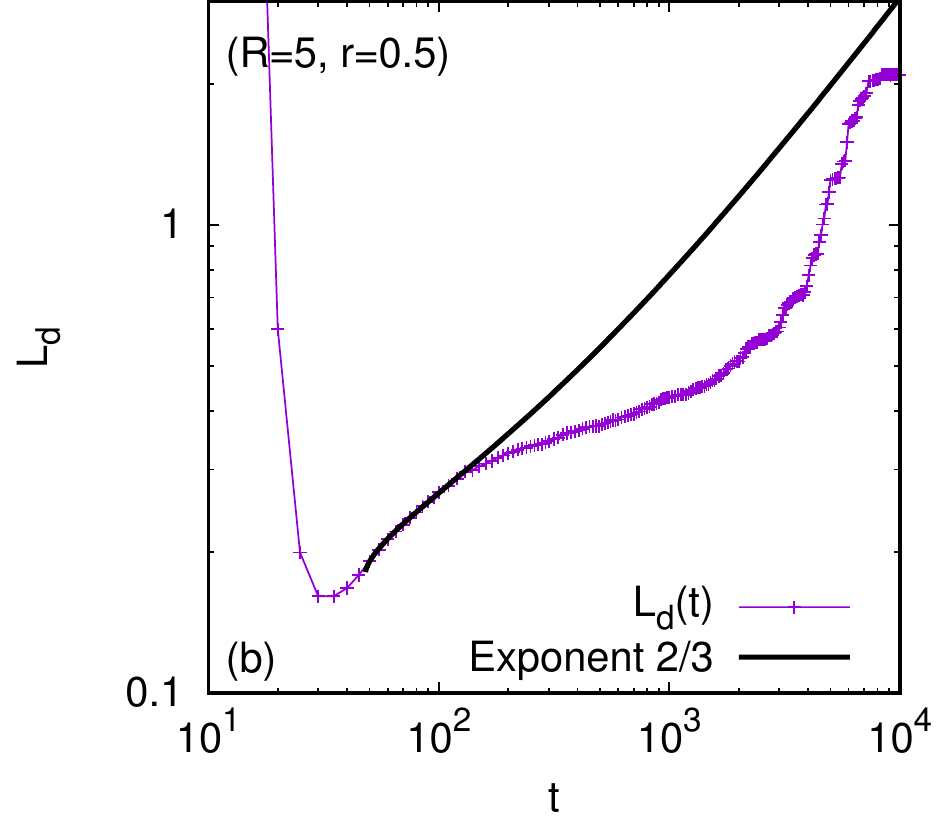}
\end{tabular}
\begin{tabular}{ccc}
 \includegraphics[width=.32\columnwidth]{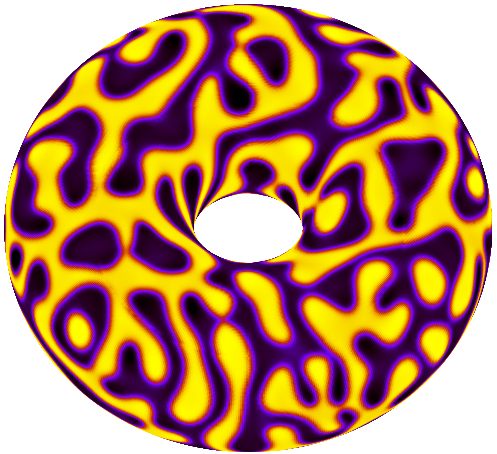} &
 \includegraphics[width=.3\columnwidth]{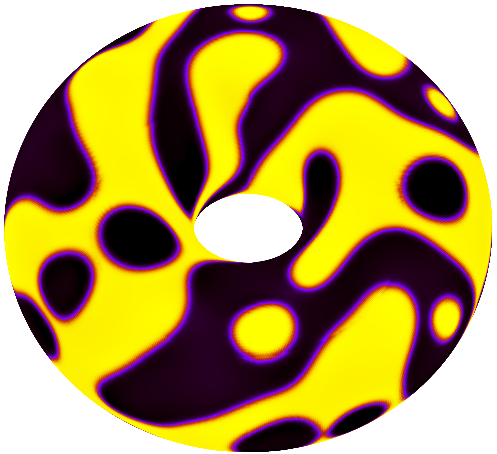} &
 \includegraphics[width=.31\columnwidth]{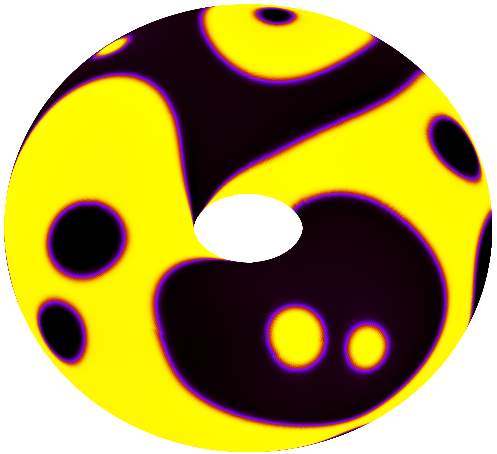}  \vspace{1 mm} \\
 (c) $t = 40$ & (d) $t = 100$ & (e) $t = 250$ \vspace{2 mm} \\
 \includegraphics[width=.3\columnwidth]{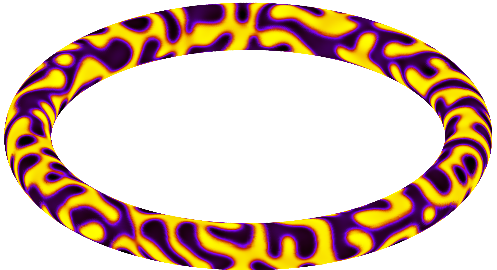} &
 \includegraphics[width=.3\columnwidth]{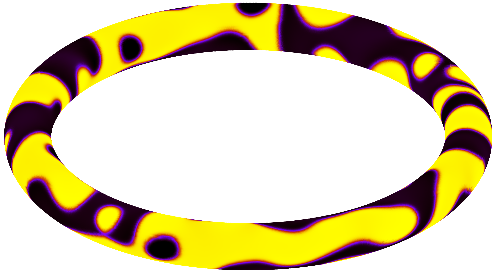} &
 \includegraphics[width=.3\columnwidth]{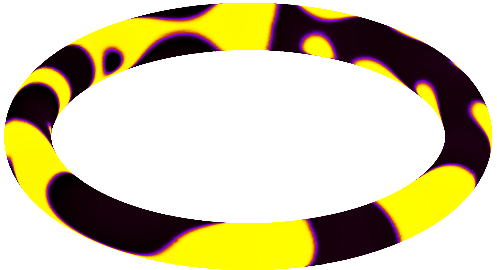}  \vspace{1 mm} \\
 (f) $t = 40$ & (g) $t = 100$ & (h) $t = 250$
\end{tabular}

\caption{
Comparing the growth of the fluid domain size $L_d(t)$ for an even mixture on 
(a) a thick torus ($R = 2$, $r = 1.25$) and (b) 
a thin torus ($R = 5$, $r = 0.5$).
(c-e) and (f-h) Snapshots of the typical fluid configurations at $t = 40$, $100$ and $250$ 
corresponding to the cases indicated in panels (a) and (b) respectively.\label{fig:tscal-ev-fat}
}
\end{center}
\end{figure}

In Fig.~\ref{fig:tscal-ev-fat} we further show  simulation results for a thicker 
($R = 2$ and $r = 1.25$; $a = 0.625$) and 
a thinner ($R = 5$ and $r = 0.5$; $a = 0.1$) torus, having a total 
area equal to the one considered at the beginning of this section.
The simulation parameters are kept the same as before, except that
for the thicker torus, the time step must 
be decreased down to $\delta t = 5 \times 10^{-5}$
since the minimum spacing along the $\varphi$ 
direction occurring on the inner equator 
is $2\pi(R-r)/N_\varphi \sim 0.00589$. Comparing 
Figs.~\ref{fig:tscal-ev}(a), \ref{fig:tscal-ev-fat}(a)
and \ref{fig:tscal-ev-fat}(b), we can further conclude that
finite size 
effects appear sooner for the thinner torus and later for the thicker one.
This further strengthens the argument that the determining lengthscale 
for the finite size effects is the circumference
in the poloidal direction,
rather than the circumference
on the inner side of the torus 
(at $\theta = \pi$), $2\pi (R-r)$. Otherwise, the thicker torus
should display finite size effects the earliest among the three geometries simulated. 

Given the fluid stripes are generally formed in the poloidal rather 
than the toroidal direction during phase separation, the
drift phenomenon reported in sub-section \ref{sec:drift:damp} for 
stripe configurations cannot be clearly visualised.
However, domain drifts for drops, as reported in 
section~\ref{sec:drift:drop}, can be seen in Figs.~\ref{fig:tscal-ev} 
and \ref{fig:tscal-ev-fat} during the late stages of the coarsening phenomenon. 
This drift phenomenon can be observed even clearer when we 
study uneven mixtures, as discussed in the next sub-section.

%During our investigations, we also considered the coarsening dynamics of an
%uneven mixture ($\phi_0 = -0.3$) on the torus having $R = 2.5$ and $r = 1$.
%The results corresponding to this case are very similar to what has already 
%been presented, thus for brevity, the corresponding discussion is 
%provided in Sec.~3 of the Supplementary Material \cite{suppl}.

\subsection{Uneven Mixtures}
\label{sec:res:growth:uneven}

\subsubsection{Cartesian Geometry}

\begin{figure}
\begin{center}
\begin{tabular}{c}
 \includegraphics[width=0.45\columnwidth]{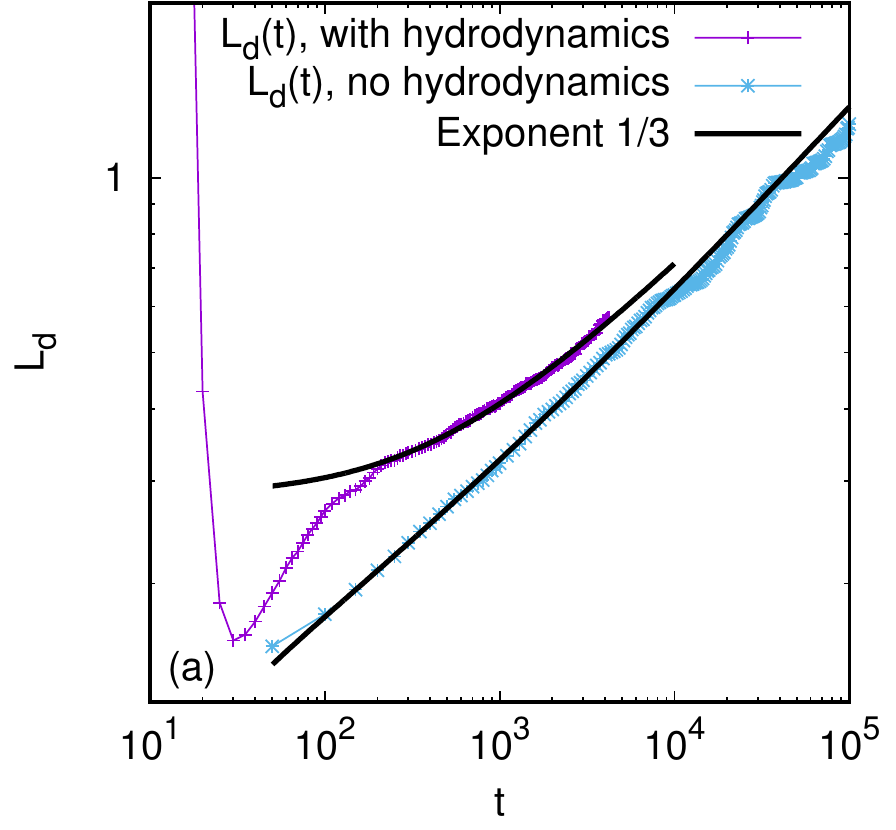} 
\end{tabular}
\begin{tabular}{cccc} 
 \includegraphics[width=0.225\columnwidth]{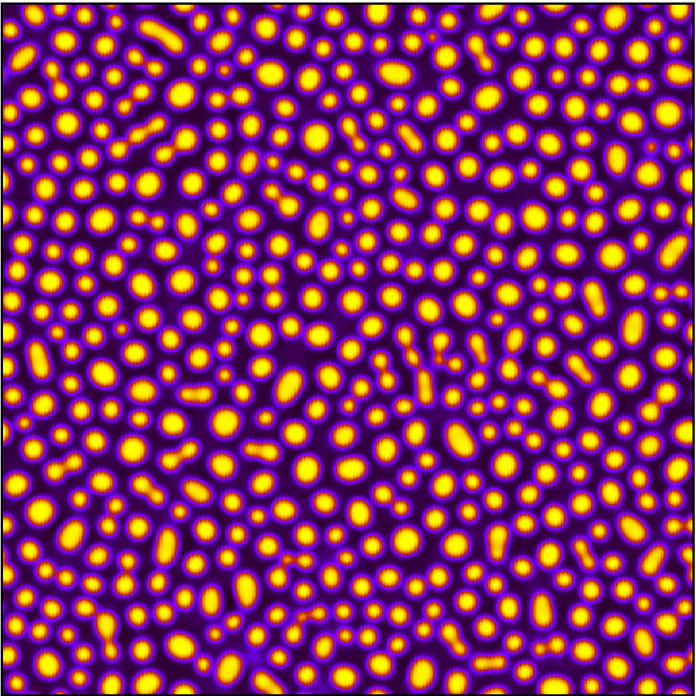} &
 \includegraphics[width=0.225\columnwidth]{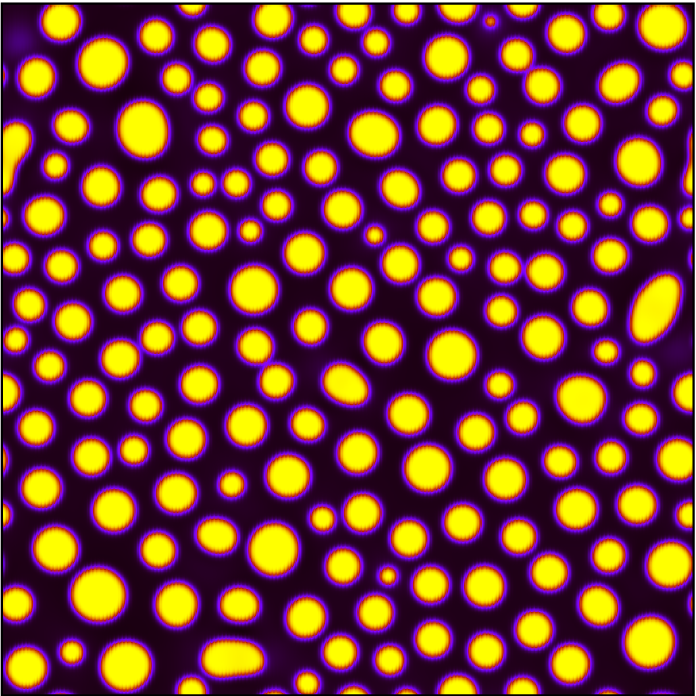}&
 \includegraphics[width=0.225\columnwidth]{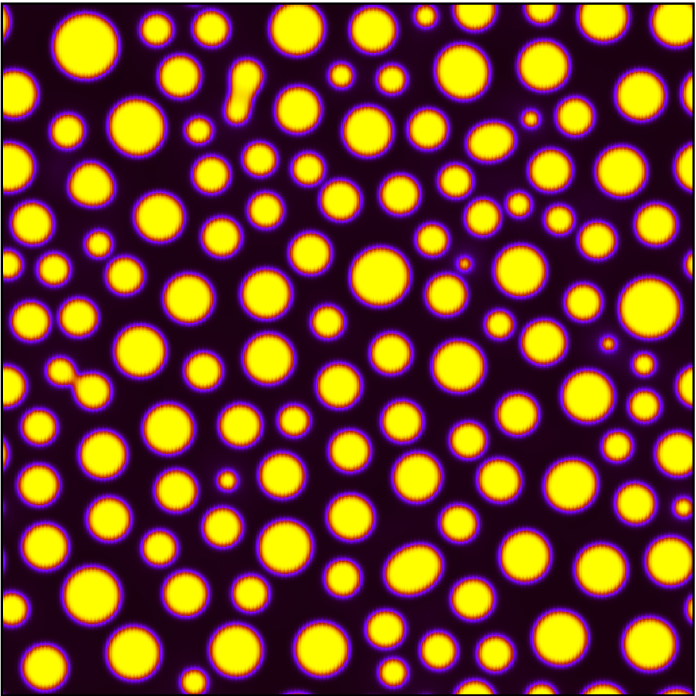}&
 \includegraphics[width=0.225\columnwidth]{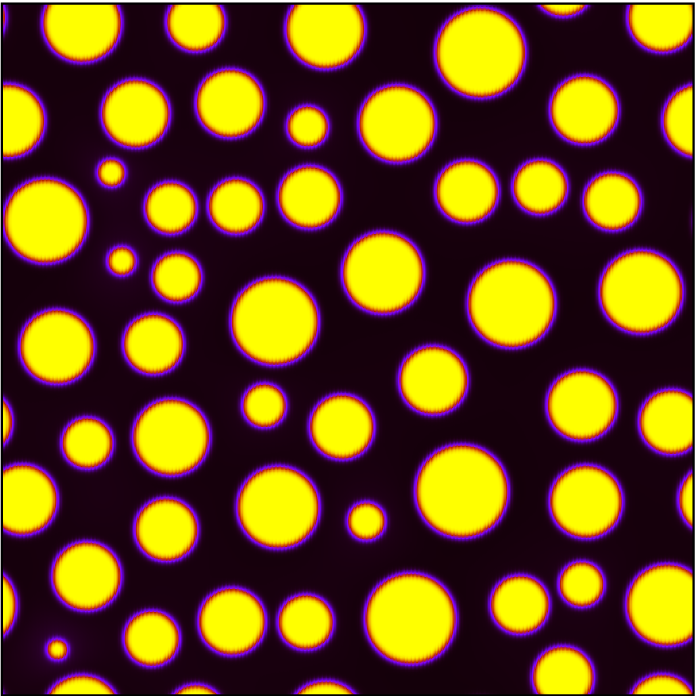} \vspace{1 mm}\\
  (b) $t = 50$ & (c) $t = 250$ & (d) $t = 500$ & (e) $t = 4000$ \vspace{1 mm} \\
 \includegraphics[width=0.225\columnwidth]{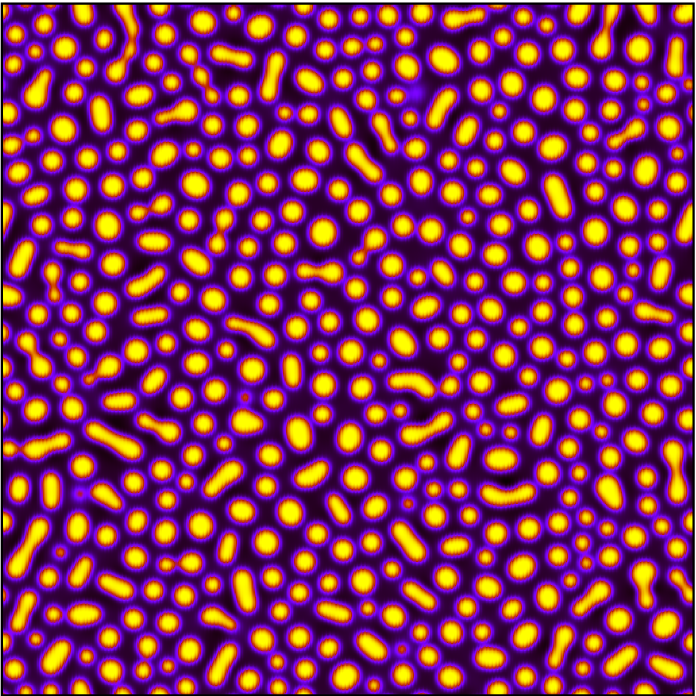} &
 \includegraphics[width=0.225\columnwidth]{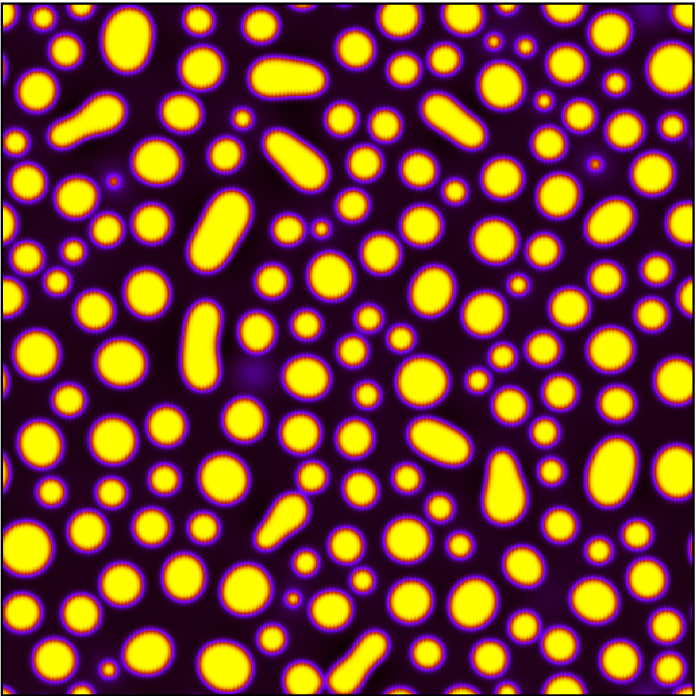}&
 \includegraphics[width=0.225\columnwidth]{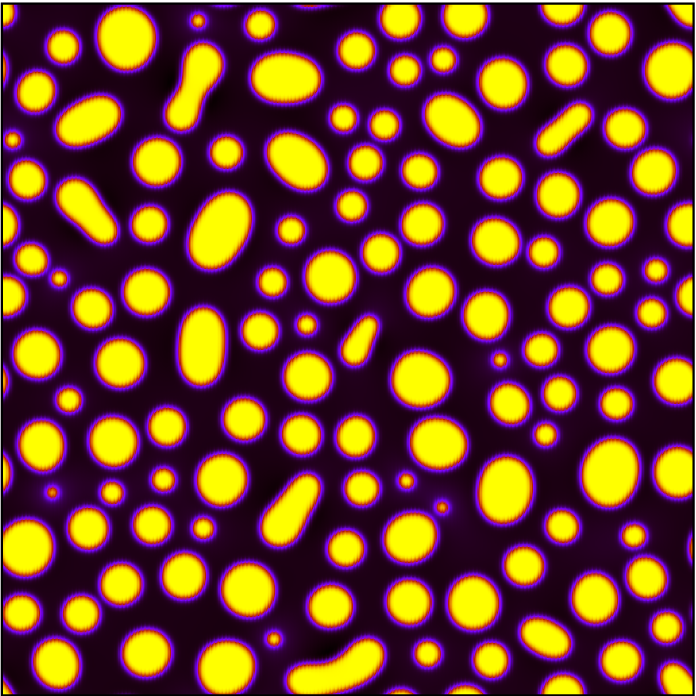}&
 \includegraphics[width=0.225\columnwidth]{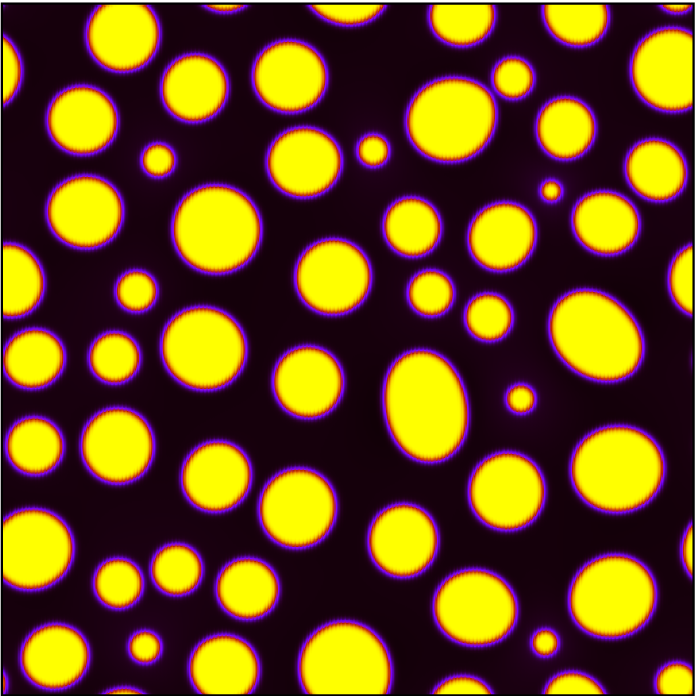} \vspace{1 mm}\\
  (f) $t = 150$ & (g) $t = 1050$ & (h) $t = 1500$ & (i) $t = 6500$
\end{tabular}

\caption{
(a) Growth of the fluid domain $L_d(t)$ for an uneven mixture ($\overline{\phi} = -0.3$) 
in two dimensions with and without hydrodynamics.
In both cases, an exponent of $1/3$ characteristic of the diffusive 
regime is observed at late times.
(b-e) Snapshots of the typical fluid configurations at times 
$t = 50$, $250$, $500$ and $4000$, corresponding to the case with hydrodynamics.
(g-j) Snapshots of the fluid configurations corresponding to the case without
hydrodynamics,
at times $t = 150$, $1050$, $1500$ and $6500$. These are selected such that the 
values of $L_d(t)$ correspond to those in panels (b-e).
\label{fig:cart_scal_uneven}}
\end{center}
\end{figure}

The simulation results for a mixture with asymmetric composition 
are shown in Fig.~\ref{fig:cart_scal_uneven}. We use the same
simulation parameters as in Fig~\ref{fig:cart_scal}, except that $\overline{\phi} = -0.3$. 
Fig.~\ref{fig:cart_scal_uneven}(a) shows how the typical domain size scales with time both
when hydrodynamics is turned on and off. Interestingly, in both 
cases we observe an exponent of $1/3$, albeit with different 
prefactors. This is in contrast to our results for the even mixtures, 
when an apparent exponent of $2/3$
is obtained with hydrodynamics. It has been suggested in the literature that the effect of 
hydrodynamics decreases as a function of the asymmetry of the mixture, though 
we do not yet know of a convincing systematic study of this effect.  
For example, \cite{wagner2001phase} showed that at high concentrations 
droplets with hydrodynamics exhibit the viscous hydrodynamic coarsening 
regime, but as droplet coalescence is reduced at lower volume fractions 
the effect of hydrodynamics diminishes. Here we observe the limit where 
the scaling is typical of that for diffusive dynamics.

The fluid configurations at various times in the simulation are shown 
in Fig.~\ref{fig:cart_scal_uneven}, panels (b-e), when hydrodynamics is 
taken into account.
These can be compared to Fig.~\ref{fig:cart_scal_uneven}, panels (f-i), 
when the advection term is 
switched off in the Cahn-Hilliard equation. The differences are mainly 
that the morphologies with hydrodynamics are coarsening faster. In the 
non-hydrodynamic simulation there are more coalescence events visible 
because the restoration of a round shape takes more time. 
Thus, while the scaling exponent is the same with and without hydrodynamics, hydrodynamics 
still plays an important role in that it allows coalescing droplets 
return to a round shape more quickly. 

\begin{figure}
\begin{center}
\begin{tabular}{c} 
\includegraphics[width=0.5\columnwidth]{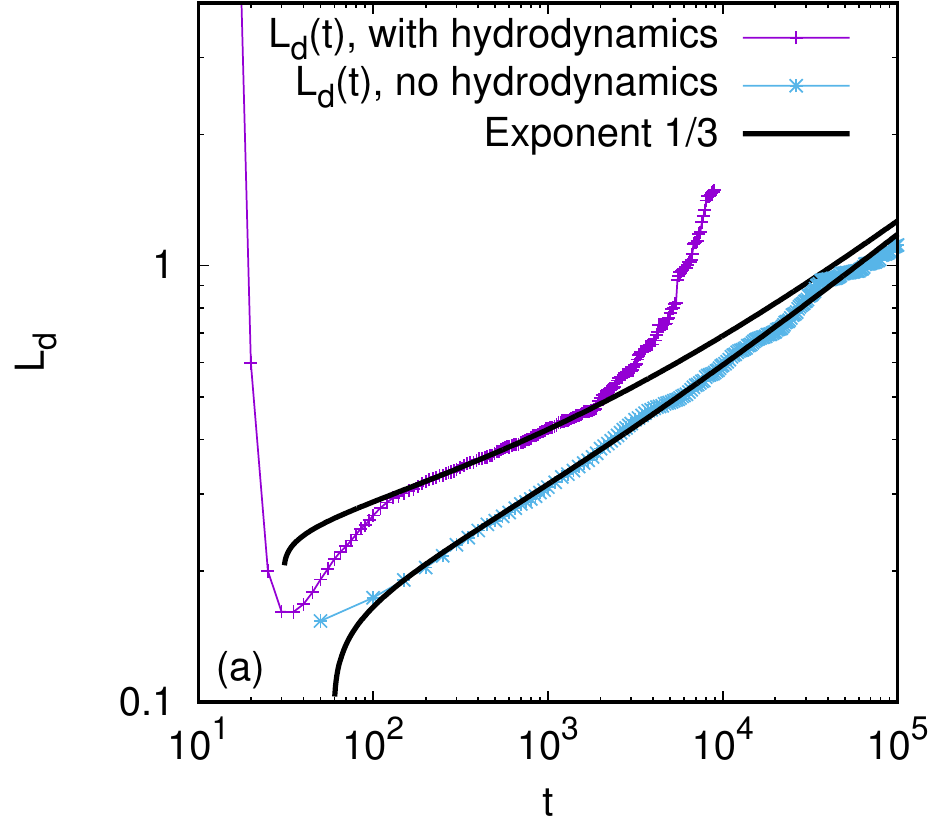} 
\end{tabular}
\begin{tabular}{ccc}
 \includegraphics[width=.3\columnwidth]{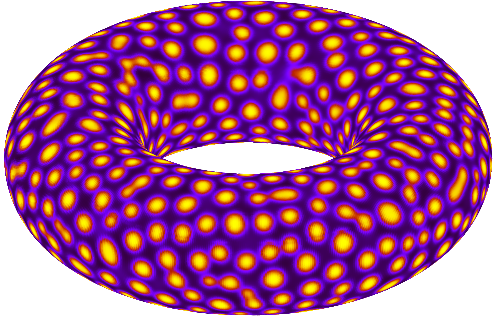} &
 \includegraphics[width=.3\columnwidth]{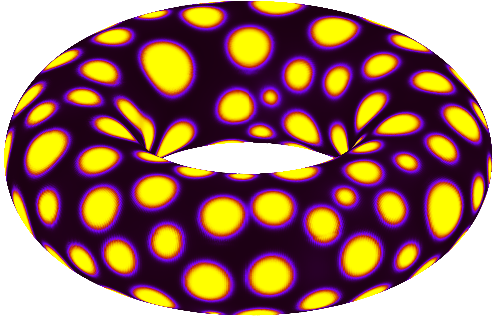} &
 \includegraphics[width=.3\columnwidth]{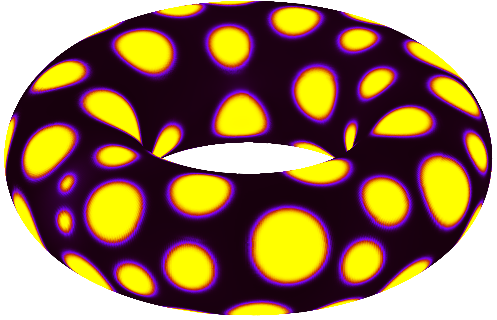} \vspace{1 mm}\\
 (b) $t = 40$ & (c) $t = 250$ & (d) $t = 1500$ \vspace{1 mm}\\
 \includegraphics[width=.3\columnwidth]{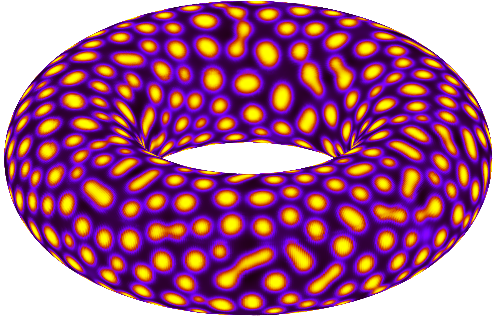} &
 \includegraphics[width=.3\columnwidth]{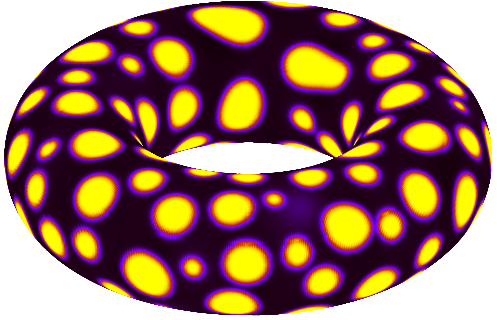} &
 \includegraphics[width=.3\columnwidth]{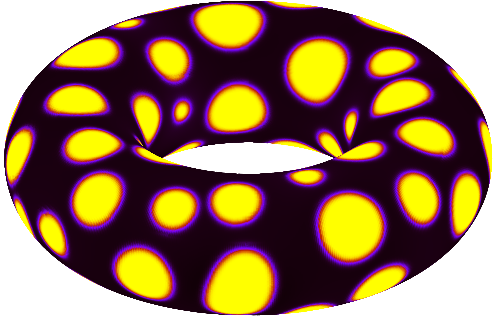} \vspace{1 mm}\\
 (e) $t = 100$ & (f) $t = 1250$ & (g) $t = 3500$
\end{tabular}
\caption{(a) 
Growth of the fluid domain size $L_d(t)$ for an uneven mixture 
on a torus with $R = 2.5$ and $r = 1$ with and without hydrodynamics. 
(b-d) Snapshots of the typical fluid configurations at $t = 40$, $250$ and $1500$ 
corresponding to the case with hydrodynamics.
(e-g) Snapshots of the typical fluid configurations at $t = 100$, $1250$ and $3500$ 
corresponding to the case without hydrodynamics. The times are chosen such that 
the values of $L_d$ match those in panles (b-d).
\label{fig:tscal-unev}
}
\end{center}
\end{figure}

\begin{figure}
\begin{center}
 \includegraphics[width=0.7\linewidth]{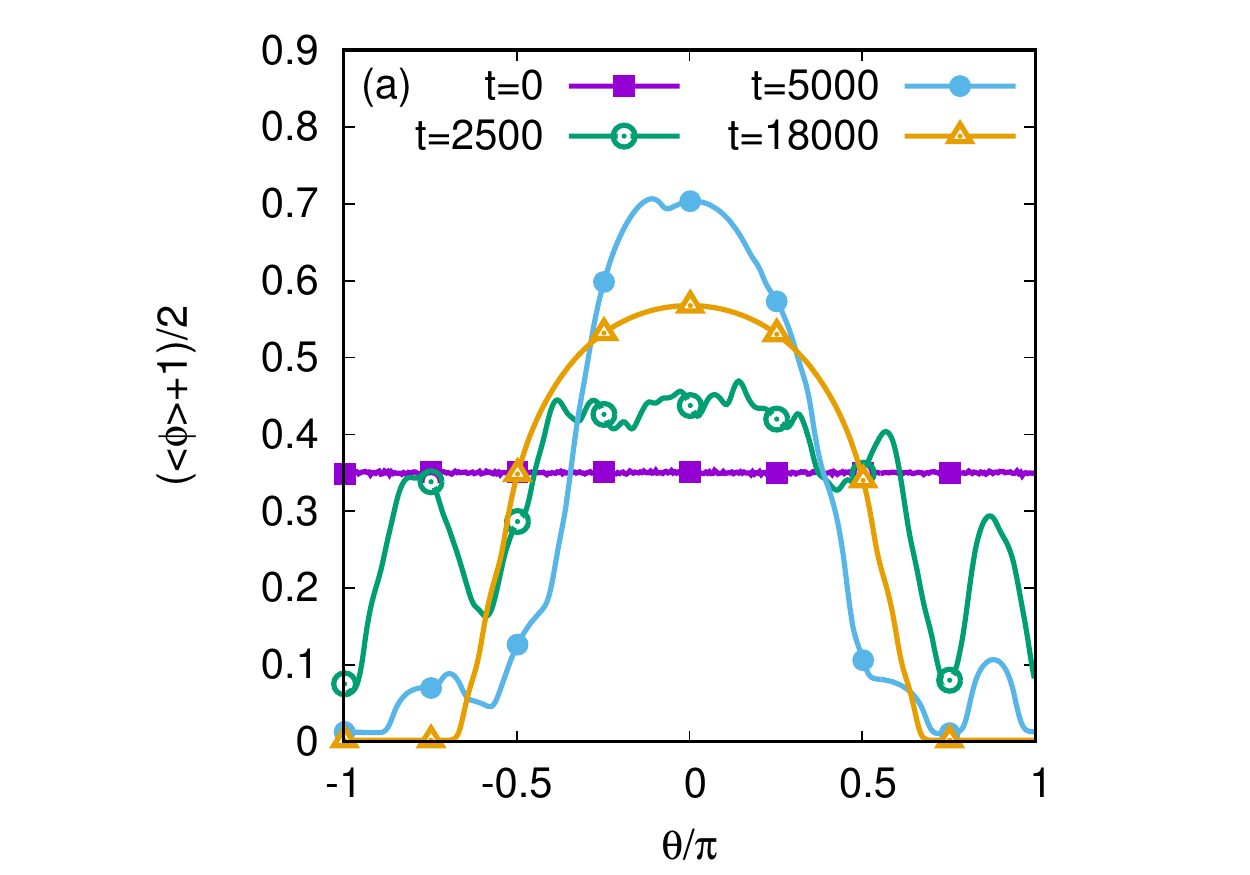}
\begin{tabular}{ccc}
 \includegraphics[width=.3\columnwidth]{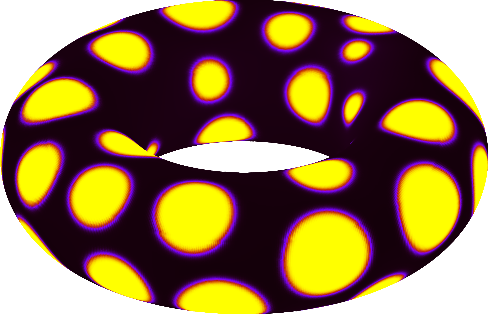} &
 \includegraphics[width=.3\columnwidth]{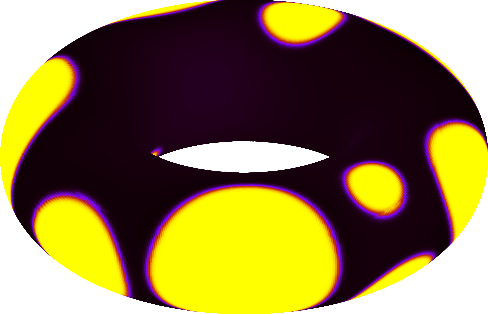} &
 \includegraphics[width=.3\columnwidth]{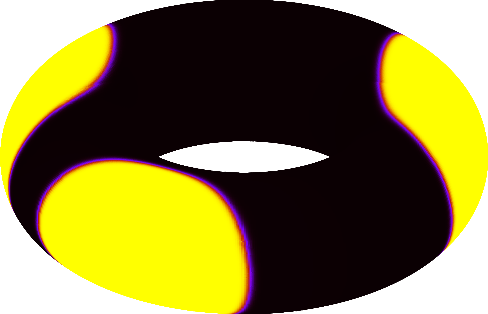} \vspace{1 mm}\\
 (b) $t = 2500$ & (c) $t = 5000$ & (d) $t = 1800$ \vspace{1 mm}\\
\end{tabular}
\caption{(a) The average distribution of the component, $(\braket{\phi} + 1)/2$, as a function of $\theta$ at various times. (b-d) Snapshots of the fluid configurations at $t=2500$, $t = 5000$ and $18000$.
\label{fig:tscal-unev-phiavg}
}
\end{center}
\end{figure}

\subsubsection{Torus Geometry} 

Here we consider a torus geometry with $R = 2.5$ and $r = 1$ 
(same geometry and simulation parameters as in Fig.~\ref{fig:tscal-ev}), and
the order parameter is initialised according to Eq.~\eqref{eq:phirand} with $\overline{\phi} = -0.3$.
The simulation results for the uneven mixture are shown in Fig. ~\ref{fig:tscal-unev}.
Quantitatively, we find a similar behaviour as for the Cartesian case.
Both when hydrodynamics is turned on and off, we observe a $1/3$ exponent in our simulations.
Similar to the even mixture shown in Fig.~\ref{fig:tscal-ev}, 
we also find that finite size effects occur earlier (smaller $L_d$) 
for the torus compared to the Cartesian geometry. As discussed in the case of even mixtures, 
this occurs when the fluid domains
start to wrap around the circle in the poloidal direction. Snapshots of the fluid configurations
during phase separation are shown in panels (b-d) and (e-g) respectively for simulations
with and without hydrodynamics.

At late times, the effect of the curvature on the domain dynamics becomes important. 
In Sec.~\ref{sec:drift:drop} we discussed how droplet domains migrate to the outer side 
of the torus. To quantify this effect during phase separation of uneven mixtures, we consider
the average of $\phi$ with respect to the azimuthal angle $\varphi$:
\begin{equation}
 \braket{\phi} = \int_0^{2\pi}  \frac{d\varphi}{2\pi} \phi(\theta,\varphi).
 \label{eq:phiavg}
\end{equation}
The discrete equivalent of the above relation is
\begin{equation}
 \braket{\phi}_q = \frac{1}{N_{\varphi}} 
 \sum_{s = 1}^{N_\varphi} \phi_{s,q}.
\end{equation}
We plot $(\braket{\phi} + 1)/2$ as a function of the poloidal angle $\theta$ at various times in
Fig.~\ref{fig:tscal-unev-phiavg}(a). 
At late times, see e.g. Fig.~\ref{fig:tscal-unev-phiavg}(c), 
the typical configuration corresponds to the majority phase ($\phi = -1$) 
forming a continuum with several large droplets of the minority phase ($\phi = +1$) 
primarily in the outer side of the torus. At $t = 18000$ [Fig.~\ref{fig:tscal-unev-phiavg}(d)], 
when the steady state is reached, the inner stripe spans $0.65\pi \lesssim \theta \lesssim 1.35\pi$. 
Our convention is to identify the stripe with the minority fluid component.
The maximum of $(\braket{\phi} + 1)/2$ is clearly reached at $\theta = 0$, 
indicating that the outer side of the torus is populated by droplets centered on $\theta = 0$.

\section{Conclusions}\label{sec:conc}

In this work we developed a vielbein lattice Boltzmann scheme to solve the hydrodynamics equations of motion of a binary fluid on an arbitrary curved surface. To illustrate the application of our vielbein lattice Boltzmann method to curved surfaces, here we focussed on the torus geometry and studied two classes of problems. First, due to the non-uniform curvature present on a torus, we showed drift motions of fluid droplets and stripes on a torus. Such dynamics are not present on a flat surface or on surfaces with uniform curvature. Interestingly the fluid droplets and stripes display preference to different regions of the torus. Fluid droplets migrate to the outer side of the torus, while fluid stripes move to the inner side of the torus. The exhibited dynamics are typical of a damped oscillatory motion. Moreover, for the fluid stripes, the corresponding dynamics can effectively be reduced to a one-dimensional problem by taking advantage of the symmetry with respect to 
the azimuthal angle. Our simulation results are in excellent agreement with the analytical predictions for the equilibrium position of the stripes, the Laplace pressure difference between
the inside and outside of the stripes, and the relaxation dynamics of the stripes towards equilibrium.

We also studied phase separation dynamics on tori of various shapes. For even mixtures, $2/3$ and $1/3$ scaling exponents characteristic of hydrodynamics and diffusive regimes are observed. In contrast, for uneven mixtures, we only observe a $1/3$ scaling exponent both when hydrodynamics is turned on and off. Compared to Cartesian geometry, we saw that finite size effects kick in earlier for the torus geometry. By comparing the results for three torus aspect ratios, we conclude 
that the determining lengthscale for the finite size effects seems to be
the perimeter in the poloidal direction, corresponding to fluid domains wrapping  around the circle in the poloidal direction.  That the stripes are observed to form in the poloidal rather than the toroidal direction prevents the observation of drift motion of fluid stripes towards the inner side of the torus during phase separation. However, the domain drifts for fluid drops to the outer side of the torus can be clearly observed at the late stage of phase separation.

While we focussed on the torus geometry, our approach can be applied to arbitrary curved geometry. Moreover, one interesting area for future work is to expand the method to account for unstructured mesh, where the geometrical objects needed for the Boltzmann equation must be evaluated numerically. A major challenge is to construct a numerical scheme which is accurate to second order or higher. Another important avenue for future investigations is to couple the hydrodynamics equations of motion with more complex dynamical equations, such as those for (active and passive) liquid crystals and viscoelastic fluids. We believe this work extends the applicability of the lattice Boltzmann approaches to a new class of problems, complex flows on curved manifolds, which are difficult to carry out using the standard lattice Boltzmann method.

{\bf Acknowledgements:} We acknowledge funding from EPSRC (HK; EP/J017566/1 and EP/P007139/1), Romanian Ministry of Research and Innovation (VEA and SB; CCCDI-UEFISCDI, project number PN-III-P1-1.2-PCCDI-2017-0371/VMS, within PNCDI III), and the EU COST action MP1305 Flowing Matter (VEA and HK; Short Term Scientific Mission 38607).
VEA gratefully acknowledges the support of NVIDIA Corporation with the donation of a 
Tesla K40 GPU used for this research. VEA and SB thank Professor 
Victor Sofonea (Romanian Academy, Timi\cb{s}oara Branch) for 
encouragement, as well as for sharing with us the GPU infrastructure 
available at the Timi\cb{s}oara Branch of the Romanian Academy.

\appendix

\section{Application of the vielbein method to the torus geometry}
\label{app:curved}

The derivation of the Boltzmann equation, Eq.~\eqref{eq:boltz_cons},
written in conservative form with respect to vielbein vector fields 
is discussed in \cite{busuioc19}. Using Eq.~\eqref{eq:boltz_cons}
as a starting point, here we present generic main 
steps required to write down the Boltzmann equation for any arbitrary
curved surface. For concreteness, we focus on the torus geometry 
in this paper.
\begin{enumerate}
\item {\it Parametrising the surface. }
As a two-dimensional manifold, a surface needs two coordinates 
$q^1$ and $q^2$ to be parametrised. In the case of a
torus of inner radius $r$ and outer radius $R$, 
the parametrisation can be chosen in terms of the 
angles $\theta \in [0, 2\pi)$ and $\varphi \in [0, 2\pi)$ as 
follows:
\begin{align}
 x =& (R + r \cos \theta) \cos \varphi,\nonumber\\
 y =& (R + r \cos \theta) \sin \varphi,\nonumber\\
 z =& r \sin \theta, \label{eq:torus}
\end{align}
and the system is periodic with respect to both of these angles.
\item {\it Writing down the line element. }
Differentiating the functions $x$, $y$ and $z$ with respect to 
$q^1$ and $q^2$ yields the formula
\begin{equation}
 ds^2 = g_{ab} dq^a dq^b, \qquad 
 g_{ab} = \delta_{ij} \frac{\partial x^i}{\partial q^a} \frac{\partial x^j}{\partial q^b},
\end{equation}
where $\{i, j\} \in \{1,2,3\}$, $1 \le a,b \le 2$
and $g_{ab}$ are the components of the metric tensor.
In the case of Eq.~\eqref{eq:torus}, the line element becomes:
\begin{equation}
 ds^2 = \left[dx^2 + dy^2 +dz^2\right]_{\rm on \, torus} =  
 (R + r \cos\theta)^2 d\varphi^2 + r^2 d\theta^2,
\end{equation}
leading to the metric tensor components
\begin{equation}
g_{\varphi\varphi} = (R + r \cos\theta)^2, \quad g_{\theta\theta} = r^2, \quad g_{\theta\varphi} = g_{\varphi\theta} = 0. 
\end{equation}
\item {\it Constructing the vielbein field.}
The vielbein vector frame consists of the 
vectors $\bm{e_\hata} = e_\hata^a \bm{\partial_a}$ which satisfy:
\begin{equation}
 g_{ab} e^a_\hata e^b_\hatb = \delta_{\hata\hatb}.
 \label{eq:vielbein}
\end{equation}
Since Eq.~\eqref{eq:vielbein} is invariant under the action of the 
orthogonal group with respect to the hatted indices, the vielbein
is defined up to an arbitrary rotation. After fixing 
the vielbein, the vielbein one-form co-frame  denoted 
via $\bm{\omega^\hata} = \omega^\hata_a \bm{dq^a}$ is uniquely fixed 
by Eq.~\eqref{eq:vielb_contr}.

For the torus geometry, the natural choice is to take
\begin{align}
 \omega^\hvarphi_\varphi =& (R + r\cos\theta), & 
 \omega^\htheta_\theta =& r, & \omega^\hvarphi_\theta = \omega^\htheta_\varphi =& 0, \nonumber\\
 e^\varphi_\hvarphi =& \frac{1}{R+ r\cos\theta}, & 
 e^\theta_\htheta =& \frac{1}{r}, & e^\theta_\hvarphi = e^\varphi_\htheta =& 0.
\end{align}
%%%%%
\item {\it Computing the Cartan coefficients}.
The commutator of two vector fields $\bm{u}$ and $\bm{v}$ is 
another vector field, denoted by $[\bm{u},\bm{v}] = (u^a \partial_a v^b - 
v^a\partial_a u^b) \bm{\partial_b}$. The contraction between the 
co-frame one-form $\bm{\omega^\hatc}$ and the commutator of the 
tetrad frame vector fields $\bm{e_\hata}$ and $\bm{e_\hatb}$ defines
the Cartan coefficient $c_{\hata\hatb}{}^{\hatc}$ \eqref{eq:cartan},
 via the following relation:
\begin{equation}
 c_{\hata\hatb}{}^{\hatc} = \braket{\bm{\omega^\hatc},[\bm{e_\hata},\bm{e_\hatb}]} = 
 \omega^\hatc_c(e_\hata^a \partial_a e_\hatb^c - 
 e_\hatb^b\partial_b e_\hata^c).
\end{equation}
In the case of the torus, the commutator of the vielbein vectors 
$\bm{e_\htheta}$ and $\bm{e_\hvarphi}$ is
\begin{equation}
 [\bm{e_\htheta}, \bm{e_\hvarphi}] = -[\bm{e_\hvarphi}, \bm{e_\htheta}] = 
 \frac{\sin\theta}{R + r\cos\theta} \bm{e_\hvarphi},
\end{equation}
leading to the Cartan coefficients 
\begin{equation}
c_{\htheta\hvarphi}{}^{\hvarphi} = -c_{\hvarphi\htheta}{}^{\hvarphi} = 
\frac{\sin\theta}{R + r\cos\theta}.
\end{equation} 
%%%%%%%%%%
\item {\it Computing the connection coefficients}.
On curved surfaces, the ordinary partial derivative operator 
$\partial_a$ must be replaced by a covariant derivative 
which ensures that the resulting vector or tensor is still 
contained in the tangent space. This is ensured starting from the 
covariant derivative of the basis vectors:
\begin{equation}
 \nabla_\hatb e_\hata = \Gamma^{\hatc}{}_{\hata\hatb} e_{\hatc}, \qquad 
 \Gamma^\hatc{}_{\hata\hatb} = \frac{1}{2} \delta^{\hatc\hatd}\left(
 c_{\hatd\hata\hatb} + c_{\hatd\hatb\hata} - c_{\hata\hatb\hatd}\right).
\end{equation}
In the case of the torus, the only non-vanishing connection 
coefficients are 
\begin{equation}
 \Gamma_{\htheta\hvarphi\hvarphi} = -\Gamma_{\hvarphi\htheta\hvarphi} = 
 \frac{\sin\theta}{R + r\cos\theta}.\label{eq:conn_torus}
\end{equation}
\item {\it Writing the Boltzmann equation.} 
Plugging Eq.~\eqref{eq:conn_torus} into Eq.~\eqref{eq:boltz_cons} 
yields the Boltzmann equation for the torus geometry,
Eq.~\eqref{eq:boltz_tor}.
\end{enumerate}

\bibliographystyle{jfm}
\bibliography{refs}

\begin{thebibliography}{62}
\expandafter\ifx\csname natexlab\endcsname\relax\def\natexlab#1{#1}\fi
\def\au#1{#1} \def\ed#1{#1} \def\yr#1{#1}\def\at#1{#1}\def\jt#1{\textit{#1}}
  \def\bt#1{#1}\def\bvol#1{\textbf{#1}} \def\vol#1{#1} \def\pg#1{#1}
  \def\publ#1{#1}\def\arxiv#1{#1}\def\org#1{#1}\def\st#1{\textit{#1}}

\bibitem[Abadi {\em et~al.\/}(2018)Abadi, Fakhari \& Rahimian]{Abadi18}
{\sc \au{Abadi, R. H.~H.}, \au{Fakhari, A.} \& \au{Rahimian, M.~H.}} \yr{2018}
  \at{Numerical simulation of three-component multiphase flows at high density
  and viscosity ratios using lattice {B}oltzmann methods}.  \jt{Phys. Rev. E}
  \bvol{97},  \pg{033312}.

\bibitem[Ambru\cb{s} {\em et~al.\/}(2019)Ambru\cb{s}, Busuioc, Wagner,
  Paillusson \& Kusumaatmaja]{suppl}
{\sc \au{Ambru\cb{s}, V.~E.}, \au{Busuioc, S.}, \au{Wagner, A.~J.},
  \au{Paillusson, F.} \& \au{Kusumaatmaja, H.}} \yr{2019} Please see the
  Supplemental Material at (URL to be inserted by publisher).

\bibitem[Arroyo \& Desimone(2009)]{Arroyo09}
{\sc \au{Arroyo, M.} \& \au{Desimone, A.}} \yr{2009}  \at{Relaxation dynamics
  of fluid membranes}.  \jt{Phys. Rev. E}  \bvol{79},  \pg{039906}.

\bibitem[Aufderhorst-Roberts {\em et~al.\/}(2017)Aufderhorst-Roberts, Chandra
  \& Connell]{Roberts17}
{\sc \au{Aufderhorst-Roberts, A.}, \au{Chandra, U.} \& \au{Connell, S.~D.}}
  \yr{2017}  \at{Three-phase coexistence in lipid membranes}.  \jt{Biophys. J.}
   \bvol{112},  \pg{313 -- 324}.

\bibitem[Bacia {\em et~al.\/}(2005)Bacia, Schwille \& Kurzchalia]{Bacia05}
{\sc \au{Bacia, K.}, \au{Schwille, P.} \& \au{Kurzchalia, T.}} \yr{2005}
  \at{Sterol structure determines the separation of phases and the curvature of
  the liquid-ordered phase in model membranes}.  \jt{Proc. Nat. Acad. Sci. USA}
   \bvol{102},  \pg{3272--3277}.

\bibitem[Baumgart {\em et~al.\/}(2003)Baumgart, Hess \& Webb]{Baumgart03}
{\sc \au{Baumgart, T.}, \au{Hess, S.~T.} \& \au{Webb, W.~W.}} \yr{2003}
  \at{Imaging coexisting fluid domains in biomembrane models coupling curvature
  and line tension}.  \jt{Nature}  \bvol{425},  \pg{821}.

\bibitem[Bray(2002)]{Bray02}
{\sc \au{Bray, A.~J.}} \yr{2002}  \at{Theory of phase ordering kinetics}.
  \jt{Adv. Phys.}  \bvol{51},  \pg{481 -- 587}.

\bibitem[Briant \& Yeomans(2004)]{Briant04}
{\sc \au{Briant, A.~J.} \& \au{Yeomans, J.~M.}} \yr{2004}  \at{Lattice
  {B}oltzmann simulations of contact line motion. {II.} binary fluids}.
  \jt{Phys. Rev. E}  \bvol{69},  \pg{031603}.

\bibitem[Busuioc \& Ambru\cb{s}(2019)]{busuioc19}
{\sc \au{Busuioc, S.} \& \au{Ambru\cb{s}, V.~E.}} \yr{2019}  \at{Lattice
  {B}oltzmann models based on the vielbein formalism for the simulation of
  flows in curvilinear geometries}.  \jt{Phys. Rev. E}  \bvol{99},
  \pg{033304}.

\bibitem[Busuioc {\em et~al.\/}(2019{\natexlab{{\em a\/}}})Busuioc,
  Ambru\cb{s}, Biciu\cb{s}c\u{a} \& Sofonea]{busuioc17arxiv}
{\sc \au{Busuioc, S.}, \au{Ambru\cb{s}, V.~E.}, \au{Biciu\cb{s}c\u{a}, T.} \&
  \au{Sofonea, V.}} \yr{2019{\natexlab{{\em a\/}}}} Two-dimensional off-lattice
  {B}oltzmann model for van der {W}aals fluids with variable temperature.
  {DOI}: 10.1016/j.camwa.2018.12.015 (ar{X}iv:1702.01690 [physics.flu-dyn]).

\bibitem[Busuioc {\em et~al.\/}(2019{\natexlab{{\em b\/}}})Busuioc,
  Kusumaatmaja \& Ambru\cb{s}]{Busuioc19bench}
{\sc \au{Busuioc, S.}, \au{Kusumaatmaja, H.} \& \au{Ambru\cb{s}, V.~E.}}
  \yr{2019{\natexlab{{\em b\/}}}} Benchmark problems for axisymmetric flows on
  the torus geometry. In preparation.

\bibitem[Camley \& Bown(2011)]{Camley11}
{\sc \au{Camley, B.~A.} \& \au{Bown, F.~L.}} \yr{2011}  \at{Dynamic scaling in
  phase separation kinetics for quasi-two-dimensional membranes}.  \jt{J. Chem.
  Phys.}  \bvol{135},  \pg{225106}.

\bibitem[Cardall {\em et~al.\/}(2013)Cardall, Endeve \& Mezzacappa]{cardall13}
{\sc \au{Cardall, C.~Y.}, \au{Endeve, E.} \& \au{Mezzacappa, A.}} \yr{2013}
  \at{Conservative general relativistic {B}oltzmann equation}.  \jt{Phys. Rev.
  D}  \bvol{88},  \pg{023011}.

\bibitem[Cicuta {\em et~al.\/}(2007)Cicuta, Keller \& Veatch]{Cicuta07}
{\sc \au{Cicuta, P.}, \au{Keller, S.~L.} \& \au{Veatch, S.~L.}} \yr{2007}
  \at{Diffusion of liquid domains in lipid bilayer membranes}.  \jt{J. Phys.
  Chem. B}  \bvol{111},  \pg{3328--3331}.

\bibitem[Dellar(2001)]{Dellar2001}
{\sc \au{Dellar, P.~J.}} \yr{2001}  \at{Bulk and shear viscosities in lattice
  {B}oltzmann equations}.  \jt{Phys. Rev. E}  \bvol{64},  \pg{031203}.

\bibitem[Denniston {\em et~al.\/}(2001)Denniston, Orlandini \&
  Yeomans]{Denniston01}
{\sc \au{Denniston, C.}, \au{Orlandini, E.} \& \au{Yeomans, J.~M.}} \yr{2001}
  \at{Lattice {B}oltzmann simulations of liquid crystal hydrodynamics}.
  \jt{Phys. Rev. E}  \bvol{63},  \pg{056702}.

\bibitem[Dimova {\em et~al.\/}(2006)Dimova, Aranda, Bezlyepkina, Nikolov, Riske
  \& Lipowsky]{Dimova06}
{\sc \au{Dimova, R.}, \au{Aranda, S.}, \au{Bezlyepkina, N.}, \au{Nikolov, V.},
  \au{Riske, K.~A.} \& \au{Lipowsky, R.}} \yr{2006}  \at{A practical guide to
  giant vesicles. {P}robing the membrane nanoregime via optical microscopy}.
  \jt{J. Phys. Condens. Matter}  \bvol{18},  \pg{S1151}.

\bibitem[Ellis {\em et~al.\/}(2012)Ellis, Maartens \& Mac-Callum]{Ellis12}
{\sc \au{Ellis, G. F.~R.}, \au{Maartens, R.} \& \au{Mac-Callum, M. A.~H.}}
  \yr{2012} {\em Relativistic Cosmology\/}.  \publ{Cambridge University Press}.

\bibitem[Fonda {\em et~al.\/}(2018)Fonda, Rinaldin, Kraft \& Giomi]{Fonda2018}
{\sc \au{Fonda, P.}, \au{Rinaldin, M.}, \au{Kraft, D.~J.} \& \au{Giomi, L.}}
  \yr{2018}  \at{Interface geometry of binary mixtures on curved substrates}.
  \jt{Phys. Rev. E}  \bvol{98},  \pg{032801}.

\bibitem[Gera \& Salac(2017)]{Gera17}
{\sc \au{Gera, P.} \& \au{Salac, D.}} \yr{2017}  \at{Stochastic phase
  segregation on surfaces}.  \jt{R. Soc. Open Sci.}  \bvol{4},  \pg{170472}.

\bibitem[Giordanelli {\em et~al.\/}(2018)Giordanelli, Mendoza \&
  Herrmann]{Giordanelli18}
{\sc \au{Giordanelli, I.}, \au{Mendoza, M.} \& \au{Herrmann, H.~J.}} \yr{2018}
  \at{Modelling electron-phonon interactions in graphene with curved space
  hydrodynamics}.  \jt{Sci. Rep.}  \bvol{8},  \pg{12545}.

\bibitem[Gunstensen {\em et~al.\/}(1991)Gunstensen, Rothman, Zaleski \&
  Zanetti]{Gunstensen91}
{\sc \au{Gunstensen, A.~K.}, \au{Rothman, D.~H.}, \au{Zaleski, S.} \&
  \au{Zanetti, G.}} \yr{1991}  \at{Lattice {B}oltzmann model of immiscible
  fluids}.  \jt{Phys. Rev. A}  \bvol{43},  \pg{4320}.

\bibitem[Gupta {\em et~al.\/}(2015)Gupta, Sbragaglia \& Scagliarini]{Gupta15}
{\sc \au{Gupta, A.}, \au{Sbragaglia, M.} \& \au{Scagliarini, A.}} \yr{2015}
  \at{Hybrid lattice {B}oltzmann/finite difference simulations of viscoelastic
  multicomponent flows in confined geometries}.  \jt{J. Comput. Phys.}
  \bvol{291},  \pg{177 -- 197}.

\bibitem[Hejranfar {\em et~al.\/}(2017)Hejranfar, Saadat \&
  Taheri]{hejranfar17pre}
{\sc \au{Hejranfar, K.}, \au{Saadat, M.~H.} \& \au{Taheri, S.}} \yr{2017}
  \at{High-order weighted essentially nonoscillatory finite-difference
  formulation of the lattice {B}oltzmann method in generalized curvilinear
  coordinates}.  \jt{Phys. Rev. E}  \bvol{95},  \pg{023314}.

\bibitem[Henkes {\em et~al.\/}(2018)Henkes, Marchetti \& Sknepnek]{Henkes18}
{\sc \au{Henkes, S.}, \au{Marchetti, M.~C.} \& \au{Sknepnek, R.}} \yr{2018}
  \at{Dynamical patterns in nematic active matter on a sphere}.  \jt{Phys. Rev.
  E}  \bvol{97},  \pg{042605}.

\bibitem[Henle \& Levine(2010)]{Henle10}
{\sc \au{Henle, M.~L.} \& \au{Levine, A.~J.}} \yr{2010}  \at{Hydrodynamics in
  curved membranes: the effect of geometry on particulate mobility}.  \jt{Phys.
  Rev. E}  \bvol{81},  \pg{011905}.

\bibitem[Howell(2003)]{Howell03}
{\sc \au{Howell, P.~D.}} \yr{2003}  \at{Surface-tension-driven flow on a moving
  curved surface}.  \jt{J. Eng. Math.}  \bvol{45},  \pg{283--308}.

\bibitem[Hu {\em et~al.\/}(2011)Hu, Weikl \& Lipowsky]{Hu11}
{\sc \au{Hu, J.}, \au{Weikl, T.} \& \au{Lipowsky, R.}} \yr{2011}  \at{Vesicles
  with multiple membrane domains}.  \jt{Soft Matter}  \bvol{7},
  \pg{6092--6102}.

\bibitem[Janssen {\em et~al.\/}(2017)Janssen, Kaiser \& L\"{o}wen]{Janssen17}
{\sc \au{Janssen, L. M.~C.}, \au{Kaiser, A.} \& \au{L\"{o}wen, H.}} \yr{2017}
  \at{Aging and rejuvenation of active matter under topological constraints}.
  \jt{Sci. Rep.}  \bvol{7},  \pg{5667}.

\bibitem[Jeong \& Kim(2015)]{Jeong15}
{\sc \au{Jeong, D.} \& \au{Kim, J.}} \yr{2015}  \at{Microphase separation
  patterns in diblock copolymers on curved surfaces using a nonlocal
  {C}ahn-{H}illiard equation}.  \jt{Eur. Phys. J. E}  \bvol{38},  \pg{117}.

\bibitem[J\"ulicher \& Lipowsky(1996)]{Julicher96}
{\sc \au{J\"ulicher, F.} \& \au{Lipowsky, R.}} \yr{1996}  \at{Shape
  transformations of vesicles with intramembrane domains}.  \jt{Phys. Rev. E}
  \bvol{53},  \pg{2670--2683}.

\bibitem[Keber {\em et~al.\/}(2014)Keber, Loiseau, Sanchez, DeCamp, Giomi,
  Bowick, Marchetti, Dogic \& Bausch]{Keber14}
{\sc \au{Keber, F.~C.}, \au{Loiseau, E.}, \au{Sanchez, T.}, \au{DeCamp, S.~J.},
  \au{Giomi, L.}, \au{Bowick, M.~J.}, \au{Marchetti, M.~C.}, \au{Dogic, Z.} \&
  \au{Bausch, A.~R.}} \yr{2014}  \at{Topology and dynamics of active nematic
  vesicles}.  \jt{Science}  \bvol{345},  \pg{1135}.

\bibitem[Kendon {\em et~al.\/}(2001)Kendon, Cates, Pagonabarraga, Desplat \&
  Bladon]{Kendon01}
{\sc \au{Kendon, V.~M.}, \au{Cates, M.~E.}, \au{Pagonabarraga, I.},
  \au{Desplat, J.~C.} \& \au{Bladon, P.}} \yr{2001}  \at{Inertial effects in
  three-dimensional spinodal decomposition of a symmetric binary fluid mixture:
  a lattice {B}oltzmann study}.  \jt{J. Fluid Mech.}  \bvol{440},
  \pg{147--203}.

\bibitem[Kr\"{u}ger {\em et~al.\/}(2017)Kr\"{u}ger, Kusumaatmaja, Kuzmin,
  Shardt, Silva \& Viggen]{KrugerBook}
{\sc \au{Kr\"{u}ger, T.}, \au{Kusumaatmaja, H.}, \au{Kuzmin, A.}, \au{Shardt,
  O.}, \au{Silva, G.} \& \au{Viggen, E.~M.}} \yr{2017} {\em Lattice Boltzmann
  Method: Principles and Practice\/}.  \publ{Springer}.

\bibitem[Li \& Wagner(2007)]{li2007symmetric}
{\sc \au{Li, Q.} \& \au{Wagner, A.~J.}} \yr{2007}  \at{Symmetric
  free-energy-based multicomponent lattice {B}oltzmann method}.  \jt{Phys. Rev.
  E}  \bvol{76},  \pg{036701}.

\bibitem[Liang {\em et~al.\/}(2016)Liang, Shi \& Chai]{Liang16}
{\sc \au{Liang, H.}, \au{Shi, B.~C.} \& \au{Chai, Z.~H.}} \yr{2016}
  \at{Lattice {B}oltzmann modeling of three-phase incompressible flows}.
  \jt{Phys. Rev. E}  \bvol{93},  \pg{013308}.

\bibitem[Liu {\em et~al.\/}(2016)Liu, Kang, Leonardi, Schmieschek, Narv\'{a}ez,
  Jones, Williams, Valocchi \& Harting]{Liu16}
{\sc \au{Liu, H.}, \au{Kang, Q.}, \au{Leonardi, C.~R.}, \au{Schmieschek, S.},
  \au{Narv\'{a}ez, A.}, \au{Jones, B.~D.}, \au{Williams, J.~R.}, \au{Valocchi,
  A.~J.} \& \au{Harting, J.}} \yr{2016}  \at{Multiphase lattice {B}oltzmann
  simulations for porous media applications: {A} review}.  \jt{Comput.
  Geosciences}  \bvol{20},  \pg{777}.

\bibitem[Liu {\em et~al.\/}(2015)Liu, Andrew, Li, Yeomans \& Wang]{Liu2015}
{\sc \au{Liu, Y.}, \au{Andrew, M.}, \au{Li, J.}, \au{Yeomans, J.~M.} \&
  \au{Wang, Z.}} \yr{2015}  \at{Symmetry breaking in drop bouncing on curved
  surfaces}.  \jt{Nature Comm.}  \bvol{6},  \pg{10034}.

\bibitem[Malaspinas {\em et~al.\/}(2010)Malaspinas, Fi\'etier \&
  Deville]{Malaspinas10}
{\sc \au{Malaspinas, O.}, \au{Fi\'etier, N.} \& \au{Deville, M.}} \yr{2010}
  \at{Lattice {B}oltzmann method for the simulation of viscoelastic fluid
  flows}.  \jt{J. Nonnewton. Fluid Mech.}  \bvol{165},  \pg{1637 -- 1653}.

\bibitem[Marenduzzo \& Orlandini(2013)]{Marenduzzo13}
{\sc \au{Marenduzzo, D.} \& \au{Orlandini, E.}} \yr{2013}  \at{Phase separation
  dynamics on curved surfaces}.  \jt{Soft Matter}  \bvol{9},  \pg{1178--1187}.

\bibitem[Mart\'i {\em et~al.\/}(2015)Mart\'i, Blandford \& Rees]{Marti15}
{\sc \au{Mart\'i, J.~M.}, \au{Blandford, R.~D.} \& \au{Rees, M.~J.}} \yr{2015}
  \at{Grid-based methods in relativistic hydrodynamics and
  magnetohydrodynamics}.  \jt{Living Rev. Comput. Astrophys.}  \bvol{1},
  \pg{3}.

\bibitem[McMahon \& Gallop(2005)]{McMahon05}
{\sc \au{McMahon, H.~T.} \& \au{Gallop, J.~L.}} \yr{2005}  \at{Membrane
  curvature and mechanisms of dynamic cell membrane remodelling}.  \jt{Nature}
  \bvol{438},  \pg{590--596}.

\bibitem[Mendoza {\em et~al.\/}(2013)Mendoza, Succi \& Herrmann]{Mendoza13}
{\sc \au{Mendoza, M.}, \au{Succi, S.} \& \au{Herrmann, H.~J.}} \yr{2013}
  \at{Flow through randomly curved manifolds}.  \jt{Sci. Rep.}  \bvol{3},
  \pg{3106}.

\bibitem[Nitschke {\em et~al.\/}(2012)Nitschke, Voigt \& Wensch]{Nitschke12}
{\sc \au{Nitschke, I.}, \au{Voigt, A.} \& \au{Wensch, J.}} \yr{2012}  \at{A
  finite element approach to incompressible two-phase flow on manifolds}.
  \jt{J. Fluid Mech.}  \bvol{708},  \pg{418–438}.

\bibitem[den Otter \& Shkulipa(2007)]{Otter07}
{\sc \au{den Otter, W.K.} \& \au{Shkulipa, S.A.}} \yr{2007}  \at{Intermonolayer
  friction and surface shear viscosity of lipid bilayer membranes}.
  \jt{Biophys. J.}  \bvol{93},  \pg{423 -- 433}.

\bibitem[Paillusson {\em et~al.\/}(2016)Paillusson, Pennington \&
  Kusumaatmaja]{Paillusson17}
{\sc \au{Paillusson, F.}, \au{Pennington, M.~R.} \& \au{Kusumaatmaja, H.}}
  \yr{2016}  \at{Phase separation on bicontinuous cubic membranes: Symmetry
  breaking, re-entrant and domain facetting}.  \jt{Phys. Rev. Lett.}
  \bvol{117},  \pg{058101}.

\bibitem[Parthasarathy {\em et~al.\/}(2006)Parthasarathy, Yu \&
  Groves]{Groves06}
{\sc \au{Parthasarathy, R.}, \au{Yu, C.-H.} \& \au{Groves, J.~T.}} \yr{2006}
  \at{Curvature-modulated phase separation in lipid bilayer membranes}.
  \jt{Langmuir}  \bvol{22},  \pg{5095--5099}.

\bibitem[Pontani {\em et~al.\/}(2013)Pontani, Haase, Raczkowska \&
  Brujic]{Pontani13}
{\sc \au{Pontani, L.-L.}, \au{Haase, M.~F.}, \au{Raczkowska, I.} \& \au{Brujic,
  J.}} \yr{2013}  \at{Immiscible lipids control the morphology of patchy
  emulsions}.  \jt{Soft Matter}  \bvol{9},  \pg{7150--7157}.

\bibitem[Ridl \& Wagner(2018)]{RidlWagner2018}
{\sc \au{Ridl, Kent~S.} \& \au{Wagner, Alexander~J.}} \yr{2018}  \at{Lattice
  boltzmann simulation of mixtures with multicomponent van der waals equation
  of state}.  \jt{Phys. Rev. E}  \bvol{98},  \pg{043305}.

\bibitem[Sadullah {\em et~al.\/}(2018)Sadullah, Semprebon \&
  Kusumaatmaja]{Sadullah18}
{\sc \au{Sadullah, M.~S.}, \au{Semprebon, C.} \& \au{Kusumaatmaja, H.}}
  \yr{2018}  \at{Drop dynamics on liquid-infused surfaces: {T}he role of the
  lubricant ridge}.  \jt{Langmuir}  \bvol{34},  \pg{8112--8118}.

\bibitem[Schwartz \& Weidner(1995)]{Schwartz95}
{\sc \au{Schwartz, L.~M.} \& \au{Weidner, D.~E.}} \yr{1995}  \at{Modeling of
  coating flows on curved surfaces}.  \jt{J. Eng. Math.}  \bvol{29},
  \pg{91--103}.

\bibitem[Seddon \& Templer(1995)]{Seddon95}
{\sc \au{Seddon, J.~M.} \& \au{Templer, R.~H.}} \yr{1995}  \at{{Polymorphism of
  Lipid-Water Systems}}.  \bt{In {\em Structure and dynamics of membranes: From
  cells to vesicles\/} (ed. \ed{R.~Lipowsky \& E.~Sackmann})},  \st{Handbook of
  Biological Physics},  \vol{vol.~1},  \pg{p. Chapter 3}.  \publ{Elsevier
  Science Publishers B.V.}

\bibitem[Semprebon {\em et~al.\/}(2016)Semprebon, Kr\"{u}ger \&
  Kusumaatmaja]{Semprebon16}
{\sc \au{Semprebon, C.}, \au{Kr\"{u}ger, T.} \& \au{Kusumaatmaja, H.}}
  \yr{2016}  \at{A ternary free energy lattice boltzmann model with tunable
  surface tensions and contact angles}.  \jt{Phys. Rev. E}  \bvol{93},
  \pg{033305}.

\bibitem[Shan \& Chen(1993)]{Shan93}
{\sc \au{Shan, X.} \& \au{Chen, H.}} \yr{1993}  \at{Lattice {B}oltzmann model
  for simulating flows with multiple phases and components}.  \jt{Phys. Rev. E}
   \bvol{47},  \pg{1815}.

\bibitem[Sofonea {\em et~al.\/}(2018)Sofonea, Biciu\cb{s}c\u{a}, Busuioc,
  Ambru\cb{s}, Gonnella \& Lamura]{sofonea18pre}
{\sc \au{Sofonea, V.}, \au{Biciu\cb{s}c\u{a}, T.}, \au{Busuioc, S.},
  \au{Ambru\cb{s}, V.~E.}, \au{Gonnella, G.} \& \au{Lamura, A.}} \yr{2018}
  \at{Corner-transport-upwind lattice {B}oltzmann model for bubble cavitation}.
   \jt{Phys. Rev. E}  \bvol{97},  \pg{023309}.

\bibitem[Spencer \& Care(2006)]{Spencer06}
{\sc \au{Spencer, T.~J.} \& \au{Care, C.~M.}} \yr{2006}  \at{Lattice
  {B}oltzmann scheme for modeling liquid-crystal dynamics: {Z}enithal bistable
  device in the presence of defect motion}.  \jt{Phys. Rev. E}  \bvol{74},
  \pg{061708}.

\bibitem[Succi(2001)]{Succi01}
{\sc \au{Succi, S.}} \yr{2001} {\em The Lattice Boltzmann Equation: For Fluid
  Dynamics and Beyond\/}.  \publ{OUP}.

\bibitem[Swift {\em et~al.\/}(1996)Swift, Orlandini, Osborn \&
  Yeomans]{Swift96}
{\sc \au{Swift, M.~R.}, \au{Orlandini, E.}, \au{Osborn, W.~R.} \& \au{Yeomans,
  J.~M.}} \yr{1996}  \at{Lattice {B}oltzmann simulations of liquid-gas and
  binary fluid systems}.  \jt{Phys. Rev. E}  \bvol{54},  \pg{5041}.

\bibitem[Varagnolo {\em et~al.\/}(2013)Varagnolo, Ferraro, Fantinel, Pierno,
  Mistura, Amati, Biferale \& Sbragaglia]{Varagnolo2013}
{\sc \au{Varagnolo, S.}, \au{Ferraro, D.}, \au{Fantinel, P.}, \au{Pierno, M.},
  \au{Mistura, G.}, \au{Amati, G.}, \au{Biferale, L.} \& \au{Sbragaglia, M.}}
  \yr{2013}  \at{Stick-slip sliding of water drops on chemically heterogeneous
  surfaces}.  \jt{Phys. Rev. Lett.}  \bvol{111},  \pg{066101}.

\bibitem[Wagner \& Cates(2001)]{wagner2001phase}
{\sc \au{Wagner, A.~J.} \& \au{Cates, M.~E.}} \yr{2001}  \at{Phase ordering of
  two-dimensional symmetric binary fluids: {A} droplet scaling state}.
  \jt{Europhys. Lett.}  \bvol{56},  \pg{556}.

\bibitem[Wagner \& Yeomans(1997)]{Wagner97}
{\sc \au{Wagner, A.~J.} \& \au{Yeomans, J.~M.}} \yr{1997}  \at{Breakdown of
  scale invariance in the coarsening of phase-separating binary fluids}.
  \jt{Phys. Rev. Lett.}  \bvol{80},  \pg{1429}.

\bibitem[W\"{o}rhwag {\em et~al.\/}(2018)W\"{o}rhwag, Semprebon, Moqaddam,
  Karlin \& Kusumaatmaja]{Wohrwag18}
{\sc \au{W\"{o}rhwag, M.}, \au{Semprebon, C.}, \au{Moqaddam, A.~M.},
  \au{Karlin, I.} \& \au{Kusumaatmaja, H.}} \yr{2018}  \at{Ternary free-energy
  entropic lattice {B}oltzmann model with high density ratio}.  \jt{Phys. Rev.
  Lett.}  \bvol{120},  \pg{234501}.

\end{thebibliography}

\end{document}